\documentclass[sigconf]{acmart} 

\AtBeginDocument{
  \providecommand\BibTeX{{
    \normalfont B\kern-0.5em{\scshape i\kern-0.15em b}\kern-0.8em\TeX
    
    }}}

\usepackage{tabularx}
\usepackage{caption}

 \author{Jordan Taylor*}
\email{jordant@andrew.cmu.edu}
 \affiliation{
   \institution{Carnegie Mellon University}
  \city{Pittsburgh}
  \state{PA}
 \country{USA}
 }

\author{Ellen Simpson*}
\orcid{0000-0003-0387-7329}
\email{ellen.simpson@colorado.edu}
 \affiliation{
   \institution{University of Colorado Boulder}
  \city{Boulder}
  \state{CO}
 \country{USA}
 }

 \author{Anh-Ton Tran*}
\email{anhton@gatech.edu}
 \affiliation{
   \institution{Georgia Institute of Technology}
  \city{Atlanta}
  \state{GA}
 \country{USA}
 }

 \author{Jed Brubaker}
\orcid{0000-0003-4826-8324}
\email{jed.brubaker@colorado.edu}
 \affiliation{
   \institution{University of Colorado Boulder}
  \city{Boulder}
  \state{CO}
 \country{USA}
 }

 \author{Sarah Fox}
\email{sarahf@andrew.cmu.edu}
 \affiliation{
   \institution{Carnegie Mellon University}
  \city{Pittsburgh}
  \state{PA}
 \country{USA}
 }

 \author{Haiyi Zhu}
\email{haiyiz@andrew.cmu.edu}
 \affiliation{
   \institution{Carnegie Mellon University}
  \city{Pittsburgh}
  \state{PA}
 \country{USA}
 }

\copyrightyear{2024}
\acmYear{2024}
\setcopyright{rightsretained}
\acmConference[CHI '24]{Proceedings of the CHI Conference on Human Factors in Computing Systems}{May 11--16, 2024}{Honolulu, HI, USA}
\acmBooktitle{Proceedings of the CHI Conference on Human Factors in Computing Systems (CHI '24), May 11--16, 2024, Honolulu, HI, USA}
\acmDOI{10.1145/3613904.3642494}
\acmISBN{979-8-4007-0330-0/24/05}

\usepackage{multirow}
\begin{document}

\title{Cruising Queer HCI on the DL: A Literature Review of LGBTQ+ People in HCI}

\begin{abstract}

LGBTQ+ people have received increased attention in HCI research, paralleling a greater emphasis on social justice in recent years. However, there has not been a systematic review of how LGBTQ+ people are researched or discussed in HCI. In this work, we review all research mentioning LGBTQ+ people across the HCI venues of CHI, CSCW, DIS, and TOCHI. Since 2014, we find a linear growth in the number of papers substantially about LGBTQ+ people and an exponential increase in the number of mentions. Research \textit{about} LGBTQ+ people tends to center experiences of being politicized, outside the norm, stigmatized, or highly vulnerable. LGBTQ+ people are typically \textit{mentioned} as a marginalized group or an area of future research. We identify gaps and opportunities for (1) research about and (2) the discussion of LGBTQ+ in HCI and provide a dataset to facilitate future Queer HCI research.

\end{abstract}



\maketitle

\section{Introduction}

Lesbian, gay, bisexual, transgender, and queer (LGBTQ+) people have received increased attention within the Human-Computer Interaction (HCI) research community as a part of the third-wave HCI paradigm \cite{harrison2007three}. This research attention parallels an expansion of LGBTQ+ people's civil rights over the past two decades, such as legalization of same-gender marriage by dozens of countries worldwide. At the same time, LGBTQ+ people still experience significant marginalization and legal discrimination.\footnote{https://www.cfr.org/article/changing-landscape-global-lgbtq-rights} Not only are LGBTQ+ people marginalized in society, but they can also be marginalized by technology design. For example, Facebook's "real name" policy harms LGBTQ+ people who use different names than those listed on government documents \cite{haimson2016constructing}. 

In recent years, "Queer HCI" has emerged as a loose contingent of HCI researchers. Prior Special Interest Groups (SIGs) at the ACM CHI conference in 2019 \cite{spiel2019queer}, 2020 \cite{devito2020queer}, and 2021 \cite{devito2021queer} have defined Queer HCI as roughly composed of three overlapping groups: (1) queer researchers regardless of what they research, (2) those researching queer people, and (3) those leveraging queer theory (e.g., playful or subversive interaction design \cite{light2011hci}). This paper focuses on one aspect of Queer HCI: research about LGBTQ+ people. Prior Queer HCI SIGs noted that there is "much ground to be covered" in research on queer populations \cite{devito2020queer, devito2021queer} and that "prior studies have largely focused on gay men ... while the queer community is, in fact, a highly diverse and heterogenous group" \cite{spiel2019queer}. Despite these observations, there has been neither a review of research involving LGBTQ+ people nor an examination of how this work has evolved. Furthermore, there has not been an examination of how HCI researchers broadly represent and discuss LGBTQ+ people. We set out to explore this topic, asking the following research questions:

\begin{itemize}
    \item[\textbf{RQ1:}] How have LGBTQ+ people been \textit{researched} in HCI?
    \item[\textbf{RQ2:}] How have LGBTQ+ people been \textit{discussed} in HCI?
    \item[\textbf{RQ3:}] How has the research about and the discussion of LGBTQ+ people in HCI changed over time? 
\end{itemize}

\noindent RQ1 seeks to understand research involving the experiences of LGBTQ+ people. To understand "LGBTQ+ people in HCI" as a particular subject in our discipline, we must consider not only research ostensibly "about" queer people but also the way HCI researchers generally talk about queer people. Therefore, RQ2 focuses on understanding how LGBTQ+ people are discussed in HCI research more generally, such as how research frames LGBTQ+ identities or rights. Finally, RQ3 addresses the temporal nature of this inquiry, looking at HCI research involving LGBTQ+ people over time while acknowledging that HCI research is situated in particular times and places \cite{haraway1988situated}.

To address these questions, we reviewed 1148 HCI publications from the ACM publication venues of TOCHI, CHI, DIS, and CSCW containing keywords related to LGBTQ+ identity from the inception of these venues through 2022. Organizing this corpus was a two-step process. First, we inductively developed a codebook that characterizes the degree to which a paper relates to LGBTQ+ people, partitioning our corpus into four subsets. Second, we analyzed these subsets using analyses derived from a Grounded Theory Literature Review (GTLR) approach \cite{wolfswinkel2013using}. We identified five genres of HCI research related to queer people. Research \textit{about} LGBTQ+ people in HCI tends to center experiences of LGBTQ+ identities and/or rights being politicized, outside the norm, stigmatized, or highly vulnerable. Additionally, we observe a small, but growing genre, of LGBTQ+ community-centered research and that HCI researchers are increasingly discussing LGBTQ+ people as a marginalized group and an area for future research. In our discussion, we present several provocations for the Queer HCI community and implications for the broader HCI community. 

This paper makes the following contributions:

\begin{itemize}
    \item \textbf{Methodology:} An approach for analyzing how a group of people or a topic is discussed in HCI at scale.
    
    \item \textbf{Data Set:} We present a public dataset of 1,148 HCI papers that discuss LGBTQ+ people, and invite other researchers to review, question, and queer our findings. \footnote{https://github.com/jtaylor351/queer\_hci\_slr}
    
    \item \textbf{Analysis of LGBTQ+ People in HCI:} We present a timeline of notable moments in Queer HCI from 1997-2022 and observations of trends in HCI researchers' discussion and study of LGBTQ+ people and issues over 26 years. 
    
    \item \textbf{Recommendations:} We provide a series of provocations and recommendations to the HCI community about how to (1) discuss LGBTQ+ people or issues and (2) do research with LGBTQ+ people.
\end{itemize}

We present the paper as follows: First, we briefly discuss language and some background on Queer HCI research. We then discuss our methodology. Next, we provide research context for Queer HCI in the form of a timeline and brief explanation of key data points. We then share our observations of how HCI researchers generally discuss queer people and identify five genres of queer-focused HCI research. Finally, we offer a discussion in the form of provocations for Queer HCI and recommendations for the broader HCI community. 

\section{Background \& A Note On Language}

For the purposes of this paper, "queer" is an umbrella term for people who are not cisgender — meaning identifying as the gender assigned at birth — and/or heterosexual. As a reclaimed slur, queer has problematic connotations for some, but for others, it is a rallying cry \cite{khayatt2002toward}. Some people subsumed under this umbrella, such as trans people, may not identify as "queer" despite the academic community labeling them as such. Like most things involving gender and sexuality, queerness is messy \cite{halberstam2011queer}. 

While gay, lesbian, or bisexual identities may come to mind when thinking about queer people, these are far from the only experiences captured by the umbrella "queer" or the acronym "LGBTQ+." "Queer" can include gay men, lesbian women, bi- and pan-sexual people, those with fluid sexualities, and people who experience no sexual (asexual) or romantic (aromantic) attraction, or take time to become attracted to people (gray-ace or demisexual people). Similarly, myriad gender experiences and expressions are captured under the label "queer" that go beyond the "transgender" represented in LGBTQ+. One could be non-binary, agender, trans masc, or trans femme. One may use the label transgender or use terms like MTF or FTM.\footnote{Male-to-Female; Female-to-Male} Moreover, indigenous understandings of gender exist outside western classification schemes, such as hijra people in South Asia or two-spirit people in North America. These lists of sexualities and genders are not exhaustive, and there may be intersections between them. For instance, non-binary lesbians and asexual people have identities that reach across multiple different minority genders and sexualities within the label of "queer." The language used to describe gender and sexual orientations changes over time. In fact, as Foucault notes, the notion that one even \textit{has} a sexual orientation is a relatively recent phenomenon \cite{foucault1990history}.

Defining queerness was also fraught at the Queer HCI SIGs, a conundrum emerging as two proposed SIGs - one on queer theory and one on queer people were asked to merge.\footnote{We know this because several authors have been involved in Queer HCI events. Throughout the paper, we note moments where we use our situated knowledge to provide important context to our findings, particularly on issues of chronology and contextual historical events in Queer HCI research and world events, such as this comment. In this case, the fourth author was involved in the planning of one of these original Queer HCI SIGs.} This resulted in several definitions of queerness and queer identity being included throughout the margins of the 2019 SIG's extended abstract \cite{spiel2019queer}. Even while writing this paper, the queer people on our research team disagreed on terminology owing to our positionalities and backgrounds, such as generational differences as the research team includes Gen-Z, Millennial, and Gen X members. We do not seek to provide guidance or settle this naming issue, nor do we believe one can or should. As a result of this linguistic unsettledness, throughout our work we refer to LGBTQ+ people using the terms referenced in the paper under discussion (e.g., LGBTQ+, LGBT, queer, gender \& sexual minority) while using more specific language to describe research about particular sub-population in the queer community (e.g., transgender, non-binary, bisexual). In our discussion and introduction, we use LGBTQ+ and queer interchangeably. 

\begin{figure*}
    \centering
    \includegraphics[scale=0.7]{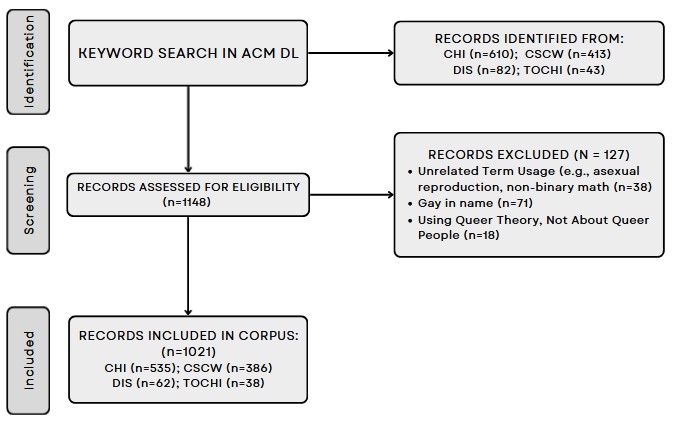}
    \caption{PRISMA Informed Process Flow demonstrating how we searched for, screened, and selected for inclusion the papers in our corpus.}
    \label{fig:flow}
    \Description[A PRISMA process flow diagram]{Figure depicting a PRISMA-Informed Process Flow in three stages: Identification, Screening, and final count of Inclusion in the corpus.}
\end{figure*}

While the Queer HCI SIGs make clear that the Queer HCI community is composed of those who study queer people and leverage queer theory, as well as researchers who happen to be queer \cite{spiel2019queer}, in this work we focus specifically on understanding representations of LGBTQ+ People in HCI research. To fully understand these representations, we study \textbf{\textit{all}} mentions of LGBTQ+ people in our chosen HCI venues, ranging from passing comments to research exclusively about LGBTQ+ people, placing them in a historical context. In the next section, we detail our methodology.

\section{Methods}
Our review focuses on HCI scholarship that was (1) published through 2022, (2) published at the most related SIGCHI venues, and (3) archived in the ACM Digital Library (DL). To construct our corpus, we followed a PRISMA-informed approach\footnote{The PRISMA allowed us to create a diagram demonstrating the research journey from development of search terms to final papers included in the review corpus \cite{shamseer2015preferred}.} to identify, screen, and include articles as this approach is helpful for planning a large literature review \cite{shamseer2015preferred}. To further organize the corpus, we inductively developed a codebook to describe the degree to which a paper concerns LGBTQ+ people and proceeded to code the entire corpus. We used this codebook to partition our data into four subsets. Afterwards, we conducted a grounded theory literature review (GTLR) inspired analysis \cite{wolfswinkel2013using} to explore patterns in how LGBTQ+ people are discussed in HCI research, which we detail below.

                    \begin{table}
    \centering

        \begin{tabular}{|c|c|c|}
            \hline queer & aromantic & gender non* \\ \hline
            lgbt* & sex with men & agender \\ \hline
            glbt & sex with women & gender fluid \\ \hline
            lesbian & women loving women & genderqueer \\ \hline
            gay & men loving men & gender minority \\ \hline
            bisexual & sexual minority & transgender \\ \hline
            pansexual & hijra & two-spirit \\ \hline
            asexual & intersex & non-binary \\ \hline
        \end{tabular}
        \captionsetup{justification=centering}
        \caption{Terms used to search the ACM Digital Library}
        \Description[A table with search terms]{A table with terms used to search the ACM Digital Library.}
        \label{tab:query_words}
          
    \end{table}

    \subsection{Identification}

    To identify relevant papers, we adopted a PRISMA-informed approach \cite{shamseer2015preferred}. We collectively developed a set of search terms informed by our wide-ranging experiences as LGBTQ+ people and generalized knowledge of research about LGBTQ+ people in HCI. The research team initially met on Zoom and brainstormed a list of search terms drawn from our collective experiences as queer people. We then searched the entire ACM Digital Library for submissions containing our initial search terms. In order to identify additional search terms, we reviewed a subset of these papers (505 total) whose titles and abstracts clearly indicated they were about LGBTQ+ people or identities. This review resulted in the addition of 10 terms that covered other ways that researchers described queer people (e.g., "women loving women") and populations we initially overlooked (e.g., "hijra"). Table \ref{tab:query_words} shows the final set of search terms.

    We searched the ACM DL for all SIGCHI publications through 2022 using our final set of terms. Given the emerging nature of this area of scholarship, we elected to include pieces published as extended abstracts, such as panels or workshop proposals, as these often reflect evolving practices and norms of the HCI community. We then decided to narrow our inclusion criteria further by limiting our analysis to a subset of venues where queer topics most commonly appear. To select these venues, we calculated the total number of publications at each venue containing our search terms. We then selected the top three: Computer-Human Interaction (CHI), Computer Supported Cooperative Work (CSCW), \footnote{CSCW changed its publication format in 2021, and as a result shows up as multiple venues in the ACM Digital Library. We aggregated CSCW publications into one venue in our analysis.} and Designing Interactive Systems (DIS). We also decided to include the ACM Transactions on Computer-Human Interaction (TOCHI) journal in our study because papers submitted to this journal can be presented at any of these conferences, although we note this is not always the case. The number of publications across each venue containing our search terms totaled 1,148.

    \subsection{Data Screening}
   
    Next, we manually examined all publications to verify their relevance to LGBTQ+ people, again drawing on a PRISMA-informed approach \cite{shamseer2015preferred}. We excluded papers that used search terms in non-applicable ways. For example, some uses of "non-binary" referred to numeric rather than queer concepts like the gender binary or non-binary people (n=37), and some search terms matched a person's name (n=71). 

                        \begin{table}
    \centering
        
            \begin{tabular}{|c|c|c|c|}
                \hline
\textbf{Venue}                                           & \textbf{\begin{tabular}[c]{@{}c@{}}Number\\ of Papers\end{tabular}} & \textbf{Timeframe} & \textbf{\begin{tabular}[c]{@{}c@{}}Year Venue\\ Founded\end{tabular}} \\ \hline
CHI      & 535     & 1986 - 2022        & 1981                \\ \hline
CSCW       & 386          & 2002 - 2022     & 1986         \\ \hline
DIS      & 62        & 2006 - 2022        & 1995            \\ \hline
TOCHI        & 38          & 1999-  2022        & 1994    \\ \hline
\begin{tabular}[c]{@{}c@{}}\textbf{Entire}\\ \textbf{Dataset}\end{tabular} & \textbf{1021}      & \textbf{1986 - 2022 }       & \textbf{--   }    \\ \hline
            \end{tabular}
            \captionsetup{justification=centering}
            \caption{Papers in our corpus by venue}
            \Description[A table papers included in corpus by ACM Venue (N=1021 papers)]{A table of papers included our corpus listed by ACM Venue: CHI (535 papers), CSCW (386 papers), DIS (62 papers), and TOCHI (38 papers)}
            \label{tab:paper_counts}
           
    \end{table}

    We also screened papers that used queer theory or queering as a methodology but otherwise had no direct relation to queer people (n=18). "Queering" as a method emerges out of queer theory and describes the analytic practice of subverting what is "normative."\footnote{Queering as a method has notable overlap with the more familiar concept in HCI of infrastructural inversion \cite{bowker1999sorting}.} However, while queering is a common method in queer literature \cite{sedgwick1993tendencies}, it is not always directly connected to queer people and their experiences. For example, one paper looked at queering input devices by placing a computer mouse in a person's underwear \cite{harrer2019mice}.\footnote{Here we make an obligatory reference to Susan Sontag's \textit{Notes on Camp} \cite{sontag2018notes}.} We removed these papers from our corpus when they did not also address LGBTQ+ experiences. Work on queer theory or queering as a method warrants an in-depth analysis that is beyond the scope of this paper but will be important to address in future work. In total, we excluded 127 papers during this phase. After screening, we had 1021 papers in our corpus for analysis (Table \ref{tab:paper_counts}).

    \begin{table*}[h]
        \centering
        \renewcommand{\arraystretch}{1.5}
                \begin{tabular}{|m{7em}|m{20em}|m{7em}|m{7em}|}
                    \hline \centering  Code for How a Paper Discusses Queer People & \centering Code Definition & \centering Number of Papers & \centering First Publication Year \tabularnewline \hline 
                    \centering Exclusively Involves (4)  & \centering Explicitly or solely about LGBTQ+ people in participant representation or discussion & \centering 73 & \centering 2014 \tabularnewline  \hline
                    \centering  Significantly Involves (3)  & \centering Significantly discusses queer issues or has a significant number of queer participants but does not necessarily center them (e.g., queer issues are one of multiple cases in the paper) &  \centering 108 & \centering 1998 \tabularnewline  \hline
                    \centering  Discusses (2)  & \centering Frames or addresses LGBTQ+ people or issues, but it is not a primary part of the paper (e.g., paper briefly discusses LGBTQ+ people as impacted by topic of study) &  \centering 450 & \centering 1997 \tabularnewline  \hline
                    \centering  Briefly Mentions (1)  & \centering Mentions LGBTQ+ people/issues but would not meet the criteria for other codes (e.g., participant demographics) &  \centering 390 & \centering 1999 \tabularnewline  \hline 
                \end{tabular}
            \caption{Description of the final codes assigned to the papers in our dataset }
            \Description[A table of number of papers in corpus organized by our coding scheme.]{A table of number of papers in corpus organized by our coding scheme: 390 briefly mention, 450 discusses, 108 significantly involves, 73 exclusively involves.}
            \label{tab:paper_data}
    \end{table*}
    
    \subsection{Data Organization}

    We inductively developed a codebook to organize the papers in our corpus. Per our research questions, we sought to differentiate between research about LGBTQ+ people (RQ1) and research merely discussing LGBTQ+ people (RQ2). As we will describe below, this proved to be more complex than a simple binary. To build our codebook, we selected a subset of publications from our corpus (n=100) and closely read each with an eye toward how the scholarship engaged LGBTQ+ people and issues. The research team initially split into separate groups of 2-3 (the first three authors and two research assistants) to code papers based on "whether they included queer people or not" (yes, no, maybe). For example, papers that mentioned "transgender" would be considered an explicit mention. Whereas papers that discussed "gender bias" without referring to a queer gender identity category would be coded as "maybe" as it would require a careful reading to determine its inclusion in the corpus. We then met to discuss similarities and disagreements, which helped us identify different ways that scholarship includes LGBTQ+ people. 

    Iterating through this initial subset surfaced important considerations for analyzing the corpus. We found that the inclusion of queer people ranges on a spectrum rather than the binary we initially anticipated. For example, a late-breaking work on AR/VR identified a unique case study for further research on non-binary embodiment \cite{freeman2015simulating}. This work did not fully include LGBTQ+ people initially but developed into a deep investigation based on Freeman and various colleagues' continued work, which eventually centered LGBTQ+ people \cite{freeman2020my,freeman2021body,freeman2022disturbing}. Our final annotation scheme\footnote{For more information about examples and non-examples for each code, please see Appendix A.} is summarized below: 

    \begin{itemize}
        \item\textbf{4 - Exclusively Involves:} Paper was explicitly or solely about LGBTQ+ people (either in participant representation or discussion)
        
        \item\textbf{3 - Significantly Involves:} Paper significantly discusses queer issues or has a significant number of queer participants but does not necessarily center them (e.g., queer issues are one of multiple cases in the paper).
        
        \item \textbf{2 - Discusses:} Paper frames or addresses LGBTQ+ people or issues, but it is not a primary part of the paper (e.g., paper briefly discusses LGBTQ+ people as impacted by topic of study).
        
        \item \textbf{1 - Briefly Mentions}: Paper mentions LGBTQ+ people/issues but would not meet the criteria for other codes (e.g., participant demographics). 
    \end{itemize}

    Once our codebook was finalized, the first three authors and a research assistant individually coded the corpus in pairs. We achieved a high inter-rater reliability weighted mean $\kappa$ of 0.88 \cite{light1971measures}, indicating strong agreement. This coding took place in independent sessions of about two hours each week over the course of three months. We then held a series of working meetings, usually lasting upwards of 4 hours a session, to reflect on places of high agreement, discuss disagreements, and settle on a final code for each paper. These meetings were also instrumental to our qualitative analysis of the corpus. We noted that our scoring was shaped by our subjectivities, such as when a lesbian team member noted the word 'lesbian' being included in a list of words associated with pornography in a paper, thus scoring the paper a 2 compared to a counterpart who marked it a 1. These sorts of disagreements were productive and helped refine our understandings. These meetings also surfaced conversations that served as early development of genres of the work, such as a pattern of framing queerness as controversial, which we discuss in Section 5 \cite{mcnee2006making}. 

    \subsection{Data Analysis}
    
    After organizing the entire corpus, we proceeded with analysis on three fronts: temporal, quantitative, and qualitative. Firstly, we conducted a descriptive quantitative analysis to identify publishing patterns over time. Getting counts of each coded group (1-4) allowed us to get counts of different LGBTQ+ identities more easily. We produced an early series of tables and visualizations to generate insights and identify patterns. For example, we plotted the corpus subdivided by our codes (1-4) temporally, which allowed us to identify notable milestones in Queer HCI scholarship. Developing this chronology provided clarity on the development of Queer HCI and how researchers shifted in the ways they describe queer people in passing or reference.

    Next, we conducted a qualitative analysis of our corpus, informed by grounded theory literature review (GTLR) \cite{wolfswinkel2013using}. For this analysis, we wanted to distinguish between Queer HCI scholarship (which we define as research specifically about or significantly involving LGBTQ+ people) and how LGBTQ+ people and experiences are represented in HCI broadly. Accordingly, our analysis for Queer HCI scholarship was based on papers coded as "4 - Exclusively Involves" or "3 - Significantly Involves." Our analysis of HCI scholarship generally was based on papers coded as "2 - Discusses" or "1 - Briefly Mentions."

    We took an analytical approach to the papers we read. General HCI scholarship about LGBTQ+ people (n=840), examined \textit{how} LGBTQ+ people were included. While classifying these papers based on the degree to which they discussed LGBTQ+ people, we also took notes on each paper in a shared spreadsheet. We then used these notes from our first pass to bucket these papers into four general and intentionally broad observations over time. We noted how the language HCI researchers use to discuss gender has shifted over time, and how a lack of inclusion of LGBTQ+ people in research increased as this language shifted to be more inclusive of varying gender expressions. We additionally noted how research would often talk about how LGBTQ+ people are marginalized along with other groups of historically marginalized communities (e.g., People of Color, women, etc.), and how there was a period in the earlier part of our corpus where researchers often framed LGBTQ+ identities as undesirable or politically controversial.

    Meanwhile, our analysis of research "3 - significantly" or "4 - exclusively" about LGBTQ+ people (181 papers) takes inspiration from DiSalvo et al.'s description of "research genres" \cite{disalvo2010mapping}. They describe genres as "emergent clusters of research that draw from similar sources, share a general problem formulation, and have similar ideas of how to approach solving those problems" \cite{disalvo2010mapping}. Here, we define genres as shared formulations or patterns in how scholarship engages queerness. To develop our genres, we organized a series of meetings where we engaged in extensive affinity diagramming, identifying shared properties around which papers could be clustered. These meetings (along with their debates) generated extensive notes and visual organizations of the corpus and served as a discursive function to elicit the genres we share in our findings.

                \begin{figure*}
    \centering
    \includegraphics[scale=0.24]{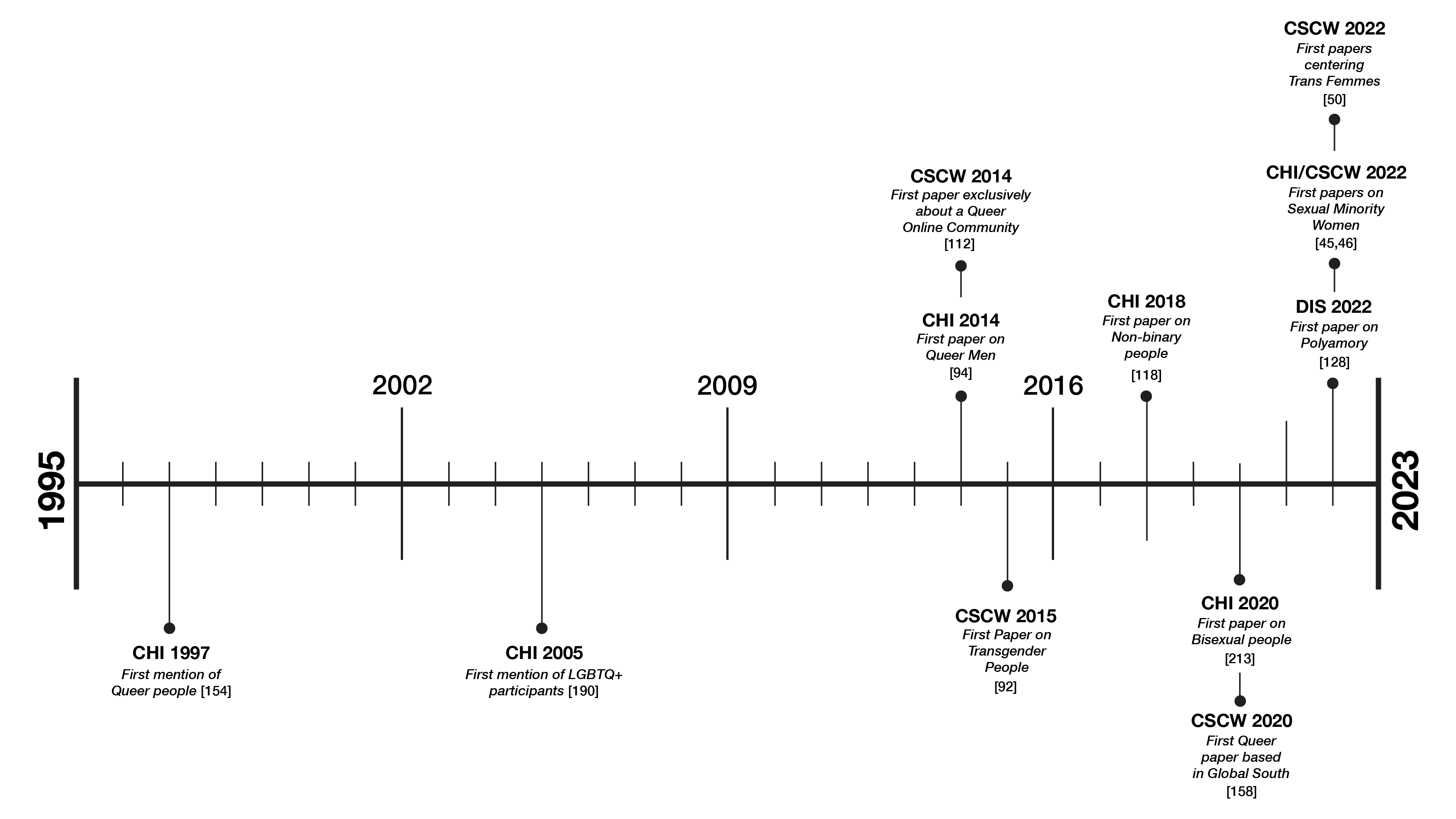}
    \caption{A Timeline of Key Events and Inclusions in HCI}
    \Description[A line key publication events in Queer HCI history.]{A timeline of key publication events in Queer HCI history from 1997 to 2022.}

    \label{fig:timeline}
\end{figure*}

\subsection{Researcher Positionality} Our research team is an inter-generational group of scholars (Gen X - Gen Z) and includes people who identify as gay, lesbian and bisexual. All authors identify as East Asian, Southeast Asian, or white, and we are all located at universities in the United States of America. Our positionalities invariably shape our analysis of queerness and queer identities through a western, academic lens. Moreover, we are speaking from positions of privilege as scholars who can openly speak about our queerness, research about LGBTQ+ people, and foreground queer issues in our field. We hope that this literature review will highlight the gaps in who is and is not accounted for within Queer HCI, which speaks to the gaps of who is and is not included in HCI research broadly. The first three authors of this work wrote the majority of this paper together over the course of two years, each contributing equally.

    \begin{figure*}
        \centering
        \includegraphics[scale=0.5]{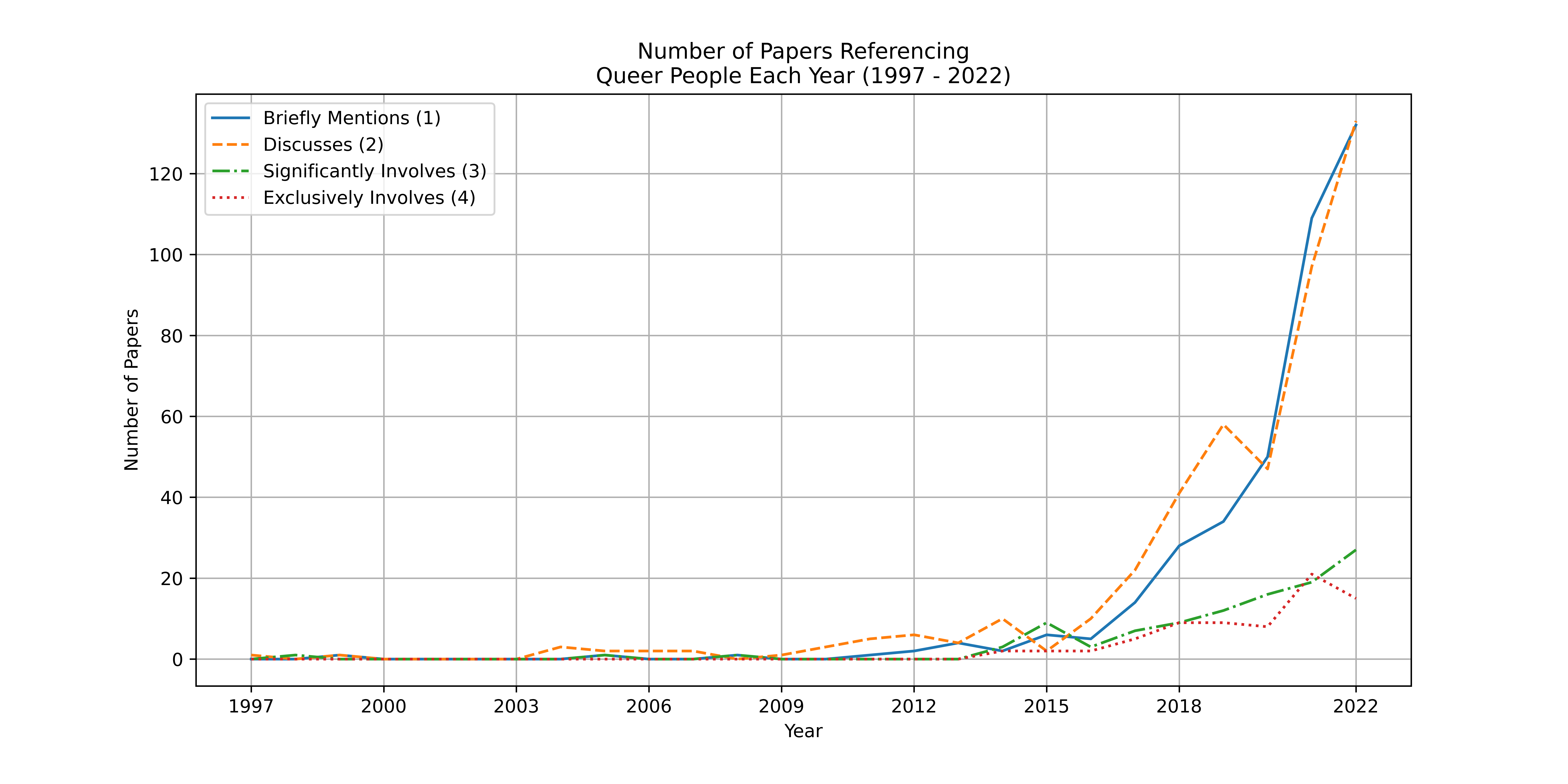}
        \caption{Number of papers in our corpus referencing queer people in each year from 1997 to 2022}
        \Description[A line graph titled "papers referencing queer people over time (1997 - 2022)".]{A line graph titled "papers referencing queer people over time (1997 - 2022)". There are four lines, one for each code (briefly mentions, discusses, significantly involves and exclusively involves). The "briefly mentions" and "discusses" lines grow exponentially starting in 2012 and the "significantly involves" and "exclusively involves" lines grow linearly beginning in 2014.}
        \label{fig:code_1_4}
    \end{figure*}

\section{Growth in LGBTQ+ Representation: A Brief Timeline of Queerness in HCI}

The representation of and research about LGBTQ+ people in HCI started slowly. Fifteen years after SIGCHI was founded in 1982, Muller et al.'s 1997 CHI publication "Toward an HCI Research and Practice Agenda Based on Human Needs and Social Responsibility" \cite{muller1997toward} included the first mention of LGBTQ+ people in our corpus. In the paper, the authors advocate for the importance of empowering marginalized communities in research and design. We found three other publications mentioning LGBTQ+ people in the late 1990s \cite{terveen1999constructing, williamson1998constructing, beeson1999closing}. However, it was not until 2005 that an empirical paper specifically mentioned having LGBTQ+ participants: a gay couple was included in a paper on technology use while relocating \cite{shklovski2005exploring}. It was not until 2014, 32 years after SIGCHI's inception, that the first papers "4 - exclusively" about queer people in our corpus were published \cite{haimson2014ddfseeks, homan2014social}. 

Research "4 - exclusively" about queer people in HCI has grown extensively in recent years (Fig. \ref{fig:code_1_4}). This growth is punctuated by a series of notable events, which we present here chronologically (Fig. \ref{fig:timeline}). The first paper centering the experiences of queer men\footnote{Here we are using queer men to capture the experiences of gay and bisexual men, and men who have sex with other men (MSM).} was published at CHI in 2014 \cite{haimson2014ddfseeks}. The first paper to center the experiences of transgender people was published at CSCW in 2015 \cite{haimson2015disclosure}, with the first paper to focus exclusively on the experiences of transfeminine people\footnote{Participants described their identities as "binary trans women, transfeminine nonbinary, nonbinary trans women and transfeminine" \cite{DeVito2022TransfemTikTok}.} published in 2022 \cite{DeVito2022TransfemTikTok}. The first paper to center the experiences of non-binary people was published at CHI in 2018 \cite{jaroszewski2018genderfluid}. The first paper to center the experiences of bisexual\footnote{Walker and DeVito's \cite{walker_more_2020} paper discusses the experiences of people who are attracted to more than one, or any, gender expression. This includes, but is not limited to, bisexuality, pansexuality and omnisexuality.} people was published at CHI in 2020 \cite{walker_more_2020}. Also in 2020, at CSCW, the first paper on the experiences of queer people living in the Global South was published: Nova et al.'s study of how Hijra in Bangladesh navigate social media ecosystems \cite{Nova2020Understanding}. The first papers about the experiences of sexual minority women\footnote{Here we are using sexual minority women to capture the experiences of lesbians and bisexual women, as well as those who fall under the term sexual minority women. Note, while many trans women may also identify as lesbian or bisexual, not all trans women are sexual minority women.} was published at CHI in 2022 \cite{cui2022so, Cui2022Wegather}. That same year, the first paper centering relationship dynamics beyond monogamy, which explored the breakup of a polyamorous queer couple, was published at DIS \cite{kinnee2022sonic}.

Considering these texts' intellectual and broader socio-political contexts, we approached the findings below with a historicist sensibility \cite{soden_historicism}. When relevant and available, we share this context. While sometimes critical, the goal of this review is not necessarily to critique. We do not — and \textit{should not} —judge decades-old research by best-practices at the time of our writing in 2023. The earliest paradigm of HCI research emphasized human factors and engineering, with the second shifting towards cognitive approaches to HCI. Therefore, it may be unsurprising that the first papers "4 - exclusively" about queer people were not published until 2014, following the rise of viewing HCI and technology as relational in the third-paradigm. This paradigm prioritized examining values brought into design and situating the user \cite{harrison2007three}. These paradigms shape HCI research. What gets published is tied to funding structures and review processes in the field. In comparing HCI to other fields, we find it prudent to note HCI lags behind other academic communities in contemporaneous understandings of gender and sexuality (e.g., the gender binary was questioned in feminist studies \cite{butler1990gender} and anthropology \cite{kulick1998travesti, nanda1990neither} since at least the late 1980s). Additionally, media scholars have studied queer online communities decades prior to HCI \cite{correll1995ethnography, gray2007websites, campbell2005outing}. 
What this tells us is that HCI's gaze on queer folks is predicated upon what the current intellectual interests of the field highlights in terms of theories and conceptualizations of HCI. As other scholars have noted, our field has a history of disjointed appropriation of theory, particularly as it moved into the third paradigm \cite{sengers2006reflective, dourish2004reflective}.

\section{How HCI Discusses LGBTQ+ People}

Here, we present trends identified in HCI research "2 - discussing" or "1 - briefly mentioning" LGBTQ+ people from 1997 to 2022. We present this section before our analysis of papers "3 - significantly" or "4 - exclusively" about LGBTQ+ people because these discussions predate and contextualize subsequent research \textit{about} queer people. By looking at how queer people are discussed in research that is not \textit{about} queer people — over a period of rapid societal changes around LGBTQ+ acceptance and civil rights — we seek to understand how LGBTQ+ people are generally represented in HCI research (RQ2) and how this representation developed over time (RQ3). 

    \subsection{Shifting Participant Demographics}

    Papers that "1 - briefly mention" LGBTQ+ people were most often included in the corpus because they reported some kind of demographic information. Most prevalent were instances of researchers reporting -- at times problematically -- participants not categorized in the male/female binary.

    While early papers in our corpus assumed a male/female binary, researchers are increasingly reporting a third category of gender alongside the "male" and "female" participant counts, such as "non-binary," "non-binary/third gender," "other," or "genderqueer." Indicative of the growing awareness of non-binary gender identities in HCI, we found multiple papers published in the 2020s reporting the absence of non-binary participants in their demographic data (e.g., "non-binary: 0"). This is possibly due to Queer HCI research on gender published in the late 2010s \cite{jaroszewski2018genderfluid, spiel2019better, scheuerman2020GenderGuidelines}. We also noticed a shift away from "male, female, non-binary" trinaries back toward a binary scheme: "male" and "not male." These papers often worked within research contexts that are embodied (e.g., menstruation) or highlight how society is gendered in a binary fashion (e.g., papers about sexism). A handful of papers tried to avoid imposing classification schemes altogether by allowing participants to describe their gender in their own words. Additionally, we note that specific gender schemes are sometimes reproduced through standardized surveys that research labs reuse in different publications, suggesting specific gender schemes may be sticky once chosen.
    
    While these attempts to be more inclusive are promising, we noticed a number of common missteps, such as reporting some participants as an "other", which can be Othering. Some works excluded transgender and/or gender non-conforming participants from the "male" and "female" counts, reporting demographic data such as: 20 male, 15 female, and 1 transgender. This gender scheme marginalizes binary transgender people by reporting them as a separate from "male" or "female." Broadly, we find researchers attempting to be more inclusive in how they discuss gender but, at times, making mistakes. For a more systematic analysis of how gender is reported in HCI over time, we direct readers to Offenwanger and colleagues' systematic review of the topic \cite{offenwanger2021diagnosing}.

    \subsection{Queerness as Political and/or "Bad"}

    In the earlier papers in our corpus, researchers often "2 - discuss" LGBTQ+ identities or rights as socially undesirable or as a controversial political topic. These papers are products of their time -- the late 1990s - mid 2010s. Many were written in U.S. contexts amid dramatic shifts in LGBTQ+ civil rights and societal acceptance of LGBTQ+ people, culminating in the 2015 legalization of "same-sex marriage" \cite{yoshino2015new}. This historical context is important given the uptick, at the time of our writing, in both homophobic and transphobic political rhetoric in the U.S. and elsewhere.

    Some papers discuss being queer as undesirable or controversial. For example, two early Human-AI Interaction papers motivated their projects by referencing a man's "gay panic" over TiVo\footnote{TiVo was a 2000s television recording device that used a recommendation algorithm to preemptively record videos a user may like.} thinking he was gay \cite{mcnee2006making, rao2009my}. Fear of being seen as gay frames LGBTQ+ identity as undesirable, as does other work discussing participants being falsely outed on social media \cite{wyche2013facebook, macbeth2013script, wisniewski2014adolescent}. Several papers also discussed how the word "gay" was often used as a pejorative \cite{lee2010receptionist, rzeszotarski2012learning, wang2014cursing}. While attitudes may be different at the time of our writing, our findings suggest that the 2000s and early 2010s may not have been a particularly gay-friendly or accepting period, which shaped how HCI researchers discussed queerness.\footnote{The members of the writing team who survived the middle and high school in the late 1990s and 2000s can confirm, it was an awful time to be gay.}
    
    Papers also discussed gay civil rights issues, such as marriage, parenting, and serving in the military \cite{kinnaird2010connect, diakopoulos2011towards, grevet2014managing, kriplean2014integrating, gulotta2012curation, farny2012anchor, shami2014understanding}. Setting the context for these works, overwhelmingly based in U.S. research institutions, is essential. Coupled with shared geography was growing interest and opportunity in social computing to study people \textit{in situ} and in real-time, using social media data. Burgeoning CSCW research used quantitative methods to examine and test social theories and phenomena at scale, from leveraging tweets for measuring positive/negative affect based on seasonality \cite{golder2011diurnal} to pulling Yelp reviews to identify linguistic structures in online sentiment \cite{jurafsky2014narrative}. Social computing research capitalized on the newly available APIs to scale research with social data, thus exposing queer issues and events to the purview of CSCW scholars.
    
    Early work on the 2016 U.S. Presidential Election mentioned LGBTQ+ rights as a contentious issue, following the advancement of marriage equality and the repeal of both the U.S. Defense of Marriage Act\footnote{The Defense of Marriage Act (DOMA) was a federal law in the U.S. enacted in 1996 that banned federal recognition of same-sex marriage through the limitation of marriage to one man and one woman. It provided states the ability to refuse to recognize same-sex marriages performed in other states.} and the U.S. Military's Don't Ask Don't Tell policy\footnote{Instituted in 1996, this policy prohibited U.S. Military personnel from discriminating against or harassing closeted or not-publicly out homosexual or bisexual service members. It also barred openly gay, lesbian, or bisexual people from military service.} during the Obama administration between 2008 and 2016 \cite{wang2017engaging, kulshrestha2017quantifying}. This is part of a broader trend we observe of growing attention paid to online political discourse, with LGBTQ+ rights being one of the topics of debate \cite{compton2016theory, wood2018rethinking, costa2018regulating, holzer2018digitally}. Likewise, some research mentioned LGBTQ+ rights in the context of fake news \cite{geeng2020fake} or censorship of materials discussing LGBTQ+ identities by librarians \cite{huang2017human} or governments \cite{bin2017internet} on political grounds. These works speak to the ways queer people are seen as immoral and, therefore, silenced by government internet filters \cite{bin2017internet} and librarians filtering our "sensitive or controversial" book topics from homophobic publics \cite{huang2017human}. We found two papers describing LGBTQ+ Wikipedia articles as controversial \cite{hube2019understanding, kuznetsov2022templates} and another two mentioning LGBTQ+ topics in the context of political ads on Facebook \cite{capozzi2021clandestino, matias2022software}. In one of these papers, the authors discuss both LGBTQ+ and veteran communities gathering advertisements mistakenly removed by Facebook because they were election-related, which was used to motivate an ad audit \cite{matias2022software}. These papers are born of a research context that allowed a particular kind of examination, and reflect a social context where queer existence is viewed as an inherently political topic of debate that wanes in and out of centrality.

    \subsection{Emphasizing Queer Marginalization}
  
    As papers discussing queer rights as controversial decreased around 2016, HCI research started to present LGBTQ+ people as marginalized and, therefore, needing inclusion or support in HCI research. We find that noting the marginalization of queer people increasingly served as a way to motivate HCI research. Put another way, discrimination toward LGBTQ+ people became rhetorically useful for HCI researchers. We see this in how instances of algorithmic discrimination toward LGBTQ+ people are often used to motivate algorithmic fairness research. Some case studies we saw frequently cited included (1) a 2011 report \cite{ananny2011curious} on the Android app store recommending Grindr\footnote{A dating app primarily used by queer men.} alongside an app for finding sex offenders \cite{hamilton2014path, dove2020monsters} and (2) a crowd audit undertaken by LGBTQ+ YouTubers to detect algorithmic bias \cite{devos2022toward, choi2022s}. 

    These papers discuss LGBTQ+ marginalization in various contexts, foreshadowing common themes that would later take center stage in Queer HCI research. Some research on marginalization looks at how queerness breaks down normative assumptions about users embedded in the design of technology \cite{feinberg2014always, brubaker2011select}, such as research on online identity management \cite{farnham2011faceted, odom2015understanding} and gender essentialism in the design of menstrual technology \cite{bardzell2015immodest}. Research also discusses the marginalization of LGBTQ+ people in everyday life, such as police brutality \cite{tseng2021digital} and intimate partner violence \cite{tseng2022care}. Other scholars pointed to marginalization in particular cultural contexts, such as in the Arab World \cite{alabdulqader2018exploring, alabdulqader2019eye}. Some emphasized that LGBTQ+ youth are a particularly marginalized group \cite{jhaver2018online, morag2022tobe, klassen2021more, badillo2021conducting, baglione2021understanding}. We also saw HCI researchers increasingly discussing queer people in conjunction with other axes of marginalization, such as work on street harassment describing LGBTQ+ people as a "traditionally marginalized" group \cite{dimond2013hollaback}. Authors also frequently mention LGBTQ+ marginalization in long lists of other axes of marginalization (e.g., gender, class, ability, race), framing these groups as "historically marginalized populations" \cite{richter2022acu}, "non-dominant groups" \cite{harrington2022all}, or "vulnerable communities" \cite{guberek2018keeping}. This work speaks to growing discussion of "marginalized people" in HCI, but in a manner that can be homogenizing.
    
    \subsection{Becoming a Limitation or Future Work}

    Due to the growing awareness of LGBTQ+ people as a marginalized social group under-considered in technology design, we find HCI researchers increasingly acknowledge that their research may not have considered or apply to LGBTQ+ people. This became more commonplace after 2016 and in particular research contexts. For example, HCI research often gendered as masculine (e.g., boardgames \cite{j2021unpacking} or e-sports \cite{li2020spontaneous}) or feminine (e.g., fertility \cite{costa2021health} or makeup \cite{li2022feels}) routinely describes the omission of LGBTQ+ people as a limitation of their work. Some mentioned using — and having to justify using — datasets or technologies that only include binary genders, such as voice technology \cite{cambre2020choice, song2020mind} and video-game-character-creation tools \cite{kao2022audio, johannes2021assessing, dechant2022don}. A large body of research, while acknowledging the limitations of their methods, used a binary gender measure to explore gender biases or inequity among researchers themselves \cite{early2018understanding, chancellor2019human} and in algorithmic \cite{ali2019discrimination, yee2021image, barlas2021see} and CSCW \cite{vashistha2019threats, dubois2020gender, foong2021understanding} systems. These researchers often included thoughtful justifications for their binary gendering, such as identifying gender biases in how people with eating disorders are described \cite{chancellor2019human}.
    
    Finally, several papers using survey data mentioned insufficient LGBTQ+ participants to draw statistically significant inferences. However, some still decided to go through the motions of analyzing data with minuscule populations, such as a study with only one non-binary participant. In contrast, Seberger et al. earnestly explained their decision to exclude data from their two non-binary respondents from their statistical analysis and, in response, called for "greater attention to the development of methods that can be effective for the inclusion of disproportionately smaller groups in research on privacy and other areas of HCI" \cite{seberger2022still}. Sometimes, a thoughtful limitation can be a meaningful call to action.

\section{Genres of Queer-Focused HCI Research}

In this section, we describe five genres of research "3 - significantly" or "4 - exclusively" about LGBTQ+ people. We find that this research focuses on LGBTQ+ people as (1) political, (2) outside the norm, (3) stigmatized, and (4) high-risk. We also identify a fifth, more nascent, genre of community-centered research. Note, these genres are not mutually exclusive.

    \subsection{Queer People as Political Subjects}

    Some research about queer people focuses on the controversial and highly politicized nature of LGBTQ+ identity. The papers "3 - significantly" about LGBTQ+ people often rely on these aspects of LGBTQ+ people as a case study for understanding social movements by leveraging social APIs and data. Around the time gay marriage was legalized in the U.S. in 2015, researchers used support for gay marriage on social media — via Twitter discourse \cite{zhang2015modeling} and adding an equals sign from the Human Rights Campaign to Facebook profile pictures \cite{state2015diffusion} — to study online social movements. Subsequent research has explored similar topics but uses LGBTQ+ rights as one of several case studies. For example, to understand the role of images in online activism, Cornet et al. studied Instagram posts related to three social movements in the U.S.: Black Lives Matter, Abortion Rights and LGBTQ+ Rights \cite{cornet2017image}. Others looked at the relationship between the inferred U.S. political party affiliation of Twitter users and discourse surrounding various political issues, such as gay rights \cite{le2017revisiting}. More recently, a paper explored direct democracy platforms to support Taiwan legalizing same-sex marriage \cite{bardzell2020join}. In sum, fights for LGBTQ+ rights served as a useful context for those interested in social movements.

    In contrast to the research "3 - significantly" about LGBTQ+ people that chooses to study queer politicization \textit{a priori}, research "4 - exclusively" about LGBTQ+ people empirically encountered the politicization of LGBTQ+ identities in the process of studying other aspects of queer experiences. For example, in their study of LGBT parents' social media experiences, Blackwell et al. find LGBT parents' everyday social media posts were perceived as incidental advocacy work for LGBT family rights during a period when these rights were in flux \cite{blackwell2016lgbt}. Likewise, other research uncovers how simply being visible online can be a form of advocacy and activism. In examining the computer security and privacy experiences of transgender people, Lerner et al., documented how transgender people regularly returned to activism, political organizing, and modeling -- being visible -- trans identity as a part of their everyday social media use \cite{Lerner2020Activism}.

    The genre of Queer HCI that frames LGBTQ+ people as controversial or political subjects is unique in that the research within it is often socially and historically situated, examining unique moments in time and advocacy for LGBTQ+ people's rights. We mark it distinct from research that frames LGBTQ+ people as vulnerable or socially stigmatized as these papers examine the political behavior of collectives in support of and by LGBTQ+ people (e.g., \cite{zhang2015modeling, Lerner2020Activism}) while also acknowledging that LGBTQ+ identity is both controversial and inherently political.

    \subsection{Queer People as Outside the Norm}

    HCI scholars have long critiqued technology researchers and designers' conception of the "user" \cite{baumer2017post}, which can be seen in work on embodiment \cite{dourish2001action, suchman1987plans} and death \cite{brubaker2010death}. Within this tradition, research on queer people often looks at how queerness breaks normative assumptions regarding users embedded in the design of technologies. For instance, research on gender transition demonstrates that the assumption that one has a single, immutable "real name" fails to meet the needs of trans people who may wish to change their name or display different names to different audiences on social media \cite{haimson2016digital}. Similarly, several studies on LGBTQ+ self-presentation (e.g., \cite{carrasco2018queer, devito2018too}) advocate for supporting selective visibility in design because the assumed isomorphism between one-account and one-self breaks down for those with heightened self-presentation needs. Other work looks at how, even when designing for queer people, normative assumptions about them can still misalign with queer experiences. For instance, design features in queer location-based dating apps assume that users will live in urban areas, failing to account for rural users \cite{hardy2017constructing}.

    A subset of this work problematizing how technologists think about people or users can be found in Queer HCI research on classification. This work builds on early HCI/CSCW research on the failures of classification systems, such as Bowker \& Star’s notions of residuality (i.e., that which falls outside classification systems) and torque (i.e., the feeling when individual biographies misalign with classification system) \cite{bowker1999sorting}. This Queer HCI research often looks specifically at how people and computers encode or classify gender, such as work on how computing systems often enforce a gendered binary \cite{hamidi2018gender, keyes2018misgendering, spiel2021they}. While some of this work focuses on potential ways computer vision \cite{Chong2021Exploring} or speech processing algorithms \cite{Rincon2021Speaking} may benefit transgender people, much of this work focuses on technological harms \cite{scheuerman2020we}. Similar inquiries have emerged around how gendered webforms enforce uncomfortable binaries for non-binary people \cite{scheuerman2021revisiting} and how non-binary people in academic survey work are often removed from datasets as 'noise' \cite{jaroszewski2018genderfluid}. These papers recommend the broader HCI community better encode gender into technological artifacts. Recent work has also explored how HCI researchers \cite{Seaborn2022PronounsPepper} and research participants \cite{seaborn2022exploring} gender robots. This work has been particularly influential in demonstrating the social construction of classification systems in HCI research, entangled with the growing emphasis on AI in HCI at the time of our writing in 2023. While much of this work focuses on the harms of falling outside classification systems, there was less work on the benefits of illegibility, such as avoiding detection.

    \subsection{Queer People as Stigmatized Subjects}

     This genre discusses the social stigma attached to being LGBTQ+ and how LGBTQ+ people manage their identities. Stigmatization is related to but distinct from marginalization. While marginalization refers to broader social structures, a stigma is an attribute that can "spoil" one's identity or is potentially discreditable in particular social contexts \cite{goffman2009stigma}. Work in this genre emphasizes that because queerness is stigmatized, LGBTQ+ people may be cautious of who they come out to. Research in this genre speaks to longstanding interests in disclosure among scholars of social computing (e.g., lying about oneself online \cite{van1996strange, donath2002identity}) and privacy (e.g., the infamous Alice and Bob metaphor \cite{rivest1978method}).
     
     The first CHI paper to focus "4 - exclusively" on LGBTQ+ people used Craigslist ads to predict HIV prevalence in cities around the U.S. \cite{haimson2014ddfseeks} and the first CSCW paper "4 - exclusively" on queer people studied depression in TrevorSpace, an online community for LGBTQ+ youth \cite{homan2014social}. Both papers mentioned similar motivations — using online communities to understand stigmatized populations that are "hard-to-reach" \cite{homan2014social}. These first studies were published in 2014 amid a growth of research in the early-to-mid 2010s using newly available social media data for health monitoring \cite{de2013predicting}. Paralleling most research about queer people in our corpus, these first works do not necessarily focus on queer experiences \textit{per se} but rather the ways queer people can fit into contemporaneous HCI research interests.

   Following these initial methodological papers, there is a significant body of work focusing on LGBTQ+ identity management across multiple venues of social computing \cite{haimson2015disclosure, blackwell2016lgbt, haimson2017social, haimson2016digital, gonzales2017prioritizing, pinter2021entering, fernandez2019don, cui2022so, pinch2021s, nova2021facebook, devito2018too, carrasco2018queer, warner2019signal, warner2020evaluating, pyle2021lgbtq}. The first study, published in 2015, focused on how trans people disclosed their gender transition\footnote{Gender transition disclosures may include sharing that one has changed their name or pronouns.} on Facebook \cite{haimson2015disclosure}. It emphasized that trans identity is not always socially accepted and may introduce stress for trans people managing that disclosure on online social platforms. Much of this identity management research also focused on the experiences of transgender people, such as self-presentation \cite{haimson2016digital} on Facebook, disclosure for crowdfunding gender-affirming healthcare \cite{gonzales2017prioritizing}, \footnote{Such as mastectomies/top surgeries for transgender men.} and disclosure of being transgender on dating apps \cite{fernandez2019don}. More recently, researchers explored the benefits and risks associated with online trans visibility \cite{pinter2021entering,DeVito2022TransfemTikTok,Lerner2020Activism}. 
    
    The first papers focusing on specific groups in the LGBTQ+ community often look at issues related to social stigma (e.g., the first papers on the experiences of bi+ \cite{walker_more_2020}, hijra \cite{Nova2020Understanding}, and lesbian/bisexual/sexual minority women \cite{Cui2022Wegather, cui2022so}). Rather than focusing on particular groups, some work has also studied the self-presentation of LGBTQ+ people writ large across various social computing contexts \cite{devito2018too, carrasco2018queer}. Beyond managing the disclosure and presentation of one's LGBTQ+ identity, some work studied the self-disclosure of other stigmatized identities or experiences in LGBTQ+ peoples' lives, such as disclosing stigmatized identities on dating apps \cite{warner2019signal, warner2020evaluating, fernandez2019don} or navigating pregnancy and disclosing pregnancy loss on social media \cite{pyle2021lgbtq, andalibi2022lgbtq}. 
    
    Similar work focuses on the privacy concerns of LGBTQ+ people, many of which were "3 - significantly" rather than "4 - exclusively" about queer people \cite{birnholtz2015weird, brubaker2016visibility, warner2018privacy, maestre2018defining, skeba2020informational, mcdonald2020politics, haque2019ulti, bussone2020trust, dym2020social, introne2021narrative, logas2022image, wu2022reasonable}. Some of this research involved privacy-conscious populations that substantially overlap with LGBTQ+ people, such as fandom members worrying about sexually explicit content being linked to their offline identity \cite{dym2020social} and people living with HIV who may worry about status disclosure \cite{warner2018privacy, maestre2018defining, bussone2020trust, introne2021narrative}. Other privacy studies incidentally encountered LGBTQ+ people, such as a study on posts in an anonymous forum \cite{birnholtz2015weird} and an ethnographic study of privacy practices in Dhaka \cite{haque2019ulti}. Meanwhile, others used queer visibility \cite{brubaker2016visibility, mcdonald2020politics} or stories of being outed by technology \cite{skeba2020informational, wu2022reasonable} as case studies for exploring privacy issues. In a literature review on privacy research with marginalized groups, LGBTQ+ people were shown to be one of the most heavily researched populations \cite{sannon_litt_review_privacy}.
    
    While the examples mentioned above meaningfully engage with specific aspects of LGBTQ+ privacy concerns, other researchers \footnote{We do not cite any papers in this paragraph because we do not want to "call out" specific authors and for purposes of citational justice \cite{kumar2021braving}.} used LGBTQ+ privacy concerns as a case study in ways that do not appear invested in the experiences of LGBTQ+ people. Some of this work treated one's LGBTQ+ status as an example of sensitive information analogous to a secret national ID number. For instance, in one work, the authors developed a classification model to identify LGBTQ+ people on social media as a case study for inferring "sensitive personal information," paying little attention to potential adverse consequences or the researchers' positionality. 

    \subsection{Queer People as Highly Vulnerable}

    An undertone in research on queer stigma or falling outside the norm is the notion that queer people are highly vulnerable to technological harm and, in turn, deserve particular research attention. However, queer people are not the only group discussed in this way. We find queer people are often "3 - significantly" included in research as one of multiple cases in research related to content moderation and demonetization, online harm, and sexual violence.

    One common "high-risk" group we found discussed alongside and intersecting with queerness is women. Much of this research looks at online harm. For example, a study of the online abuse experiences and coping practices of women in India, Pakistan, and Bangladesh deliberately sought to include LGBTQ+ women participants  \cite{sambasivan2019they}. Similarly, research exploring the experiences "Black women and femmes" on social media details the experiences and impacts of online harm while also attending to healing and joy \cite{musgrave2022experiences}. This research acknowledges a distinct overlap between LGBTQ+ people and women's experiences, deliberately seeking out these experiences to ensure they are documented. Other research frames LGBTQ+ people as a distinct group alongside women, facing unique risks, such as research on women and LGBTQ+ people's decisions to participate in India's \#MeToo movement against sexual harassment on social media \cite{moitra2021parsing}, which finds, in contrast to cisgender heterosexual women, LGBTQ+ participants "fall through the cracks" of sexual harassment laws. Similarly, Furlo et al. studied "dating app users identifying as LGBTQIA+ or women" because these communities experience "disproportionate risk of sexual violence" \cite{furlo2021rethinking}. Although not limited to women or LGBTQ+ people, Zytko et al. deliberately recruited a large sample of LGBTQ+ participants to study sexual consent on the app Tinder for similar reasons \cite{zytko2021computer}. 
    
    Queer people are also discussed in conjunction with other groups, such as BIPOC people, in research on content moderation and, relatedly, algorithmic harm. The earliest work in this area looks at both gender and sexuality biases in data annotation \cite{Otterbacher2015Crowdsourcing}. More recently, in 2021, both Simpson \& Semaan's research on LGBTQ+ TikTok users \cite{simpson2021you} and Karizat et al.'s research on marginalized TikTok users generally \cite{karizat2021algorithmic} explored how TikTok's recommendation algorithms can privilege certain identity performances over others. Another example of algorithmic harm is YouTube's content moderation algorithm, which was shown to demonetize the videos of LGBTQ+ creators \cite{shen2021everyday, alkhatib2019street}, which are one of several groups, including BIPOC and political conservatives, to disproportionately have their online content removed \cite{haimson2021disproportionate}. Research into demonetization has also explored algorithmic audits by content creators \cite{shen2021everyday} and ways to introduce algorithmic transparency following demonetization of user-generated content \cite{Dunna2022demonetization, Kingsley2022Give}.
    
    Researchers differ in how they discuss LGBTQ+ people alongside other marginalized groups. Some contend with the specific circumstances of LGBTQ+ people, such as how Vaccaro et al. conducted separate participatory design workshops with BIPOC, LGBTQ+, and artist social media users based on prior work suggesting these groups are negatively impacted by content moderation decisions \cite{vaccaro2021contestability}. Others homogenize queer people's experiences with other social groups into a vague category of "marginalized groups."
    
    As a consequence of viewing LGBTQ+ people as a high-risk "marginalized group," HCI researchers are increasingly interested in supporting this community. Researchers are also beginning to study how to do research with "marginalized people." Liang et al. outline tensions conducting HCI research with marginalized people by interviewing HCI scholars working in these contexts \cite{liang2021embracing}. Next, we explore at a body of research that examines queerness as it is understood from the community's perspective, addressing some of the concerns raised by Liang et al. surrounding extractive research engagements with marginalized groups.

    \subsection{Community-Centered Research}

    We identified works that detail researcher reflexivity explicitly coming from queer communities and researchers themselves. Often, these works engaged with gender and sexuality, whether as a focal point or as issues entangled within a larger area of interest. Most emblematic of queer-specific reflexivity are the reflections of the Queer SIGs \cite{devito2020queer,devito2021queer}, which we described earlier, negotiating what it means to do queer research and be a queer researcher. Beyond a collective reflection of Queer HCI, we see personal reflections specific to particular queer identities, such as non-binary experiences of "casual violence" in the field \cite{spiel2019patching}. These papers, SIGs, and abstracts contour the burgeoning space of Queer HCI scholarship that is community-centered or designed for and by queer communities.

    Early works on exploring queer communities focused on intersex (1998) \cite{williamson1998constructing} and genderqueer (2008) \cite{irani2008situated} online communities. However, these early groups were not objects of study but rather a means to explore other HCI concepts, such as interaction in virtual worlds \cite{irani2008situated}. Following these initial encounters, early work on queer community building centered on creating safe places. For instance, Beirl et al. designed a mobile application to "improve safe access to gendered toilets" \cite{beirl2017gotyourback}, and Scheuerman et al. studied how trans and non-binary people use technology to "find, create, and navigate safe spaces" \cite{scheuerman2018safe}, both physically and virtually. More recently Acena et al. extended conversations around the design of LGBTQ+ safe places into the liminal space between virtual and physical occupied by virtual reality \cite{acena2021my}.
    
    In recent research, we found a shift toward emphasizing futuring or designing with queer people intertwined with community building, such as designing an online community by and for trans people \cite{haimson2020trans}. We also see an emphasis on creating queer futures and narratives to push back against dominant understandings of LGBTQ+ identity in research on transformative fandom \cite{dym2019coming} and TikTok \cite{simpson2021you}. In 2022, Cui et al. \cite{cui2022so} explored relationship and community building of sexual minority women (SMW) in China on location-based SMW dating apps. Similarly, Hardy and Lindtner \cite{hardy2017constructing} detailed how rural gay, bisexual, and queer men use queer location-based apps to construct communities. However, sometimes community building can be fraught, as research into the intra-community marginalization of bi+ people in LGBTQ+ online spaces demonstrates \cite{walker_more_2020}. Additionally, some recent scholars are using participatory methods to design technologies with trans people \cite{ahmed2021online, haimson2020designing} and rural LGBTQ+ communities \cite{hardy2019participatory, hardy2022lgbtq}. This genre is growing but remains limited.

\section{Provocations for Queer HCI}

    In this section, we provide three provocations for Queer HCI researchers. We identify an opportunity for the field to be more specific on the populations they study. We urge scholars to expand their inquiry beyond forms of marginalization queer communities face and to instead consider other aspects of queer life, such as queer joy. Finally, we call on scholars to not simply use queer people as a means to advance more general HCI ends. Rather, when we study with groups like queer folks, we must also ensure our research answers questions these communities need answered.
    
    \subsection{Can We Be (More) Specific?}

    We found a significant body of research about "LGBTQ+ people" but comparatively less research on the particularities of certain populations. While queer people are all marginalized by heteronormative axes of domination \cite{collins1990black}, the way this marginalization takes place is situated \cite{haraway1988situated} in the particular lived experiences of those with different queer identities. This gets complicated as queer identities cannot be thought of as mutually exclusive and tend to overlap in various ways. For example, Spiel et al. point out: "Non-binary people are rarely considered by technologies or technologists, and often subsumed under binary trans experiences on the rare occasions when [they] are discussed" \cite{spiel2019patching}. Individual queer experiences may shift over time as people come out, try on identities, discard them, and find themselves anew.

    There is very little HCI research on \textit{specific} queer populations. We found one paper centered on polyamorous queer people \cite{kinnee2022sonic}, one centered on bisexual people \cite{walker_more_2020}, and two centered on sexual minority women \cite{cui2022so, Cui2022Wegather}. Clearly, there is "much ground to be covered" in research on queer populations \cite{devito2020queer, devito2021queer}, and specificity in Queer HCI research is a way we can be held accountable as a research community to cover this ground. Of course, this tension is not unique to the Queer HCI community. Burrell \& Toyama describe similar tensions among Information and Communication Technologies for Development researchers over how much to emphasize cultural differences versus commonalities \cite{burrell2009constitutes}. We encourage Queer HCI researchers to reckon with differences and commonalities within and across LGBTQ+ identities — which will require troubling the categories of \textit{queer} or \textit{LGBTQ+} people.

    At the same time, we acknowledge that there can be politically strategic reasons for simplifying how we sometimes discuss queer people \cite{danius1993interview}. Advocating for particular policies (e.g., calling attention to algorithmic discrimination \cite{devos2022toward}) may require presenting LGBTQ+ people as a unified, marginalized front. However, in making queer people legible to outsiders, we must not lose sight of differences within the queer community. We suggest that greater specificity — a deeper understanding of highly particular, intersectional \cite{rankin2019straighten, crenshaw1990mapping} experiences — will help Queer HCI researchers embrace the nuances of our community. In doing so, Queer HCI researchers can center how communities understand themselves rather than how queerness is understood by outsiders — decentering the dominant to build a research agenda toward justice.

    \subsection{Must Our Research Always Be About Trauma?}

    We found that a large body of research about LGBTQ+ people focuses on social stigma, marginalization, and harm. This focus centers LGBTQ+ experiences in relation to dominant social power structures. For example, while coming out or identity disclosure is often a lifelong project for queer people, it is only a small part of LGBTQ+ people's lives. Meanwhile, identity disclosure is one of the must heavily studied aspects of queer people's lives in HCI research. By focusing so narrowly, this research perpetuates a narrative of the "queer experience" as a trauma that we, as queer people, must endure. This resembles recent critiques of trauma or deficit focused research in BIPOC \cite{to_everyday_dis_2023} or otherwise marginalized communities \cite{pei2019we}. Fortunately, we also found recent Queer HCI research beginning to explore futuring and community building, but this research direction is nascent. This leads us to the provocation: what would research that studies the experiences of everyday LGBTQ+ life look like? Can we study joy, sex, or pleasure? More broadly, what would it look like to research how queer people understand themselves and experience the world (e.g., the everyday) rather than how queer people are understood by others (e.g., as a marginalized community)?

    Our provocation to move beyond trauma should not be understood as a naive or overly-optimistic proposal. We need only look to our own experiences as queer researchers to know that LGBTQ+ people \textit{do} experience social stigma and marginalization. These experiences warrant continued research attention from the Queer HCI community. However, we want to encourage additional research about the broader scope of LGBTQ+ people's lives beyond deficits, echoing similar suggestions by assets-based design \cite{wong2020culture}, ICT4D \cite{arora2012leisure}, BIPOC \cite{to_everyday_dis_2023} HCI researchers that members of marginalized groups are more than their marginalization. We also acknowledge that there is not one single "dominant" view to decenter but rather many intersecting dominances. For instance, Walker \& DeVito's research on power dynamics within the LGBTQ+ community focuses on the particular ways bisexual people experience domination \cite{walker_more_2020}. As Crenshaw reminds us, one of the limitations of identity politics is that "it frequently conflates or ignores intragroup differences" \cite{crenshaw1990mapping}. Furthermore, we recognize that Queer HCI research is contingent, taking place in a system "not created for or by those who are minoritized and marginalized" \cite{erete2021can}. One may experience pressure to hegemonically represent LGBTQ+ people because this is how those who fund and review Queer HCI research view LGBTQ+ people. Undermining the conventional, hegemonic views of LGBTQ+ people in HCI will not be easy, but we have hope. Queer people have long been adept at troubling the status quo \cite{light2011hci}.

     \subsection{Means or Ends: Why Are We Studying Queer People?}

    We should study queer people in HCI, but we must also be upfront about what our objectives are. Building on critical theory \cite{ahmed2019use}, HCI researchers are increasingly interrogating the value system of "usefulness" that persist in practices such as human-centered design \cite{lin2021techniques}. What's the use of studying queer people in HCI? Across Section 6, we detailed how aspects of queer experiences are useful to certain types of HCI research. Breaking down categories is useful for AI research. Coming out is useful for studying privacy and online-self disclosure. The political nature of queer existence is useful for studying online social movements. Across these instances we find that HCI researchers are able to extract value from narrow aspects of queer lives that fit within existing HCI research agendas.
    
    Interrogating how HCI researchers use queer people — and toward what ends — lies in interrogating whether queer people are a research focus or an application area in HCI \cite{disalvo2010mapping}? Here, we draw from a distinction DiSalvo et al. made in their literature review of sustainable HCI \cite{disalvo2010mapping}. They found that while many works engage with sustainability as their central research focus "with tools and methods chosen or adapted as appropriate to address concerns about sustainability" \cite{disalvo2010mapping}, some start with an interest in "particular tools and methods" and use sustainability as an "application domain to test out those tools and methods." Similarly, some research in our corpus centered on queer people as a research focus (e.g., \cite{haimson2020designing, walker_more_2020, carrasco2018queer, Cui2022Wegather}), while others used queer experiences \cite{vaccaro2021contestability} or queer rights \cite{zhang2015modeling} as an application area to, for instance, understand content moderation \cite{vaccaro2021contestability} or predict policy changes \cite{zhang2015modeling}. 

    In research using queer people as an application area, we noted reliance on aspects of queer experiences that are useful to HCI researchers. Gay rights movements in the U.S. proved to be a useful case for HCI researchers interested in online social movements. The socio-political context motivating this work fit neatly into HCI researchers' contemporaneous interest in sensemaking through social media. Both privacy \cite{nissenbaum_internet_2017} and social computing researchers \cite{donath2002identity} have long been interested in studying self-disclosure and managing secrets. In these works, queer people make for useful subjects because they are assumed to have secret information to manage (i.e. coming out). These works make important theoretical contributions to HCI in the various sub-fields outlined above. However, the aspects of queer lives most useful to HCI researchers represent only a small slice of queer experiences. We call for more Queer HCI research studying what queer people care about in relation to technology.
     
    We encourage future research on particular populations in HCI to interrogate and consider whether research about these groups are \textit{about} them or rather about aspects of these populations' experiences that are most useful to HCI researchers. One response may be "refusing" such research, or carefully interrogating research to ensure its use to "overstudied Others" \cite{tuck2014refusingresearch}. We encourage those researching queer people — or any other group — to critically reflect on whether this population is the focus of their research or being used as an application area to look at some other phenomena of interest to HCI researchers. Based on our findings about the ways queer people are used in HCI research, we further discuss the role of technology in HCI research below.

\section{Recommendations for Broader HCI}

While Queer HCI is clearly a part of the HCI community, in this section we provide broader recommendations for any HCI researcher engaging with or discussing queer people.
    
\subsection{How to Discuss Queer People}
   
    HCI researchers have made significant improvements in how they discuss gender. However, even researchers with the best intentions may not always get it right. We encourage HCI researchers and computing  educators broadly to familiarize themselves with existing Queer HCI research related to collecting and reporting gender (e.g., \cite{jaroszewski2018genderfluid, spiel2019better, scheuerman2021revisiting}). We cannot provide strict guidelines on dealing with gender in HCI because every research context is different \cite{haraway1988situated}. However, in qualitative research, we encourage researchers to let participants self-identify their gender, possibly using an open-text box. In quantitative research, we generally encourage authors to use multi-option boxes (as described by \cite{spiel2019better}) and consider domain and context (as described by \cite{scheuerman2021revisiting}). Lastly, like Seberger et al. \cite{seberger2022still}, we call for future work on methods for working with small populations to prevent trans and non-binary people from being filtered out of survey research related to gender \cite{jaroszewski2018genderfluid}.
        
    HCI researchers often discuss LGBTQ+ civil rights in the context of political debates or online polarization, reaching a peak around the mid-2010s. However, an implicit assumption is embedded into much of this work that "both sides" of debates over LGBTQ+ people have merit, particularly in research on polarization that treats "far-right" and "far-left" biases as equally undesirable. We do not suggest that researchers should avoid studying LGBTQ+ civil rights. However, we encourage researchers to acknowledge that research is political and to take explicit political stances in favor of LGBTQ+ people, much like recent calls for data scientists to acknowledge they are political actors \cite{green2021data}. This discussion is particularly salient because, at the time of our writing in 2023, LGBTQ+ civil rights are under attack, for instance, by politicians in the United States,\footnote{https://www.aclu.org/legislative-attacks-on-lgbtq-rights-2024} the United Kingdom,\footnote{https://www.cnn.com/2022/07/17/uk/uk-conservative-leadership-trans-intl-gbr} and Uganda.\footnote{https://www.npr.org/2023/05/29/1178718092/uganda-anti-gay-law} Much like the calls for "explicit anti-racist actions" \cite{ogbonnaya2020critical} in HCI, we call on HCI researchers to stand up for LGBTQ+ civil rights, thus stripping the veneer of \textit{"neutral objectivity"} from HCI research. There is no room for "level-headed" compromise. A "neutral" position on gay rights is not that queer people can have a few rights "as a treat." In doing so, we call for placing debates over LGBTQ+ civil rights beyond the Overton window of "respectable" debate in HCI research.
    
    \subsection{On the Limitations of Limitations} 
    
    In recent years, we found a dramatic rise in papers mentioning LGBTQ+ people as a limitation or area for future work. It is heartening that HCI researchers are beginning to recognize that their particular research may not extend to all "humans." As scholars like Suchman \cite{suchman1987plans} and Dourish \cite{dourish2001action} remind us, HCI is always situated within a particular time and place. Nevertheless, there are limitations to the growing discussion of queer people in limitation sections. A limitation section should not allow authors \textit{carte blanche} to ignore LGBTQ+ people or absolve authors from critique. We encourage researchers to do research \textit{with} LGBTQ+ people in contexts where they are routinely relegated to limitations or future work sections, such as the design of menstrual technology. 

    Similarly, we encourage authors to use limitations sections to frankly consider the bounds of their research rather than pay lip service to marginalized communities or evade criticism. In this regard, Seberger et al.'s research is exemplary \cite{seberger2022still}. Rather than performing a statistical analysis on a population of non-binary people far too small to be significant in a gesture of performative allyship — like some other papers in our corpus — the authors decided not to run this analysis. Instead, they described this as a limitation of their work.

    While well intended, the issues we draw out with limitation sections contours a larger issue in HCI. When we write these sections to pay lip service, the impact at best is to avoid critique and at worst obfuscates bad scientific practices. Therefore, much like Erete et al. condemn acts of "performative anti-racism" by HCI researchers \cite{erete2021can}, we encourage HCI scholars to consider whether their discussion of queer people in limitations or future work sections benefits them or benefits the queer people in their study.

    \subsection{On the Limitations of Marginalization} 
    
    Although the earliest mention of LGBTQ+ people discussed queer people as a marginalized group, this was not widely acknowledged in the HCI community until the late 2010s. We found LGBTQ+ marginalization is often used to rhetorically motivate research alongside other paradigmatic marginalized groups. In doing so, HCI researchers often discuss queer people — as well as members of other marginalized populations — as a vague, interchangeable Other, which may occlude the experiences of particular groups and experiences at the intersections of these groups. HCI research tends to assume all "marginalized people" are the same. Instead, what if our null hypothesis was that all "marginalized people" are different until proven otherwise?

    Moreover, while LGBTQ+ people are certainly marginalized, we caution HCI researchers against only viewing LGBTQ+ people and Queer HCI research through this lens. Although much of the early Queer HCI research focuses on stigma and marginalization, there is more the broader HCI community can learn from queer people and Queer HCI research. For instance, the messiness of LGBTQ+ identities denaturalizes the taken-for-granted ways people — LGBTQ+ and otherwise — are encoded in computing systems \cite{brubaker2011select, scheuerman2020we, scheuerman2021revisiting}. We find the growing attention paid to LGBTQ+ people in HCI promising. However, we encourage HCI researchers to reckon with the limitations of understanding research about and by marginalized people solely through the lens of marginalization.
    
    Finally, we acknowledge that there can be politically expedient reason for simplifying the way we talk about social groups when interests are aligned, such as building broad environmentalist social movements by bringing disparate parties together \cite{dourish2008points}. At the same time, Mohanty cautions against mistaking "discursively consensual homogeneity" of, in her case, women broadly defined for "historically specific material realities" \cite{mohanty1988under}. We should be deliberate about when to center particular experiences of social groups and when to invoke broader subjects, such as "marginalized" or "LGBTQ+" people. We must ask ourselves when and for whom it is expedient to broadly frame social groups in such manners.

    \subsection{Oops I Made It About the Technology Again: The Need for Decentering Technology}

    Our findings demonstrate how research trends — the latest methods and shiniest technologies — drive HCI research (queer or otherwise). This parallels Soden et al.'s critique of presentism in CSCW \cite{soden_historicism} and anthropological critiques of techno-centrism in early HCI research \cite{suchman1987plans, dourish2006implications}. When Big Data and social data mining were made more accessible in the mid-2010s, researchers fit queer people in this research \cite{lee2010receptionist, rzeszotarski2012learning, wang2014cursing, compton2016theory, wood2018rethinking, costa2018regulating, holzer2018digitally}. At the time of our writing, we similarly see a rise in Queer HCI research related to AI \cite{scheuerman2019computers}. While fitting into whatever is "chic" in HCI may open possibilities by trojan-horsing particular social groups through contemporaneous trends, it impacts how HCI understands these groups. A substantial portion of research on queer people used particular methodologies HCI researchers were/are particularly interested in. However, we found little work guided by queer peoples' everyday experiences, problems, and pleasures.

    Some may see an argument here for decentering technology akin to Dourish's critique of implications for design \cite{dourish2006implications}. While we certainly support such proposals, the relationship between technology and human experiences has changed since the early 2000s. Given that it is no longer possible to consider life in the absence of computing, we see an opportunity to more forcefully start with human-experience, confident that technology will inevitably play an important role and be worthy of design. We advocate for a research agenda that decenters technology in-and-of-itself in HCI research, especially as they entangle particular social groups, such as Queer People.

    In advocating for more research decentering technology, we are not critiquing Queer HCI research concerning whatever the shiny technology \textit{du jour} is, such as AI. At the time of our writing (December 2023), techno-hucksters are pulling the god trick \cite{haraway1988situated} with Generative AI models, such as GPT-4. HCI scholars can elevate subaltern views on technologies to explain how "universal" views are, in fact, situated. Queer people and other marginalized populations are ready means to demonstrating the social construction of technology as these groups so often fall outside initial designs. This strategic use of these groups by HCI researchers plays a crucial role in checking power.
    We encourage HCI scholars to consider how their technical expertise can serve forms of community accountability, where the tools we build are a means to an end rather than an end in and of itself.

    \begin{table}[]
\begin{tabular}{lcl}

\textbf{Community} & \begin{tabular}[c]{@{}c@{}} \textbf{Paper}\\ \textbf{Count}\end{tabular} & \textbf{Citations} \\

\hline

\textbf{Lesbians}                    & \textbf{2}                      &    \\
... \textit{Sexual Minority Women }        & 1                               & \cite{cui2022so} \\
... \textit{Lesbians and Bisexuals  }      & 1                               & \cite{Cui2022Wegather} \\
... \textit{Lesbians Exclusively  }        & 0                               & --- \\

\textbf{Gay Men}                     & \textbf{6}                      &   \\
... \textit{MSM}                           & 2                               & \cite{haimson2014ddfseeks, warner2020evaluating} \\
... \textit{Gay Men}                       & 1                               & \cite{Wang2022GayDating} \\
... \textit{Sexual-Minority Men}           & 1                               & \cite{taylor2017social} \\
... \textit{Gay and Bisexual Men}          & 1                               & \cite{warner2019signal} \\
... \textit{Gay, Bisexual and Queer Men}   & 1                               & \cite{hardy2017constructing} \\

\textbf{Bisexuals}                   & \textbf{1}                      & \cite{walker_more_2020} \\
... \textit{Bi Women Exclusively}        & 0                               & ---\\
... \textit{Bi Men Exclusively  }        & 0                               & --- \\
... \textit{Bi Non-Binary People Exclusively} & 0 & ---\\

\textbf{QTBIPOC}                     & \textbf{2}                               & \cite{rizvi2022qtbipoc,starks2019designing} \\

\textbf{Asexual / Aromantic People}                    & 0                               & ---  \\

\textbf{Polyamorous People}          & 1  &                     \cite{kinnee2022sonic}\\

\textbf{Transgender People}          & \textbf{36}                     &   \\
... \textit{Trans People Broadly }            & 25                              &   \\
... \textit{Non-Binary People}   & 4*                              & \cite{scheuerman2021revisiting,spiel2021they,jaroszewski2018genderfluid,spiel2019patching}  \\
... \textit{Non-Western (Gender Minority)}   & 4                               & \cite{nova2021facebook, Nova2020Understanding, moitra2021negotiating,Chong2021Exploring} \\
... \textit{Transfemmes Exclusively }         & 1                               & \cite{DeVito2022TransfemTikTok} \\
\begin{tabular}[c]{@{}c@{}} \textit{...Trans Women \&}\\ \textit{   Non-Binary BIPOC}\end{tabular}  & 1                               & \cite{starks2019designing} \\
... \textit{Trans Men Exclusively }           & 1                               & \cite{gonzales2017prioritizing}    \\

\textbf{LGBTQ+}                      & \textbf{28}                     &    \\
... \textit{LGBTQ+ People Broadly }           & 15                              &    \\
... \textit{LGBTQ+ Families }                 & 3                               & \cite{blackwell2016lgbt,andalibi2022lgbtq,pyle2021lgbtq}  \\
... \textit{LGBTQ+ Youth }                    & 3                               & \cite{homan2014social,gatehouse2018troubling,pinch2021s} \\
... \textit{Rural LBGTQ+ Experiences }        & 3                               & \cite{hardy2022lgbtq,hardy2019participatory,hardy2017constructing} \\
... \textit{Non-Western (Sexual Minority) }            & 3                               & \cite{Cui2022Wegather, cui2022so,Wang2022GayDating} \\
... \textit{LGBTQ+ Sex Workers }                  & 1                               & \cite{Hamilton2022SexWork} \\
... \textit{Older LGBTQ+ People}        & 0                               & --- \\
... \textit{LGBTQ+ Sex or Pleasure} & 0 & --- \\

\end{tabular}
\caption{Prior Work and Gaps in Research Exclusively About LGBTQ+ Populations in HCI. Note, if a particular population is not in this table then we did not find it in our corpus. We note a few glaring omissions in the table.}
\Description[A table of prior work and gaps in research exclusively about LGBTQ+ populations in HCI.]{A table of prior work and gaps in research exclusively about LGBTQ+ populations in HCI.}
\label{tab:Gaps}
\end{table}

\section{Future Work}

    The HCI research community has made substantial improvements in queer representation since the first papers "4 - exclusively" about LGBTQ+ people were published in 2014. However, there is still much work to be done. As seen in Table \ref{tab:Gaps}, most of the 73 papers exclusively about queer people have looked at the broad categories of LGBTQ+ (n=15) or Transgender (n=25) people. Less work focuses on the particular experience of subgroups (e.g., Lesbians (2), Gays (6), and Bisexuals (1)) and intersections (e.g., QTBIPOC (2), rural (3), youth (3), non-western (7)) of the LGBTQ+ community. We note (Table \ref{tab:Gaps}) that there is a need for more particular queer research in HCI.\footnote{There are technically five papers that discuss non-binary identity, but for citational justice purposes, we have chosen to exclude one paper.} There is no research exclusively about lesbian experiences,\footnote{Here we take "lesbian" to mean a person who identifies as a lesbian - including cisgender and transgender women, as well as non-binary people — who have distinct experiences from bisexual or heterosexual women. In the papers that discuss queer women, lesbian and bisexual women are discussed together, which is why we make this distinction.} nor is there any research on asexual or aromatic people. There is no research on older LGBTQ+ adults, and few papers discuss kink or other fringe and queer-adjacent sub-communities (e.g., furries \cite{bardzell2014lonely}) in queer ways. There is also little examination of queer pornography, queer sex, and how queer sex workers navigate, engage in, and use technologies \cite{uttarapong2022social}.  
    
    There are also limitations to our methods that should be addressed in future work. One cannot understand Queer HCI by studying published texts alone. We cannot know why authors choose to center queer people as a research focus versus an application area. Authors may wish to focus on queer experiences as a research area but worry their work will not be funded or published without treating queer people as a case study to justify "generalizability." HCI researchers must be able to treat queer people as ends in themselves rather than only a means to an end. Therefore, future work should study the systems within which Queer HCI research is produced.
    
    Moreover, studying only published texts may lead to a survivorship bias. We do not know which papers did not get published. There have been Queer HCI researchers long before the first paper "4 - exclusively" about queer people was published in 2014. Future work should seek to disentangle what has yet to be said from what researchers cannot or could not say. As Foucault notes, silence is a discourse itself \cite{foucault1990history}. There are silences in the archives \cite{adler2017cruising} of Queer HCI. Much like Soden et al.'s recent advocacy for historical methods \cite{soden_historicism}, we encourage future historical research on Queer HCI beyond textual analysis, such as oral histories and studies of Queer HCI ephemera (e.g., conference flyers).

    Our decision to study queerness in HCI among a handful of venues also came at the expense of a deeper understanding of Queer HCI across more venues. As a result, we failed to include other prominent conferences in HCI (\textit{e.g.,} FAccT, PDC, GROUP, UIST, UbiComp, VIS, SOUPS), games conferences (\textit{e.g.,} FDG, DiGRA), and prominent media studies journals (\textit{e.g.,} New Media \& Society, First Monday, Social Media + Society). While research centered "4 - exclusively" on queer people did not begin until 2014, research on LGBTQ+ social media use was published in New Media \& Society as early as 2005 \cite{campbell2005outing}. Future work should explore Queer HCI research in these different venues, potentially exploring similarities or differences between venues or publishers.

\section{Conclusion}

In this paper, we compiled a queer archive of HCI. We sought to understand how the HCI community, Queer HCI included, engages in research involving queer people over time. Despite a handful of earlier works, we find Queer HCI publications began, in earnest, in 2014 when the first papers exclusively about queer people were published \cite{homan2014social, haimson2014ddfseeks}. Since then, there has been an exponential growth in the number of research papers discussing LGBTQ+ people. However, we found HCI research only focuses on some aspects of queer people's lives, such as coming out or breaking classification systems, with little attention paid to joy and the everyday. We also found that research on queer people tends to study "LGBTQ+" or "trans" people generally, with less attention paid to specific sub-groups and intersections. We hope our work can stimulate more thoughtful discussions of queer people across the HCI community and more particular and joyful research about queer people.

\begin{acks}
We would like to thank Kelly Wang, Ashley Milton, Harkiran Saluja, and Sriya Mupparaju for their research assistance. We would also like to thank Oliver Haimson, Amy Bruckman, and Carl DiSalvo for feedback during the writing process; as well as the reviewers this year and last for their thoughtful suggestions and critique. We would also like to thank Morgan Klaus Scheuerman, Blakeley H. Payne, Rose Chang, Angie Goffredi, Katta Spiel, and Mary Gray for emotional support and being sounding boards during our writing process. This work was supported in part by the Atlanta Interdisciplinary AI (AIAI) Network, sponsored by the Mellon Foundation. Finally, we would like to thank Queer HCI Researchers everywhere, both those who came before us and those who will come after. This work would not have been possible without those who have worked both through scholarship and social support to ensure that Queer HCI can exist. 
\end{acks}

\bibliographystyle{ACM-Reference-Format}
\bibliography{refs}


\begin{thebibliography}{230}


\ifx \showCODEN    \undefined \def \showCODEN     #1{\unskip}     \fi
\ifx \showDOI      \undefined \def \showDOI       #1{#1}\fi
\ifx \showISBNx    \undefined \def \showISBNx     #1{\unskip}     \fi
\ifx \showISBNxiii \undefined \def \showISBNxiii  #1{\unskip}     \fi
\ifx \showISSN     \undefined \def \showISSN      #1{\unskip}     \fi
\ifx \showLCCN     \undefined \def \showLCCN      #1{\unskip}     \fi
\ifx \shownote     \undefined \def \shownote      #1{#1}          \fi
\ifx \showarticletitle \undefined \def \showarticletitle #1{#1}   \fi
\ifx \showURL      \undefined \def \showURL       {\relax}        \fi
\providecommand\bibfield[2]{#2}
\providecommand\bibinfo[2]{#2}
\providecommand\natexlab[1]{#1}
\providecommand\showeprint[2][]{arXiv:#2}

\bibitem[\protect\citeauthoryear{Acena and Freeman}{Acena and Freeman}{2021}]%
        {acena2021my}
\bibfield{author}{\bibinfo{person}{Dane Acena} {and} \bibinfo{person}{Guo Freeman}.} \bibinfo{year}{2021}\natexlab{}.
\newblock \showarticletitle{“In My Safe Space”: Social Support for LGBTQ Users in Social Virtual Reality}. In \bibinfo{booktitle}{\emph{Extended Abstracts of the 2021 CHI Conference on Human Factors in Computing Systems}}. \bibinfo{pages}{1--6}.
\newblock


\bibitem[\protect\citeauthoryear{Adler}{Adler}{2017}]%
        {adler2017cruising}
\bibfield{author}{\bibinfo{person}{Melissa Adler}.} \bibinfo{year}{2017}\natexlab{}.
\newblock \bibinfo{booktitle}{\emph{Cruising the library: Perversities in the organization of knowledge}}.
\newblock \bibinfo{publisher}{Fordham Univ Press}.
\newblock


\bibitem[\protect\citeauthoryear{Ahmed, Kok, Howard, and Still}{Ahmed et~al\mbox{.}}{2021}]%
        {ahmed2021online}
\bibfield{author}{\bibinfo{person}{Alex~A Ahmed}, \bibinfo{person}{Bryan Kok}, \bibinfo{person}{Coranna Howard}, {and} \bibinfo{person}{Klew Still}.} \bibinfo{year}{2021}\natexlab{}.
\newblock \showarticletitle{Online community-based design of free and open source software for transgender voice training}.
\newblock \bibinfo{journal}{\emph{Proceedings of the ACM on Human-Computer Interaction}} \bibinfo{volume}{4}, \bibinfo{number}{CSCW3} (\bibinfo{year}{2021}), \bibinfo{pages}{1--27}.
\newblock


\bibitem[\protect\citeauthoryear{Ahmed}{Ahmed}{2019}]%
        {ahmed2019use}
\bibfield{author}{\bibinfo{person}{Sara Ahmed}.} \bibinfo{year}{2019}\natexlab{}.
\newblock \bibinfo{booktitle}{\emph{What's the use?: On the uses of use}}.
\newblock \bibinfo{publisher}{Duke University Press}.
\newblock


\bibitem[\protect\citeauthoryear{Alabdulqader, Lazem, Khamis, and Dray}{Alabdulqader et~al\mbox{.}}{2018}]%
        {alabdulqader2018exploring}
\bibfield{author}{\bibinfo{person}{Ebtisam Alabdulqader}, \bibinfo{person}{Shaimaa Lazem}, \bibinfo{person}{Mohamed Khamis}, {and} \bibinfo{person}{Susan~M Dray}.} \bibinfo{year}{2018}\natexlab{}.
\newblock \showarticletitle{Exploring participatory design methods to engage with Arab communities}. In \bibinfo{booktitle}{\emph{Extended Abstracts of the 2018 CHI Conference on Human Factors in Computing Systems}}. \bibinfo{pages}{1--8}.
\newblock


\bibitem[\protect\citeauthoryear{Alabdulqader, Lazem, Nassir, Saleh, Armouch, and Dray}{Alabdulqader et~al\mbox{.}}{2019}]%
        {alabdulqader2019eye}
\bibfield{author}{\bibinfo{person}{Ebtisam Alabdulqader}, \bibinfo{person}{Shaimaa Lazem}, \bibinfo{person}{Soud Nassir}, \bibinfo{person}{Mennatallah Saleh}, \bibinfo{person}{Sara Armouch}, {and} \bibinfo{person}{Susan Dray}.} \bibinfo{year}{2019}\natexlab{}.
\newblock \showarticletitle{With an eye to the future: Hci practice and research in the Arab world1}. In \bibinfo{booktitle}{\emph{Extended abstracts of the 2019 chi conference on human factors in computing systems}}. \bibinfo{pages}{1--9}.
\newblock


\bibitem[\protect\citeauthoryear{Ali, Sapiezynski, Bogen, Korolova, Mislove, and Rieke}{Ali et~al\mbox{.}}{2019}]%
        {ali2019discrimination}
\bibfield{author}{\bibinfo{person}{Muhammad Ali}, \bibinfo{person}{Piotr Sapiezynski}, \bibinfo{person}{Miranda Bogen}, \bibinfo{person}{Aleksandra Korolova}, \bibinfo{person}{Alan Mislove}, {and} \bibinfo{person}{Aaron Rieke}.} \bibinfo{year}{2019}\natexlab{}.
\newblock \showarticletitle{Discrimination through optimization: How Facebook's Ad delivery can lead to biased outcomes}.
\newblock \bibinfo{journal}{\emph{Proceedings of the ACM on human-computer interaction}} \bibinfo{volume}{3}, \bibinfo{number}{CSCW} (\bibinfo{year}{2019}), \bibinfo{pages}{1--30}.
\newblock


\bibitem[\protect\citeauthoryear{Alkhatib and Bernstein}{Alkhatib and Bernstein}{2019}]%
        {alkhatib2019street}
\bibfield{author}{\bibinfo{person}{Ali Alkhatib} {and} \bibinfo{person}{Michael Bernstein}.} \bibinfo{year}{2019}\natexlab{}.
\newblock \showarticletitle{Street-level algorithms: A theory at the gaps between policy and decisions}. In \bibinfo{booktitle}{\emph{Proceedings of the 2019 CHI Conference on Human Factors in Computing Systems}}. \bibinfo{pages}{1--13}.
\newblock


\bibitem[\protect\citeauthoryear{Ananny}{Ananny}{2011}]%
        {ananny2011curious}
\bibfield{author}{\bibinfo{person}{Mike Ananny}.} \bibinfo{year}{2011}\natexlab{}.
\newblock \showarticletitle{The curious connection between apps for gay men and sex offenders}.
\newblock \bibinfo{journal}{\emph{The Atlantic}}  \bibinfo{volume}{14} (\bibinfo{year}{2011}).
\newblock


\bibitem[\protect\citeauthoryear{Andalibi, Lacombe-Duncan, Roosevelt, Wojciechowski, and Giniel}{Andalibi et~al\mbox{.}}{2022}]%
        {andalibi2022lgbtq}
\bibfield{author}{\bibinfo{person}{Nazanin Andalibi}, \bibinfo{person}{Ashley Lacombe-Duncan}, \bibinfo{person}{Lee Roosevelt}, \bibinfo{person}{Kylie Wojciechowski}, {and} \bibinfo{person}{Cameron Giniel}.} \bibinfo{year}{2022}\natexlab{}.
\newblock \showarticletitle{LGBTQ Persons’ Use of Online Spaces to Navigate Conception, Pregnancy, and Pregnancy Loss: An Intersectional Approach}.
\newblock \bibinfo{journal}{\emph{ACM Trans. Comput.-Hum. Interact.}} \bibinfo{volume}{29}, \bibinfo{number}{1}, Article \bibinfo{articleno}{2} (\bibinfo{date}{jan} \bibinfo{year}{2022}), \bibinfo{numpages}{46}~pages.
\newblock
\showISSN{1073-0516}
\urldef\tempurl%
\url{https://doi.org/10.1145/3474362}
\showDOI{\tempurl}


\bibitem[\protect\citeauthoryear{Arora}{Arora}{2012}]%
        {arora2012leisure}
\bibfield{author}{\bibinfo{person}{Payal Arora}.} \bibinfo{year}{2012}\natexlab{}.
\newblock \showarticletitle{The leisure divide: Can the ‘Third World’come out to play?}
\newblock \bibinfo{journal}{\emph{Information Development}} \bibinfo{volume}{28}, \bibinfo{number}{2} (\bibinfo{year}{2012}), \bibinfo{pages}{93--101}.
\newblock


\bibitem[\protect\citeauthoryear{Badillo-Urquiola, Shea, Agha, Lediaeva, and Wisniewski}{Badillo-Urquiola et~al\mbox{.}}{2021}]%
        {badillo2021conducting}
\bibfield{author}{\bibinfo{person}{Karla Badillo-Urquiola}, \bibinfo{person}{Zachary Shea}, \bibinfo{person}{Zainab Agha}, \bibinfo{person}{Irina Lediaeva}, {and} \bibinfo{person}{Pamela Wisniewski}.} \bibinfo{year}{2021}\natexlab{}.
\newblock \showarticletitle{Conducting risky research with teens: co-designing for the ethical treatment and protection of adolescents}.
\newblock \bibinfo{journal}{\emph{Proceedings of the ACM on Human-Computer Interaction}} \bibinfo{volume}{4}, \bibinfo{number}{CSCW3} (\bibinfo{year}{2021}), \bibinfo{pages}{1--46}.
\newblock


\bibitem[\protect\citeauthoryear{Baglione, Clemens, Maestre, Min, Dahl, and Shih}{Baglione et~al\mbox{.}}{2021}]%
        {baglione2021understanding}
\bibfield{author}{\bibinfo{person}{Anna~N Baglione}, \bibinfo{person}{Michael~Paul Clemens}, \bibinfo{person}{Juan~F Maestre}, \bibinfo{person}{Aehong Min}, \bibinfo{person}{Luke Dahl}, {and} \bibinfo{person}{Patrick~C Shih}.} \bibinfo{year}{2021}\natexlab{}.
\newblock \showarticletitle{Understanding the Technological Practices and Needs of Music Therapists}.
\newblock \bibinfo{journal}{\emph{Proceedings of the ACM on Human-Computer Interaction}} \bibinfo{volume}{5}, \bibinfo{number}{CSCW1} (\bibinfo{year}{2021}), \bibinfo{pages}{1--25}.
\newblock


\bibitem[\protect\citeauthoryear{Bardzell, Bardzell, and Koefoed~Hansen}{Bardzell et~al\mbox{.}}{2015}]%
        {bardzell2015immodest}
\bibfield{author}{\bibinfo{person}{Jeffrey Bardzell}, \bibinfo{person}{Shaowen Bardzell}, {and} \bibinfo{person}{Lone Koefoed~Hansen}.} \bibinfo{year}{2015}\natexlab{}.
\newblock \showarticletitle{Immodest proposals: Research through design and knowledge}. In \bibinfo{booktitle}{\emph{Proceedings of the 33rd annual ACM conference on human factors in computing systems}}. \bibinfo{pages}{2093--2102}.
\newblock


\bibitem[\protect\citeauthoryear{Bardzell, Bardzell, Zhang, and Pace}{Bardzell et~al\mbox{.}}{2014}]%
        {bardzell2014lonely}
\bibfield{author}{\bibinfo{person}{Jeffrey Bardzell}, \bibinfo{person}{Shaowen Bardzell}, \bibinfo{person}{Guo Zhang}, {and} \bibinfo{person}{Tyler Pace}.} \bibinfo{year}{2014}\natexlab{}.
\newblock \showarticletitle{The lonely raccoon at the ball: Designing for intimacy, sociability, and selfhood}. In \bibinfo{booktitle}{\emph{Proceedings of the SIGCHI Conference on Human Factors in Computing Systems}}. \bibinfo{pages}{3943--3952}.
\newblock


\bibitem[\protect\citeauthoryear{Bardzell, Freeman, Bardzell, and Chen}{Bardzell et~al\mbox{.}}{2020}]%
        {bardzell2020join}
\bibfield{author}{\bibinfo{person}{Jeffrey Bardzell}, \bibinfo{person}{Guo Freeman}, \bibinfo{person}{Shaowen Bardzell}, {and} \bibinfo{person}{Pei-ying Chen}.} \bibinfo{year}{2020}\natexlab{}.
\newblock \showarticletitle{Join. Love: A sociotechnical genealogy of the legalization of same-sex marriage}. In \bibinfo{booktitle}{\emph{Proceedings of the 2020 CHI Conference on Human Factors in Computing Systems}}. \bibinfo{pages}{1--13}.
\newblock


\bibitem[\protect\citeauthoryear{Barlas, Kyriakou, Guest, Kleanthous, and Otterbacher}{Barlas et~al\mbox{.}}{2021}]%
        {barlas2021see}
\bibfield{author}{\bibinfo{person}{Pinar Barlas}, \bibinfo{person}{Kyriakos Kyriakou}, \bibinfo{person}{Olivia Guest}, \bibinfo{person}{Styliani Kleanthous}, {and} \bibinfo{person}{Jahna Otterbacher}.} \bibinfo{year}{2021}\natexlab{}.
\newblock \showarticletitle{To" see" is to stereotype: Image tagging algorithms, gender recognition, and the accuracy-fairness trade-off}.
\newblock \bibinfo{journal}{\emph{Proceedings of the ACM on Human-Computer Interaction}} \bibinfo{volume}{4}, \bibinfo{number}{CSCW3} (\bibinfo{year}{2021}), \bibinfo{pages}{1--31}.
\newblock


\bibitem[\protect\citeauthoryear{Baumer and Brubaker}{Baumer and Brubaker}{2017}]%
        {baumer2017post}
\bibfield{author}{\bibinfo{person}{Eric~PS Baumer} {and} \bibinfo{person}{Jed~R Brubaker}.} \bibinfo{year}{2017}\natexlab{}.
\newblock \showarticletitle{Post-userism}. In \bibinfo{booktitle}{\emph{Proceedings of the 2017 CHI Conference on Human Factors in Computing Systems}}. \bibinfo{pages}{6291--6303}.
\newblock


\bibitem[\protect\citeauthoryear{Beeson}{Beeson}{1999}]%
        {beeson1999closing}
\bibfield{author}{\bibinfo{person}{Ann Beeson}.} \bibinfo{year}{1999}\natexlab{}.
\newblock \showarticletitle{Closing plenary: Civil rights in cyberspace: How online free speech restrictions will inhibit online diversity}. In \bibinfo{booktitle}{\emph{CHI'99 Extended Abstracts on Human Factors in Computing Systems}}. \bibinfo{pages}{98--99}.
\newblock


\bibitem[\protect\citeauthoryear{Beirl, Zeitlin, Chan, Loh, and Zhong}{Beirl et~al\mbox{.}}{2017}]%
        {beirl2017gotyourback}
\bibfield{author}{\bibinfo{person}{Diana Beirl}, \bibinfo{person}{Anya Zeitlin}, \bibinfo{person}{Jerald Chan}, \bibinfo{person}{Kai Ip~Alvin Loh}, {and} \bibinfo{person}{Xiaodi Zhong}.} \bibinfo{year}{2017}\natexlab{}.
\newblock \showarticletitle{GotYourBack: An internet of toilets for the trans* community}. In \bibinfo{booktitle}{\emph{Proceedings of the 2017 CHI Conference Extended Abstracts on Human Factors in Computing Systems}}. \bibinfo{pages}{39--45}.
\newblock


\bibitem[\protect\citeauthoryear{Bin~Morshed, Dye, Ahmed, and Kumar}{Bin~Morshed et~al\mbox{.}}{2017}]%
        {bin2017internet}
\bibfield{author}{\bibinfo{person}{Mehrab Bin~Morshed}, \bibinfo{person}{Michaelanne Dye}, \bibinfo{person}{Syed~Ishtiaque Ahmed}, {and} \bibinfo{person}{Neha Kumar}.} \bibinfo{year}{2017}\natexlab{}.
\newblock \showarticletitle{When the internet goes down in Bangladesh}. In \bibinfo{booktitle}{\emph{Proceedings of the 2017 ACM conference on computer supported cooperative work and social computing}}. \bibinfo{pages}{1591--1604}.
\newblock


\bibitem[\protect\citeauthoryear{Birnholtz, Merola, and Paul}{Birnholtz et~al\mbox{.}}{2015}]%
        {birnholtz2015weird}
\bibfield{author}{\bibinfo{person}{Jeremy Birnholtz}, \bibinfo{person}{Nicholas Aaron~Ross Merola}, {and} \bibinfo{person}{Arindam Paul}.} \bibinfo{year}{2015}\natexlab{}.
\newblock \showarticletitle{"Is it Weird to Still Be a Virgin" Anonymous, Locally Targeted Questions on Facebook Confession Boards}. In \bibinfo{booktitle}{\emph{Proceedings of the 33rd annual ACM conference on human factors in computing systems}}. \bibinfo{pages}{2613--2622}.
\newblock


\bibitem[\protect\citeauthoryear{Blackwell, Hardy, Ammari, Veinot, Lampe, and Schoenebeck}{Blackwell et~al\mbox{.}}{2016}]%
        {blackwell2016lgbt}
\bibfield{author}{\bibinfo{person}{Lindsay Blackwell}, \bibinfo{person}{Jean Hardy}, \bibinfo{person}{Tawfiq Ammari}, \bibinfo{person}{Tiffany Veinot}, \bibinfo{person}{Cliff Lampe}, {and} \bibinfo{person}{Sarita Schoenebeck}.} \bibinfo{year}{2016}\natexlab{}.
\newblock \showarticletitle{LGBT parents and social media: Advocacy, privacy, and disclosure during shifting social movements}. In \bibinfo{booktitle}{\emph{Proceedings of the 2016 CHI conference on human factors in computing systems}}. \bibinfo{pages}{610--622}.
\newblock


\bibitem[\protect\citeauthoryear{Bowker and Star}{Bowker and Star}{1999}]%
        {bowker1999sorting}
\bibfield{author}{\bibinfo{person}{Geoffrey Bowker} {and} \bibinfo{person}{Susan~Leigh Star}.} \bibinfo{year}{1999}\natexlab{}.
\newblock \bibinfo{booktitle}{\emph{Sorting Things Out: Classification and its Concequences}}.
\newblock \bibinfo{publisher}{Citeseer}.
\newblock


\bibitem[\protect\citeauthoryear{Brubaker and Hayes}{Brubaker and Hayes}{2011}]%
        {brubaker2011select}
\bibfield{author}{\bibinfo{person}{Jed~R Brubaker} {and} \bibinfo{person}{Gillian~R Hayes}.} \bibinfo{year}{2011}\natexlab{}.
\newblock \showarticletitle{SELECT* FROM USER: infrastructure and socio-technical representation}. In \bibinfo{booktitle}{\emph{Proceedings of the ACM 2011 conference on Computer supported cooperative work}}. \bibinfo{pages}{369--378}.
\newblock


\bibitem[\protect\citeauthoryear{Brubaker, Kaye, Schoenebeck, and Vertesi}{Brubaker et~al\mbox{.}}{2016}]%
        {brubaker2016visibility}
\bibfield{author}{\bibinfo{person}{Jed~R Brubaker}, \bibinfo{person}{Jofish Kaye}, \bibinfo{person}{Sarita Schoenebeck}, {and} \bibinfo{person}{Janet Vertesi}.} \bibinfo{year}{2016}\natexlab{}.
\newblock \showarticletitle{Visibility in digital space: Controlling personal information online}. In \bibinfo{booktitle}{\emph{Proceedings of the 19th ACM conference on computer supported cooperative work and social computing companion}}. \bibinfo{pages}{184--187}.
\newblock


\bibitem[\protect\citeauthoryear{Brubaker and Vertesi}{Brubaker and Vertesi}{2010}]%
        {brubaker2010death}
\bibfield{author}{\bibinfo{person}{Jed~R Brubaker} {and} \bibinfo{person}{Janet Vertesi}.} \bibinfo{year}{2010}\natexlab{}.
\newblock \showarticletitle{Death and the social network}. In \bibinfo{booktitle}{\emph{Proc. CHI Workshop on Death and the Digital}}. \bibinfo{pages}{1--4}.
\newblock


\bibitem[\protect\citeauthoryear{Burrell and Toyama}{Burrell and Toyama}{2009}]%
        {burrell2009constitutes}
\bibfield{author}{\bibinfo{person}{Jenna Burrell} {and} \bibinfo{person}{Kentaro Toyama}.} \bibinfo{year}{2009}\natexlab{}.
\newblock \showarticletitle{What constitutes good ICTD research?}
\newblock \bibinfo{journal}{\emph{Information Technologies \& International Development}} \bibinfo{volume}{5}, \bibinfo{number}{3} (\bibinfo{year}{2009}), \bibinfo{pages}{pp--82}.
\newblock


\bibitem[\protect\citeauthoryear{Bussone, Kasadha, Stumpf, Durrant, Tariq, Gibbs, Lloyd, and Bird}{Bussone et~al\mbox{.}}{2020}]%
        {bussone2020trust}
\bibfield{author}{\bibinfo{person}{Adrian Bussone}, \bibinfo{person}{Bakita Kasadha}, \bibinfo{person}{Simone Stumpf}, \bibinfo{person}{Abigail~C Durrant}, \bibinfo{person}{Shema Tariq}, \bibinfo{person}{Jo Gibbs}, \bibinfo{person}{Karen~C Lloyd}, {and} \bibinfo{person}{Jon Bird}.} \bibinfo{year}{2020}\natexlab{}.
\newblock \showarticletitle{Trust, identity, privacy, and security considerations for designing a peer data sharing platform between people living with HIV}.
\newblock \bibinfo{journal}{\emph{Proceedings of the ACM on Human-Computer Interaction}} \bibinfo{volume}{4}, \bibinfo{number}{CSCW2} (\bibinfo{year}{2020}), \bibinfo{pages}{1--27}.
\newblock


\bibitem[\protect\citeauthoryear{Butler}{Butler}{1990}]%
        {butler1990gender}
\bibfield{author}{\bibinfo{person}{Judith Butler}.} \bibinfo{year}{1990}\natexlab{}.
\newblock \bibinfo{booktitle}{\emph{Gender Trouble: Feminism and the Subversion of Identity}}.
\newblock \bibinfo{publisher}{Routledge}.
\newblock


\bibitem[\protect\citeauthoryear{Cambre, Colnago, Maddock, Tsai, and Kaye}{Cambre et~al\mbox{.}}{2020}]%
        {cambre2020choice}
\bibfield{author}{\bibinfo{person}{Julia Cambre}, \bibinfo{person}{Jessica Colnago}, \bibinfo{person}{Jim Maddock}, \bibinfo{person}{Janice Tsai}, {and} \bibinfo{person}{Jofish Kaye}.} \bibinfo{year}{2020}\natexlab{}.
\newblock \showarticletitle{Choice of voices: A large-scale evaluation of text-to-speech voice quality for long-form content}. In \bibinfo{booktitle}{\emph{Proceedings of the 2020 CHI Conference on Human Factors in Computing Systems}}. \bibinfo{pages}{1--13}.
\newblock


\bibitem[\protect\citeauthoryear{Campbell}{Campbell}{2005}]%
        {campbell2005outing}
\bibfield{author}{\bibinfo{person}{John~Edward Campbell}.} \bibinfo{year}{2005}\natexlab{}.
\newblock \showarticletitle{Outing PlanetOut: Surveillance, gay marketing and internet affinity portals}.
\newblock \bibinfo{journal}{\emph{New Media \& Society}} \bibinfo{volume}{7}, \bibinfo{number}{5} (\bibinfo{year}{2005}), \bibinfo{pages}{663--683}.
\newblock


\bibitem[\protect\citeauthoryear{Capozzi, De~Francisci~Morales, Mejova, Monti, Panisson, and Paolotti}{Capozzi et~al\mbox{.}}{2021}]%
        {capozzi2021clandestino}
\bibfield{author}{\bibinfo{person}{Arthur Capozzi}, \bibinfo{person}{Gianmarco De~Francisci~Morales}, \bibinfo{person}{Yelena Mejova}, \bibinfo{person}{Corrado Monti}, \bibinfo{person}{Andr{\'e} Panisson}, {and} \bibinfo{person}{Daniela Paolotti}.} \bibinfo{year}{2021}\natexlab{}.
\newblock \showarticletitle{Clandestino or Rifugiato? Anti-immigration Facebook Ad Targeting in Italy}. In \bibinfo{booktitle}{\emph{Proceedings of the 2021 CHI Conference on Human Factors in Computing Systems}}. \bibinfo{pages}{1--15}.
\newblock


\bibitem[\protect\citeauthoryear{Carrasco and Kerne}{Carrasco and Kerne}{2018}]%
        {carrasco2018queer}
\bibfield{author}{\bibinfo{person}{Matthew Carrasco} {and} \bibinfo{person}{Andruid Kerne}.} \bibinfo{year}{2018}\natexlab{}.
\newblock \showarticletitle{Queer visibility: Supporting LGBT+ selective visibility on social media}. In \bibinfo{booktitle}{\emph{Proceedings of the 2018 CHI Conference on Human Factors in Computing Systems}}. \bibinfo{pages}{1--12}.
\newblock


\bibitem[\protect\citeauthoryear{Chancellor, Baumer, and De~Choudhury}{Chancellor et~al\mbox{.}}{2019}]%
        {chancellor2019human}
\bibfield{author}{\bibinfo{person}{Stevie Chancellor}, \bibinfo{person}{Eric~PS Baumer}, {and} \bibinfo{person}{Munmun De~Choudhury}.} \bibinfo{year}{2019}\natexlab{}.
\newblock \showarticletitle{Who is the" human" in human-centered machine learning: The case of predicting mental health from social media}.
\newblock \bibinfo{journal}{\emph{Proceedings of the ACM on Human-Computer Interaction}} \bibinfo{volume}{3}, \bibinfo{number}{CSCW} (\bibinfo{year}{2019}), \bibinfo{pages}{1--32}.
\newblock


\bibitem[\protect\citeauthoryear{Choi, Lee, and Hong}{Choi et~al\mbox{.}}{2022}]%
        {choi2022s}
\bibfield{author}{\bibinfo{person}{Dasom Choi}, \bibinfo{person}{Uichin Lee}, {and} \bibinfo{person}{Hwajung Hong}.} \bibinfo{year}{2022}\natexlab{}.
\newblock \showarticletitle{“It’s not wrong, but I’m quite disappointed”: Toward an Inclusive Algorithmic Experience for Content Creators with Disabilities}. In \bibinfo{booktitle}{\emph{CHI Conference on Human Factors in Computing Systems}}. \bibinfo{pages}{1--19}.
\newblock


\bibitem[\protect\citeauthoryear{Chong, Maudet, Harima, and Igarashi}{Chong et~al\mbox{.}}{2021}]%
        {Chong2021Exploring}
\bibfield{author}{\bibinfo{person}{Toby Chong}, \bibinfo{person}{Nolwenn Maudet}, \bibinfo{person}{Katsuki Harima}, {and} \bibinfo{person}{Takeo Igarashi}.} \bibinfo{year}{2021}\natexlab{}.
\newblock \showarticletitle{Exploring a Makeup Support System for Transgender Passing Based on Automatic Gender Recognition}. In \bibinfo{booktitle}{\emph{Proceedings of the 2021 CHI Conference on Human Factors in Computing Systems}} (Yokohama, Japan) \emph{(\bibinfo{series}{CHI '21})}. \bibinfo{publisher}{Association for Computing Machinery}, \bibinfo{address}{New York, NY, USA}, Article \bibinfo{articleno}{568}, \bibinfo{numpages}{13}~pages.
\newblock
\showISBNx{9781450380966}
\urldef\tempurl%
\url{https://doi.org/10.1145/3411764.3445364}
\showDOI{\tempurl}


\bibitem[\protect\citeauthoryear{Collins}{Collins}{1990}]%
        {collins1990black}
\bibfield{author}{\bibinfo{person}{Patricia~Hill Collins}.} \bibinfo{year}{1990}\natexlab{}.
\newblock \showarticletitle{Black feminist thought in the matrix of domination}.
\newblock \bibinfo{journal}{\emph{Black feminist thought: Knowledge, consciousness, and the politics of empowerment}} \bibinfo{volume}{138}, \bibinfo{number}{1990} (\bibinfo{year}{1990}), \bibinfo{pages}{221--238}.
\newblock


\bibitem[\protect\citeauthoryear{Compton}{Compton}{2016}]%
        {compton2016theory}
\bibfield{author}{\bibinfo{person}{Ryan Compton}.} \bibinfo{year}{2016}\natexlab{}.
\newblock \showarticletitle{Theory driven community analytics and influence on community success}. In \bibinfo{booktitle}{\emph{Proceedings of the 19th ACM Conference on Computer Supported Cooperative Work and Social Computing Companion}}. \bibinfo{pages}{135--138}.
\newblock


\bibitem[\protect\citeauthoryear{Cornet, Hall, Cafaro, and Brady}{Cornet et~al\mbox{.}}{2017}]%
        {cornet2017image}
\bibfield{author}{\bibinfo{person}{Victor~P Cornet}, \bibinfo{person}{Natalie~K Hall}, \bibinfo{person}{Francesco Cafaro}, {and} \bibinfo{person}{Erin~L Brady}.} \bibinfo{year}{2017}\natexlab{}.
\newblock \showarticletitle{How image-based social media websites support social movements}. In \bibinfo{booktitle}{\emph{Proceedings of the 2017 CHI conference extended abstracts on Human Factors in Computing Systems}}. \bibinfo{pages}{2473--2479}.
\newblock


\bibitem[\protect\citeauthoryear{Correll}{Correll}{1995}]%
        {correll1995ethnography}
\bibfield{author}{\bibinfo{person}{Shelley Correll}.} \bibinfo{year}{1995}\natexlab{}.
\newblock \showarticletitle{The ethnography of an electronic bar: The lesbian cafe}.
\newblock \bibinfo{journal}{\emph{Journal of contemporary ethnography}} \bibinfo{volume}{24}, \bibinfo{number}{3} (\bibinfo{year}{1995}), \bibinfo{pages}{270--298}.
\newblock


\bibitem[\protect\citeauthoryear{Costa, Jung, Czerwinski, Guimbreti{\`e}re, Le, and Choudhury}{Costa et~al\mbox{.}}{2018}]%
        {costa2018regulating}
\bibfield{author}{\bibinfo{person}{Jean Costa}, \bibinfo{person}{Malte~F Jung}, \bibinfo{person}{Mary Czerwinski}, \bibinfo{person}{Fran{\c{c}}ois Guimbreti{\`e}re}, \bibinfo{person}{Trinh Le}, {and} \bibinfo{person}{Tanzeem Choudhury}.} \bibinfo{year}{2018}\natexlab{}.
\newblock \showarticletitle{Regulating feelings during interpersonal conflicts by changing voice self-perception}. In \bibinfo{booktitle}{\emph{Proceedings of the 2018 CHI Conference on Human Factors in Computing Systems}}. \bibinfo{pages}{1--13}.
\newblock


\bibitem[\protect\citeauthoryear{Costa~Figueiredo and Chen}{Costa~Figueiredo and Chen}{2021}]%
        {costa2021health}
\bibfield{author}{\bibinfo{person}{Mayara Costa~Figueiredo} {and} \bibinfo{person}{Yunan Chen}.} \bibinfo{year}{2021}\natexlab{}.
\newblock \showarticletitle{Health data in fertility care: an ecological perspective}. In \bibinfo{booktitle}{\emph{Proceedings of the 2021 CHI Conference on Human Factors in Computing Systems}}. \bibinfo{pages}{1--17}.
\newblock


\bibitem[\protect\citeauthoryear{Crenshaw}{Crenshaw}{1990}]%
        {crenshaw1990mapping}
\bibfield{author}{\bibinfo{person}{Kimberle Crenshaw}.} \bibinfo{year}{1990}\natexlab{}.
\newblock \showarticletitle{Mapping the margins: Intersectionality, identity politics, and violence against women of color}.
\newblock \bibinfo{journal}{\emph{Stan. L. Rev.}}  \bibinfo{volume}{43} (\bibinfo{year}{1990}), \bibinfo{pages}{1241}.
\newblock


\bibitem[\protect\citeauthoryear{Cui, Yamashita, and Lee}{Cui et~al\mbox{.}}{2022a}]%
        {Cui2022Wegather}
\bibfield{author}{\bibinfo{person}{Yichao Cui}, \bibinfo{person}{Naomi Yamashita}, {and} \bibinfo{person}{Yi-Chieh Lee}.} \bibinfo{year}{2022}\natexlab{a}.
\newblock \showarticletitle{"We Gather Together We Collaborate Together": Exploring the Challenges and Strategies of Chinese Lesbian and Bisexual Women's Online Communities on Weibo}.
\newblock \bibinfo{journal}{\emph{Proc. ACM Hum.-Comput. Interact.}} \bibinfo{volume}{6}, \bibinfo{number}{CSCW2}, Article \bibinfo{articleno}{423} (\bibinfo{date}{nov} \bibinfo{year}{2022}), \bibinfo{numpages}{31}~pages.
\newblock
\urldef\tempurl%
\url{https://doi.org/10.1145/3555148}
\showDOI{\tempurl}


\bibitem[\protect\citeauthoryear{Cui, Yamashita, Liu, and Lee}{Cui et~al\mbox{.}}{2022b}]%
        {cui2022so}
\bibfield{author}{\bibinfo{person}{Yichao Cui}, \bibinfo{person}{Naomi Yamashita}, \bibinfo{person}{Mingjie Liu}, {and} \bibinfo{person}{Yi-Chieh Lee}.} \bibinfo{year}{2022}\natexlab{b}.
\newblock \showarticletitle{“So Close, yet So Far”: Exploring Sexual-minority Women’s Relationship-building via Online Dating in China}. In \bibinfo{booktitle}{\emph{CHI Conference on Human Factors in Computing Systems}}.
\newblock


\bibitem[\protect\citeauthoryear{Danius, Jonsson, and Spivak}{Danius et~al\mbox{.}}{1993}]%
        {danius1993interview}
\bibfield{author}{\bibinfo{person}{Sara Danius}, \bibinfo{person}{Stefan Jonsson}, {and} \bibinfo{person}{Gayatri~Chakravorty Spivak}.} \bibinfo{year}{1993}\natexlab{}.
\newblock \showarticletitle{An Interview with Gayatri Chakravorty Spivak}.
\newblock \bibinfo{journal}{\emph{boundary 2}} \bibinfo{volume}{20}, \bibinfo{number}{2} (\bibinfo{year}{1993}), \bibinfo{pages}{24--50}.
\newblock


\bibitem[\protect\citeauthoryear{De~Choudhury, Gamon, Counts, and Horvitz}{De~Choudhury et~al\mbox{.}}{2013}]%
        {de2013predicting}
\bibfield{author}{\bibinfo{person}{Munmun De~Choudhury}, \bibinfo{person}{Michael Gamon}, \bibinfo{person}{Scott Counts}, {and} \bibinfo{person}{Eric Horvitz}.} \bibinfo{year}{2013}\natexlab{}.
\newblock \showarticletitle{Predicting depression via social media}. In \bibinfo{booktitle}{\emph{Proceedings of the international AAAI conference on web and social media}}, Vol.~\bibinfo{volume}{7}. \bibinfo{pages}{128--137}.
\newblock


\bibitem[\protect\citeauthoryear{Dechant, Welsch, Frommel, and Mandryk}{Dechant et~al\mbox{.}}{2022}]%
        {dechant2022don}
\bibfield{author}{\bibinfo{person}{Martin~Johannes Dechant}, \bibinfo{person}{Robin Welsch}, \bibinfo{person}{Julian Frommel}, {and} \bibinfo{person}{Regan~L Mandryk}.} \bibinfo{year}{2022}\natexlab{}.
\newblock \showarticletitle{(Don’t) stand by me: How trait psychopathy and NPC emotion influence player perceptions, verbal responses, and movement behaviours in a gaming task}. In \bibinfo{booktitle}{\emph{CHI Conference on Human Factors in Computing Systems}}. \bibinfo{pages}{1--17}.
\newblock


\bibitem[\protect\citeauthoryear{DeVito}{DeVito}{2022}]%
        {DeVito2022TransfemTikTok}
\bibfield{author}{\bibinfo{person}{Michael~Ann DeVito}.} \bibinfo{year}{2022}\natexlab{}.
\newblock \showarticletitle{How Transfeminine TikTok Creators Navigate the Algorithmic Trap of Visibility Via Folk Theorization}.
\newblock \bibinfo{journal}{\emph{Proc. ACM Hum.-Comput. Interact.}} \bibinfo{volume}{6}, \bibinfo{number}{CSCW2}, Article \bibinfo{articleno}{380} (\bibinfo{date}{Nov} \bibinfo{year}{2022}), \bibinfo{numpages}{31}~pages.
\newblock


\bibitem[\protect\citeauthoryear{DeVito, Lustig, Simpson, Allison, Chuanromanee, Spiel, Ko, Rode, Dym, Muller, et~al\mbox{.}}{DeVito et~al\mbox{.}}{2021}]%
        {devito2021queer}
\bibfield{author}{\bibinfo{person}{Michael~Ann DeVito}, \bibinfo{person}{Caitlin Lustig}, \bibinfo{person}{Ellen Simpson}, \bibinfo{person}{Kimberley Allison}, \bibinfo{person}{Tya Chuanromanee}, \bibinfo{person}{Katta Spiel}, \bibinfo{person}{Amy Ko}, \bibinfo{person}{Jennifer Rode}, \bibinfo{person}{Brianna Dym}, \bibinfo{person}{Michael Muller}, {et~al\mbox{.}}} \bibinfo{year}{2021}\natexlab{}.
\newblock \showarticletitle{Queer in HCI: Strengthening the Community of LGBTQIA+ Researchers and Research}. In \bibinfo{booktitle}{\emph{Extended Abstracts of the 2021 CHI Conference on Human Factors in Computing Systems}}. \bibinfo{pages}{1--3}.
\newblock


\bibitem[\protect\citeauthoryear{DeVito, Walker, and Birnholtz}{DeVito et~al\mbox{.}}{2018}]%
        {devito2018too}
\bibfield{author}{\bibinfo{person}{Michael~Ann DeVito}, \bibinfo{person}{Ashley~Marie Walker}, {and} \bibinfo{person}{Jeremy Birnholtz}.} \bibinfo{year}{2018}\natexlab{}.
\newblock \showarticletitle{'Too Gay for Facebook' Presenting LGBTQ+ Identity Throughout the Personal Social Media Ecosystem}.
\newblock \bibinfo{journal}{\emph{Proceedings of the ACM on Human-Computer Interaction}} \bibinfo{volume}{2}, \bibinfo{number}{CSCW} (\bibinfo{year}{2018}), \bibinfo{pages}{1--23}.
\newblock


\bibitem[\protect\citeauthoryear{DeVito, Walker, Lustig, Ko, Spiel, Ahmed, Allison, Scheuerman, Dym, Brubaker, et~al\mbox{.}}{DeVito et~al\mbox{.}}{2020}]%
        {devito2020queer}
\bibfield{author}{\bibinfo{person}{Michael~Ann DeVito}, \bibinfo{person}{Ashley~Marie Walker}, \bibinfo{person}{Caitlin Lustig}, \bibinfo{person}{Amy~J Ko}, \bibinfo{person}{Katta Spiel}, \bibinfo{person}{Alex~A Ahmed}, \bibinfo{person}{Kimberley Allison}, \bibinfo{person}{Morgan Scheuerman}, \bibinfo{person}{Briana Dym}, \bibinfo{person}{Jed~R Brubaker}, {et~al\mbox{.}}} \bibinfo{year}{2020}\natexlab{}.
\newblock \showarticletitle{Queer in HCI: supporting LGBTQIA+ researchers and research across domains}. In \bibinfo{booktitle}{\emph{Extended Abstracts of the 2020 CHI Conference on Human Factors in Computing Systems}}. \bibinfo{pages}{1--4}.
\newblock


\bibitem[\protect\citeauthoryear{DeVos, Dhabalia, Shen, Holstein, and Eslami}{DeVos et~al\mbox{.}}{2022}]%
        {devos2022toward}
\bibfield{author}{\bibinfo{person}{Alicia DeVos}, \bibinfo{person}{Aditi Dhabalia}, \bibinfo{person}{Hong Shen}, \bibinfo{person}{Kenneth Holstein}, {and} \bibinfo{person}{Motahhare Eslami}.} \bibinfo{year}{2022}\natexlab{}.
\newblock \showarticletitle{Toward User-Driven Algorithm Auditing: Investigating users’ strategies for uncovering harmful algorithmic behavior}. In \bibinfo{booktitle}{\emph{CHI Conference on Human Factors in Computing Systems}}. \bibinfo{pages}{1--19}.
\newblock


\bibitem[\protect\citeauthoryear{Diakopoulos and Naaman}{Diakopoulos and Naaman}{2011}]%
        {diakopoulos2011towards}
\bibfield{author}{\bibinfo{person}{Nicholas Diakopoulos} {and} \bibinfo{person}{Mor Naaman}.} \bibinfo{year}{2011}\natexlab{}.
\newblock \showarticletitle{Towards quality discourse in online news comments}. In \bibinfo{booktitle}{\emph{Proceedings of the ACM 2011 conference on Computer supported cooperative work}}. \bibinfo{pages}{133--142}.
\newblock


\bibitem[\protect\citeauthoryear{Dimond, Dye, LaRose, and Bruckman}{Dimond et~al\mbox{.}}{2013}]%
        {dimond2013hollaback}
\bibfield{author}{\bibinfo{person}{Jill~P Dimond}, \bibinfo{person}{Michaelanne Dye}, \bibinfo{person}{Daphne LaRose}, {and} \bibinfo{person}{Amy~S Bruckman}.} \bibinfo{year}{2013}\natexlab{}.
\newblock \showarticletitle{Hollaback! The role of storytelling online in a social movement organization}. In \bibinfo{booktitle}{\emph{Proceedings of the 2013 Conference on Computer Supported Cooperative Work}}. \bibinfo{pages}{477--490}.
\newblock


\bibitem[\protect\citeauthoryear{DiSalvo, Sengers, and Brynjarsd{\'o}ttir}{DiSalvo et~al\mbox{.}}{2010}]%
        {disalvo2010mapping}
\bibfield{author}{\bibinfo{person}{Carl DiSalvo}, \bibinfo{person}{Phoebe Sengers}, {and} \bibinfo{person}{Hr{\"o}nn Brynjarsd{\'o}ttir}.} \bibinfo{year}{2010}\natexlab{}.
\newblock \showarticletitle{Mapping the landscape of sustainable HCI}. In \bibinfo{booktitle}{\emph{Proceedings of the SIGCHI conference on human factors in computing systems}}. \bibinfo{pages}{1975--1984}.
\newblock


\bibitem[\protect\citeauthoryear{Donath}{Donath}{2002}]%
        {donath2002identity}
\bibfield{author}{\bibinfo{person}{Judith~S Donath}.} \bibinfo{year}{2002}\natexlab{}.
\newblock \bibinfo{booktitle}{\emph{Identity and deception in the virtual community}}.
\newblock \bibinfo{publisher}{Routledge}.
\newblock


\bibitem[\protect\citeauthoryear{Dourish}{Dourish}{2001}]%
        {dourish2001action}
\bibfield{author}{\bibinfo{person}{Paul Dourish}.} \bibinfo{year}{2001}\natexlab{}.
\newblock \bibinfo{booktitle}{\emph{Where the action is}}.
\newblock \bibinfo{publisher}{MIT press Cambridge}.
\newblock


\bibitem[\protect\citeauthoryear{Dourish}{Dourish}{2006}]%
        {dourish2006implications}
\bibfield{author}{\bibinfo{person}{Paul Dourish}.} \bibinfo{year}{2006}\natexlab{}.
\newblock \showarticletitle{Implications for design}. In \bibinfo{booktitle}{\emph{Proceedings of the SIGCHI conference on Human Factors in computing systems}}. \bibinfo{pages}{541--550}.
\newblock


\bibitem[\protect\citeauthoryear{Dourish}{Dourish}{2008}]%
        {dourish2008points}
\bibfield{author}{\bibinfo{person}{Paul Dourish}.} \bibinfo{year}{2008}\natexlab{}.
\newblock \showarticletitle{Points of persuasion: Strategic essentialism and environmental sustainability}. In \bibinfo{booktitle}{\emph{Persuasive Pervasive Technology and Environmental Sustainability, Workshop at Pervasive}}, Vol.~\bibinfo{volume}{2008}.
\newblock


\bibitem[\protect\citeauthoryear{Dourish, Finlay, Sengers, and Wright}{Dourish et~al\mbox{.}}{2004}]%
        {dourish2004reflective}
\bibfield{author}{\bibinfo{person}{Paul Dourish}, \bibinfo{person}{Janet Finlay}, \bibinfo{person}{Phoebe Sengers}, {and} \bibinfo{person}{Peter Wright}.} \bibinfo{year}{2004}\natexlab{}.
\newblock \showarticletitle{Reflective HCI: Towards a critical technical practice}. In \bibinfo{booktitle}{\emph{CHI'04 extended abstracts on Human factors in computing systems}}. \bibinfo{pages}{1727--1728}.
\newblock


\bibitem[\protect\citeauthoryear{Dove and Fayard}{Dove and Fayard}{2020}]%
        {dove2020monsters}
\bibfield{author}{\bibinfo{person}{Graham Dove} {and} \bibinfo{person}{Anne-Laure Fayard}.} \bibinfo{year}{2020}\natexlab{}.
\newblock \showarticletitle{Monsters, metaphors, and machine learning}. In \bibinfo{booktitle}{\emph{Proceedings of the 2020 CHI Conference on Human Factors in Computing Systems}}. \bibinfo{pages}{1--17}.
\newblock


\bibitem[\protect\citeauthoryear{Dubois, Maftouni, Chilana, McGrenere, and Bunt}{Dubois et~al\mbox{.}}{2020}]%
        {dubois2020gender}
\bibfield{author}{\bibinfo{person}{Patrick Marcel~Joseph Dubois}, \bibinfo{person}{Mahya Maftouni}, \bibinfo{person}{Parmit~K Chilana}, \bibinfo{person}{Joanna McGrenere}, {and} \bibinfo{person}{Andrea Bunt}.} \bibinfo{year}{2020}\natexlab{}.
\newblock \showarticletitle{Gender Differences in Graphic Design Q\&As: How Community and Site Characteristics Contribute to Gender Gaps in Answering Questions}.
\newblock \bibinfo{journal}{\emph{Proceedings of the ACM on Human-Computer Interaction}} \bibinfo{volume}{4}, \bibinfo{number}{CSCW2} (\bibinfo{year}{2020}), \bibinfo{pages}{1--26}.
\newblock


\bibitem[\protect\citeauthoryear{Dunna, Keith, Zuckerman, Vallina-Rodriguez, O'Connor, and Nithyanand}{Dunna et~al\mbox{.}}{2022}]%
        {Dunna2022demonetization}
\bibfield{author}{\bibinfo{person}{Arun Dunna}, \bibinfo{person}{Katherine~A. Keith}, \bibinfo{person}{Ethan Zuckerman}, \bibinfo{person}{Narseo Vallina-Rodriguez}, \bibinfo{person}{Brendan O'Connor}, {and} \bibinfo{person}{Rishab Nithyanand}.} \bibinfo{year}{2022}\natexlab{}.
\newblock \showarticletitle{Paying Attention to the Algorithm Behind the Curtain: Bringing Transparency to YouTube's Demonetization Algorithms}.
\newblock \bibinfo{journal}{\emph{Proc. ACM Hum.-Comput. Interact.}} \bibinfo{volume}{6}, \bibinfo{number}{CSCW2}, Article \bibinfo{articleno}{318} (\bibinfo{date}{nov} \bibinfo{year}{2022}), \bibinfo{numpages}{31}~pages.
\newblock
\urldef\tempurl%
\url{https://doi.org/10.1145/3555209}
\showDOI{\tempurl}


\bibitem[\protect\citeauthoryear{Dym, Brubaker, Fiesler, and Semaan}{Dym et~al\mbox{.}}{2019}]%
        {dym2019coming}
\bibfield{author}{\bibinfo{person}{Brianna Dym}, \bibinfo{person}{Jed~R Brubaker}, \bibinfo{person}{Casey Fiesler}, {and} \bibinfo{person}{Bryan Semaan}.} \bibinfo{year}{2019}\natexlab{}.
\newblock \showarticletitle{" Coming Out Okay" Community Narratives for LGBTQ Identity Recovery Work}.
\newblock \bibinfo{journal}{\emph{Proceedings of the ACM on Human-Computer Interaction}} \bibinfo{volume}{3}, \bibinfo{number}{CSCW} (\bibinfo{year}{2019}), \bibinfo{pages}{1--28}.
\newblock


\bibitem[\protect\citeauthoryear{Dym and Fiesler}{Dym and Fiesler}{2020}]%
        {dym2020social}
\bibfield{author}{\bibinfo{person}{Brianna Dym} {and} \bibinfo{person}{Casey Fiesler}.} \bibinfo{year}{2020}\natexlab{}.
\newblock \showarticletitle{Social norm vulnerability and its consequences for privacy and safety in an online community}.
\newblock \bibinfo{journal}{\emph{Proceedings of the ACM on Human-Computer Interaction}} \bibinfo{volume}{4}, \bibinfo{number}{CSCW2} (\bibinfo{year}{2020}), \bibinfo{pages}{1--24}.
\newblock


\bibitem[\protect\citeauthoryear{Early, Hammer, Hofmann, Rode, Wong, and Mankoff}{Early et~al\mbox{.}}{2018}]%
        {early2018understanding}
\bibfield{author}{\bibinfo{person}{Kirstin Early}, \bibinfo{person}{Jessica Hammer}, \bibinfo{person}{Megan~Kelly Hofmann}, \bibinfo{person}{Jennifer~A Rode}, \bibinfo{person}{Anna Wong}, {and} \bibinfo{person}{Jennifer Mankoff}.} \bibinfo{year}{2018}\natexlab{}.
\newblock \showarticletitle{Understanding gender equity in author order assignment}.
\newblock \bibinfo{journal}{\emph{Proceedings of the ACM on Human-Computer Interaction}} \bibinfo{volume}{2}, \bibinfo{number}{CSCW} (\bibinfo{year}{2018}), \bibinfo{pages}{1--21}.
\newblock


\bibitem[\protect\citeauthoryear{Erete, Rankin, and Thomas}{Erete et~al\mbox{.}}{2021}]%
        {erete2021can}
\bibfield{author}{\bibinfo{person}{Sheena Erete}, \bibinfo{person}{Yolanda~A Rankin}, {and} \bibinfo{person}{Jakita~O Thomas}.} \bibinfo{year}{2021}\natexlab{}.
\newblock \showarticletitle{I can't breathe: Reflections from Black women in CSCW and HCI}.
\newblock \bibinfo{journal}{\emph{Proceedings of the ACM on Human-Computer Interaction}} \bibinfo{volume}{4}, \bibinfo{number}{CSCW3} (\bibinfo{year}{2021}), \bibinfo{pages}{1--23}.
\newblock


\bibitem[\protect\citeauthoryear{Farnham and Churchill}{Farnham and Churchill}{2011}]%
        {farnham2011faceted}
\bibfield{author}{\bibinfo{person}{Shelly~D Farnham} {and} \bibinfo{person}{Elizabeth~F Churchill}.} \bibinfo{year}{2011}\natexlab{}.
\newblock \showarticletitle{Faceted identity, faceted lives: social and technical issues with being yourself online}. In \bibinfo{booktitle}{\emph{Proceedings of the ACM 2011 conference on Computer supported cooperative work}}. \bibinfo{pages}{359--368}.
\newblock


\bibitem[\protect\citeauthoryear{Farny, Jennex, Olsen, and Rodriguez}{Farny et~al\mbox{.}}{2012}]%
        {farny2012anchor}
\bibfield{author}{\bibinfo{person}{Jacob Farny}, \bibinfo{person}{Matthew Jennex}, \bibinfo{person}{Rebekah Olsen}, {and} \bibinfo{person}{Melissa Rodriguez}.} \bibinfo{year}{2012}\natexlab{}.
\newblock \showarticletitle{Anchor: connecting sailors to home}.
\newblock In \bibinfo{booktitle}{\emph{CHI'12 Extended Abstracts on Human Factors in Computing Systems}}. \bibinfo{pages}{1267--1272}.
\newblock


\bibitem[\protect\citeauthoryear{Feinberg, Carter, and Bullard}{Feinberg et~al\mbox{.}}{2014}]%
        {feinberg2014always}
\bibfield{author}{\bibinfo{person}{Melanie Feinberg}, \bibinfo{person}{Daniel Carter}, {and} \bibinfo{person}{Julia Bullard}.} \bibinfo{year}{2014}\natexlab{}.
\newblock \showarticletitle{Always somewhere, never there: using critical design to understand database interactions}. In \bibinfo{booktitle}{\emph{Proceedings of the SIGCHI Conference on Human Factors in Computing Systems}}. \bibinfo{pages}{1941--1950}.
\newblock


\bibitem[\protect\citeauthoryear{Fernandez and Birnholtz}{Fernandez and Birnholtz}{2019}]%
        {fernandez2019don}
\bibfield{author}{\bibinfo{person}{Julia~R Fernandez} {and} \bibinfo{person}{Jeremy Birnholtz}.} \bibinfo{year}{2019}\natexlab{}.
\newblock \showarticletitle{" I Don't Want Them to Not Know" Investigating Decisions to Disclose Transgender Identity on Dating Platforms}.
\newblock \bibinfo{journal}{\emph{Proceedings of the ACM on Human-Computer Interaction}} \bibinfo{volume}{3}, \bibinfo{number}{CSCW} (\bibinfo{year}{2019}), \bibinfo{pages}{1--21}.
\newblock


\bibitem[\protect\citeauthoryear{Foong and Gerber}{Foong and Gerber}{2021}]%
        {foong2021understanding}
\bibfield{author}{\bibinfo{person}{Eureka Foong} {and} \bibinfo{person}{Elizabeth Gerber}.} \bibinfo{year}{2021}\natexlab{}.
\newblock \showarticletitle{Understanding Gender Differences in Pricing Strategies in Online Labor Marketplaces}. In \bibinfo{booktitle}{\emph{Proceedings of the 2021 CHI Conference on Human Factors in Computing Systems}}. \bibinfo{pages}{1--16}.
\newblock


\bibitem[\protect\citeauthoryear{Foucault}{Foucault}{1990}]%
        {foucault1990history}
\bibfield{author}{\bibinfo{person}{Michel Foucault}.} \bibinfo{year}{1990}\natexlab{}.
\newblock \bibinfo{booktitle}{\emph{The history of sexuality: An introduction}}.
\newblock \bibinfo{publisher}{Vintage}.
\newblock


\bibitem[\protect\citeauthoryear{Freeman, Bardzell, Bardzell, and Herring}{Freeman et~al\mbox{.}}{2015}]%
        {freeman2015simulating}
\bibfield{author}{\bibinfo{person}{Guo Freeman}, \bibinfo{person}{Jeffrey Bardzell}, \bibinfo{person}{Shaowen Bardzell}, {and} \bibinfo{person}{Susan~C Herring}.} \bibinfo{year}{2015}\natexlab{}.
\newblock \showarticletitle{Simulating marriage: Gender roles and emerging intimacy in an online game}. In \bibinfo{booktitle}{\emph{Proceedings of the 18th ACM Conference on Computer Supported Cooperative Work \& Social Computing}}. \bibinfo{pages}{1191--1200}.
\newblock


\bibitem[\protect\citeauthoryear{Freeman and Maloney}{Freeman and Maloney}{2021}]%
        {freeman2021body}
\bibfield{author}{\bibinfo{person}{Guo Freeman} {and} \bibinfo{person}{Divine Maloney}.} \bibinfo{year}{2021}\natexlab{}.
\newblock \showarticletitle{Body, avatar, and me: The presentation and perception of self in social virtual reality}.
\newblock \bibinfo{journal}{\emph{Proceedings of the ACM on Human-Computer Interaction}} \bibinfo{volume}{4}, \bibinfo{number}{CSCW3} (\bibinfo{year}{2021}), \bibinfo{pages}{1--27}.
\newblock


\bibitem[\protect\citeauthoryear{Freeman, Zamanifard, Maloney, and Acena}{Freeman et~al\mbox{.}}{2022}]%
        {freeman2022disturbing}
\bibfield{author}{\bibinfo{person}{Guo Freeman}, \bibinfo{person}{Samaneh Zamanifard}, \bibinfo{person}{Divine Maloney}, {and} \bibinfo{person}{Dane Acena}.} \bibinfo{year}{2022}\natexlab{}.
\newblock \showarticletitle{Disturbing the Peace: Experiencing and Mitigating Emerging Harassment in Social Virtual Reality}.
\newblock \bibinfo{journal}{\emph{Proceedings of the ACM on Human-Computer Interaction}} \bibinfo{volume}{6}, \bibinfo{number}{CSCW1} (\bibinfo{year}{2022}), \bibinfo{pages}{1--30}.
\newblock


\bibitem[\protect\citeauthoryear{Freeman, Zamanifard, Maloney, and Adkins}{Freeman et~al\mbox{.}}{2020}]%
        {freeman2020my}
\bibfield{author}{\bibinfo{person}{Guo Freeman}, \bibinfo{person}{Samaneh Zamanifard}, \bibinfo{person}{Divine Maloney}, {and} \bibinfo{person}{Alexandra Adkins}.} \bibinfo{year}{2020}\natexlab{}.
\newblock \showarticletitle{My body, my avatar: How people perceive their avatars in social virtual reality}. In \bibinfo{booktitle}{\emph{Extended Abstracts of the 2020 CHI Conference on Human Factors in Computing Systems}}. \bibinfo{pages}{1--8}.
\newblock


\bibitem[\protect\citeauthoryear{Furlo, Gleason, Feun, and Zytko}{Furlo et~al\mbox{.}}{2021}]%
        {furlo2021rethinking}
\bibfield{author}{\bibinfo{person}{Nicholas Furlo}, \bibinfo{person}{Jacob Gleason}, \bibinfo{person}{Karen Feun}, {and} \bibinfo{person}{Douglas Zytko}.} \bibinfo{year}{2021}\natexlab{}.
\newblock \showarticletitle{Rethinking Dating Apps as Sexual Consent Apps: A New Use Case for AI-Mediated Communication}. In \bibinfo{booktitle}{\emph{Companion Publication of the 2021 Conference on Computer Supported Cooperative Work and Social Computing}}. \bibinfo{pages}{53--56}.
\newblock


\bibitem[\protect\citeauthoryear{Gatehouse, Wood, Briggs, Pickles, and Lawson}{Gatehouse et~al\mbox{.}}{2018}]%
        {gatehouse2018troubling}
\bibfield{author}{\bibinfo{person}{Cally Gatehouse}, \bibinfo{person}{Matthew Wood}, \bibinfo{person}{Jo Briggs}, \bibinfo{person}{James Pickles}, {and} \bibinfo{person}{Shaun Lawson}.} \bibinfo{year}{2018}\natexlab{}.
\newblock \showarticletitle{Troubling vulnerability: Designing with LGBT young people's ambivalence towards hate crime reporting}. In \bibinfo{booktitle}{\emph{Proceedings of the 2018 CHI Conference on Human Factors in Computing Systems}}. \bibinfo{pages}{1--13}.
\newblock


\bibitem[\protect\citeauthoryear{Geeng, Yee, and Roesner}{Geeng et~al\mbox{.}}{2020}]%
        {geeng2020fake}
\bibfield{author}{\bibinfo{person}{Christine Geeng}, \bibinfo{person}{Savanna Yee}, {and} \bibinfo{person}{Franziska Roesner}.} \bibinfo{year}{2020}\natexlab{}.
\newblock \showarticletitle{Fake news on Facebook and Twitter: Investigating how people (don't) investigate}. In \bibinfo{booktitle}{\emph{Proceedings of the 2020 CHI conference on human factors in computing systems}}. \bibinfo{pages}{1--14}.
\newblock


\bibitem[\protect\citeauthoryear{Goffman}{Goffman}{2009}]%
        {goffman2009stigma}
\bibfield{author}{\bibinfo{person}{Erving Goffman}.} \bibinfo{year}{2009}\natexlab{}.
\newblock \bibinfo{booktitle}{\emph{Stigma: Notes on the management of spoiled identity}}.
\newblock \bibinfo{publisher}{Simon and Schuster}.
\newblock


\bibitem[\protect\citeauthoryear{Golder and Macy}{Golder and Macy}{2011}]%
        {golder2011diurnal}
\bibfield{author}{\bibinfo{person}{Scott~A Golder} {and} \bibinfo{person}{Michael~W Macy}.} \bibinfo{year}{2011}\natexlab{}.
\newblock \showarticletitle{Diurnal and seasonal mood vary with work, sleep, and daylength across diverse cultures}.
\newblock \bibinfo{journal}{\emph{Science}} \bibinfo{volume}{333}, \bibinfo{number}{6051} (\bibinfo{year}{2011}), \bibinfo{pages}{1878--1881}.
\newblock


\bibitem[\protect\citeauthoryear{Gonzales and Fritz}{Gonzales and Fritz}{2017}]%
        {gonzales2017prioritizing}
\bibfield{author}{\bibinfo{person}{Amy Gonzales} {and} \bibinfo{person}{Nicole Fritz}.} \bibinfo{year}{2017}\natexlab{}.
\newblock \showarticletitle{Prioritizing flexibility and intangibles: Medical crowdfunding for stigmatized individuals}. In \bibinfo{booktitle}{\emph{Proceedings of the 2017 CHI conference on human factors in computing systems}}. \bibinfo{pages}{2371--2375}.
\newblock


\bibitem[\protect\citeauthoryear{Gray}{Gray}{2007}]%
        {gray2007websites}
\bibfield{author}{\bibinfo{person}{Mary~L Gray}.} \bibinfo{year}{2007}\natexlab{}.
\newblock \showarticletitle{From websites to Wal-Mart: Youth, identity work, and the queering of boundary publics in Small Town, USA}.
\newblock \bibinfo{journal}{\emph{American Studies}} \bibinfo{volume}{48}, \bibinfo{number}{2} (\bibinfo{year}{2007}), \bibinfo{pages}{49--59}.
\newblock


\bibitem[\protect\citeauthoryear{Green}{Green}{2021}]%
        {green2021data}
\bibfield{author}{\bibinfo{person}{Ben Green}.} \bibinfo{year}{2021}\natexlab{}.
\newblock \showarticletitle{Data science as political action: Grounding data science in a politics of justice}.
\newblock \bibinfo{journal}{\emph{Journal of Social Computing}} \bibinfo{volume}{2}, \bibinfo{number}{3} (\bibinfo{year}{2021}), \bibinfo{pages}{249--265}.
\newblock


\bibitem[\protect\citeauthoryear{Grevet, Terveen, and Gilbert}{Grevet et~al\mbox{.}}{2014}]%
        {grevet2014managing}
\bibfield{author}{\bibinfo{person}{Catherine Grevet}, \bibinfo{person}{Loren~G Terveen}, {and} \bibinfo{person}{Eric Gilbert}.} \bibinfo{year}{2014}\natexlab{}.
\newblock \showarticletitle{Managing political differences in social media}. In \bibinfo{booktitle}{\emph{Proceedings of the 17th ACM conference on Computer supported cooperative work \& social computing}}. \bibinfo{pages}{1400--1408}.
\newblock


\bibitem[\protect\citeauthoryear{Guberek, McDonald, Simioni, Mhaidli, Toyama, and Schaub}{Guberek et~al\mbox{.}}{2018}]%
        {guberek2018keeping}
\bibfield{author}{\bibinfo{person}{Tamy Guberek}, \bibinfo{person}{Allison McDonald}, \bibinfo{person}{Sylvia Simioni}, \bibinfo{person}{Abraham~H Mhaidli}, \bibinfo{person}{Kentaro Toyama}, {and} \bibinfo{person}{Florian Schaub}.} \bibinfo{year}{2018}\natexlab{}.
\newblock \showarticletitle{Keeping a low profile? Technology, risk and privacy among undocumented immigrants}. In \bibinfo{booktitle}{\emph{Proceedings of the 2018 CHI conference on human factors in computing systems}}. \bibinfo{pages}{1--15}.
\newblock


\bibitem[\protect\citeauthoryear{Gulotta, Faste, and Mankoff}{Gulotta et~al\mbox{.}}{2012}]%
        {gulotta2012curation}
\bibfield{author}{\bibinfo{person}{Rebecca Gulotta}, \bibinfo{person}{Haakon Faste}, {and} \bibinfo{person}{Jennifer Mankoff}.} \bibinfo{year}{2012}\natexlab{}.
\newblock \showarticletitle{Curation, provocation, and digital identity: risks and motivations for sharing provocative images online}. In \bibinfo{booktitle}{\emph{Proceedings of the SIGCHI Conference on Human Factors in Computing Systems}}. \bibinfo{pages}{387--390}.
\newblock


\bibitem[\protect\citeauthoryear{Haimson}{Haimson}{2017}]%
        {haimson2017social}
\bibfield{author}{\bibinfo{person}{Oliver~L Haimson}.} \bibinfo{year}{2017}\natexlab{}.
\newblock \showarticletitle{The social complexities of transgender identity disclosure on social network sites}. In \bibinfo{booktitle}{\emph{Proceedings of the 2017 CHI Conference Extended Abstracts on Human Factors in Computing Systems}}. \bibinfo{pages}{280--285}.
\newblock


\bibitem[\protect\citeauthoryear{Haimson, Brubaker, Dombrowski, and Hayes}{Haimson et~al\mbox{.}}{2015}]%
        {haimson2015disclosure}
\bibfield{author}{\bibinfo{person}{Oliver~L Haimson}, \bibinfo{person}{Jed~R Brubaker}, \bibinfo{person}{Lynn Dombrowski}, {and} \bibinfo{person}{Gillian~R Hayes}.} \bibinfo{year}{2015}\natexlab{}.
\newblock \showarticletitle{Disclosure, stress, and support during gender transition on Facebook}. In \bibinfo{booktitle}{\emph{Proceedings of the 18th ACM conference on computer supported cooperative work \& social computing}}. \bibinfo{pages}{1176--1190}.
\newblock


\bibitem[\protect\citeauthoryear{Haimson, Brubaker, Dombrowski, and Hayes}{Haimson et~al\mbox{.}}{2016}]%
        {haimson2016digital}
\bibfield{author}{\bibinfo{person}{Oliver~L. Haimson}, \bibinfo{person}{Jed~R. Brubaker}, \bibinfo{person}{Lynn Dombrowski}, {and} \bibinfo{person}{Gillian~R. Hayes}.} \bibinfo{year}{2016}\natexlab{}.
\newblock \showarticletitle{Digital Footprints and Changing Networks During Online Identity Transitions}. In \bibinfo{booktitle}{\emph{Proceedings of the 2016 CHI Conference on Human Factors in Computing Systems}} (San Jose, California, USA) \emph{(\bibinfo{series}{CHI '16})}. \bibinfo{publisher}{Association for Computing Machinery}, \bibinfo{address}{New York, NY, USA}, \bibinfo{pages}{2895–2907}.
\newblock
\showISBNx{9781450333627}
\urldef\tempurl%
\url{https://doi.org/10.1145/2858036.2858136}
\showDOI{\tempurl}


\bibitem[\protect\citeauthoryear{Haimson, Brubaker, and Hayes}{Haimson et~al\mbox{.}}{2014}]%
        {haimson2014ddfseeks}
\bibfield{author}{\bibinfo{person}{Oliver~L Haimson}, \bibinfo{person}{Jed~R Brubaker}, {and} \bibinfo{person}{Gillian~R Hayes}.} \bibinfo{year}{2014}\natexlab{}.
\newblock \showarticletitle{DDFSeeks same: sexual health-related language in online personal ads for men who have sex with men}. In \bibinfo{booktitle}{\emph{Proceedings of the SIGCHI Conference on Human Factors in Computing Systems}}. \bibinfo{pages}{1615--1624}.
\newblock


\bibitem[\protect\citeauthoryear{Haimson, Buss, Weinger, Starks, Gorrell, and Baron}{Haimson et~al\mbox{.}}{2020a}]%
        {haimson2020trans}
\bibfield{author}{\bibinfo{person}{Oliver~L Haimson}, \bibinfo{person}{Justin Buss}, \bibinfo{person}{Zu Weinger}, \bibinfo{person}{Denny~L Starks}, \bibinfo{person}{Dykee Gorrell}, {and} \bibinfo{person}{Briar~Sweetbriar Baron}.} \bibinfo{year}{2020}\natexlab{a}.
\newblock \showarticletitle{Trans Time: Safety, Privacy, and Content Warnings on a Transgender-Specific Social Media Site}.
\newblock \bibinfo{journal}{\emph{Proceedings of the ACM on Human-Computer Interaction}} \bibinfo{volume}{4}, \bibinfo{number}{CSCW2} (\bibinfo{year}{2020}), \bibinfo{pages}{1--27}.
\newblock


\bibitem[\protect\citeauthoryear{Haimson, Delmonaco, Nie, and Wegner}{Haimson et~al\mbox{.}}{2021}]%
        {haimson2021disproportionate}
\bibfield{author}{\bibinfo{person}{Oliver~L Haimson}, \bibinfo{person}{Daniel Delmonaco}, \bibinfo{person}{Peipei Nie}, {and} \bibinfo{person}{Andrea Wegner}.} \bibinfo{year}{2021}\natexlab{}.
\newblock \showarticletitle{Disproportionate removals and differing content moderation experiences for conservative, transgender, and black social media users: Marginalization and moderation gray areas}.
\newblock \bibinfo{journal}{\emph{Proceedings of the ACM on Human-Computer Interaction}} \bibinfo{volume}{5}, \bibinfo{number}{CSCW2} (\bibinfo{year}{2021}), \bibinfo{pages}{1--35}.
\newblock


\bibitem[\protect\citeauthoryear{Haimson, Gorrell, Starks, and Weinger}{Haimson et~al\mbox{.}}{2020b}]%
        {haimson2020designing}
\bibfield{author}{\bibinfo{person}{Oliver~L Haimson}, \bibinfo{person}{Dykee Gorrell}, \bibinfo{person}{Denny~L Starks}, {and} \bibinfo{person}{Zu Weinger}.} \bibinfo{year}{2020}\natexlab{b}.
\newblock \showarticletitle{Designing trans technology: Defining challenges and envisioning community-centered solutions}. In \bibinfo{booktitle}{\emph{Proceedings of the 2020 CHI Conference on Human Factors in Computing Systems}}. \bibinfo{pages}{1--13}.
\newblock


\bibitem[\protect\citeauthoryear{Haimson and Hoffmann}{Haimson and Hoffmann}{2016}]%
        {haimson2016constructing}
\bibfield{author}{\bibinfo{person}{Oliver~L Haimson} {and} \bibinfo{person}{Anna~Lauren Hoffmann}.} \bibinfo{year}{2016}\natexlab{}.
\newblock \showarticletitle{Constructing and enforcing" authentic" identity online: Facebook, real names, and non-normative identities}.
\newblock \bibinfo{journal}{\emph{First Monday}} (\bibinfo{year}{2016}).
\newblock


\bibitem[\protect\citeauthoryear{Halberstam}{Halberstam}{2011}]%
        {halberstam2011queer}
\bibfield{author}{\bibinfo{person}{Jack Halberstam}.} \bibinfo{year}{2011}\natexlab{}.
\newblock \showarticletitle{The queer art of failure}.
\newblock In \bibinfo{booktitle}{\emph{The queer art of failure}}. \bibinfo{publisher}{Duke University Press}.
\newblock


\bibitem[\protect\citeauthoryear{Hamidi, Scheuerman, and Branham}{Hamidi et~al\mbox{.}}{2018}]%
        {hamidi2018gender}
\bibfield{author}{\bibinfo{person}{Foad Hamidi}, \bibinfo{person}{Morgan~Klaus Scheuerman}, {and} \bibinfo{person}{Stacy~M Branham}.} \bibinfo{year}{2018}\natexlab{}.
\newblock \showarticletitle{Gender recognition or gender reductionism? The social implications of embedded gender recognition systems}. In \bibinfo{booktitle}{\emph{Proceedings of the 2018 chi conference on human factors in computing systems}}. \bibinfo{pages}{1--13}.
\newblock


\bibitem[\protect\citeauthoryear{Hamilton, Karahalios, Sandvig, and Eslami}{Hamilton et~al\mbox{.}}{2014}]%
        {hamilton2014path}
\bibfield{author}{\bibinfo{person}{Kevin Hamilton}, \bibinfo{person}{Karrie Karahalios}, \bibinfo{person}{Christian Sandvig}, {and} \bibinfo{person}{Motahhare Eslami}.} \bibinfo{year}{2014}\natexlab{}.
\newblock \showarticletitle{A path to understanding the effects of algorithm awareness}.
\newblock In \bibinfo{booktitle}{\emph{CHI'14 extended abstracts on human factors in computing systems}}. \bibinfo{pages}{631--642}.
\newblock


\bibitem[\protect\citeauthoryear{Hamilton, Barakat, and Redmiles}{Hamilton et~al\mbox{.}}{2022}]%
        {Hamilton2022SexWork}
\bibfield{author}{\bibinfo{person}{Vaughn Hamilton}, \bibinfo{person}{Hanna Barakat}, {and} \bibinfo{person}{Elissa~M. Redmiles}.} \bibinfo{year}{2022}\natexlab{}.
\newblock \showarticletitle{Risk, Resilience and Reward: Impacts of Shifting to Digital Sex Work}.
\newblock \bibinfo{journal}{\emph{Proc. ACM Hum.-Comput. Interact.}} \bibinfo{volume}{6}, \bibinfo{number}{CSCW2}, Article \bibinfo{articleno}{537} (\bibinfo{date}{nov} \bibinfo{year}{2022}), \bibinfo{numpages}{37}~pages.
\newblock
\urldef\tempurl%
\url{https://doi.org/10.1145/3555650}
\showDOI{\tempurl}


\bibitem[\protect\citeauthoryear{Haque, Saha, Rahman, and Ahmed}{Haque et~al\mbox{.}}{2019}]%
        {haque2019ulti}
\bibfield{author}{\bibinfo{person}{SM~Taiabul Haque}, \bibinfo{person}{Pratyasha Saha}, \bibinfo{person}{Muhammad~Sajidur Rahman}, {and} \bibinfo{person}{Syed~Ishtiaque Ahmed}.} \bibinfo{year}{2019}\natexlab{}.
\newblock \showarticletitle{Of Ulti,'hajano', and" Matachetar otanetak datam" Exploring Local Practices of Exchanging Confidential and Sensitive Information in Urban Bangladesh}.
\newblock \bibinfo{journal}{\emph{Proceedings of the ACM on Human-Computer Interaction}} \bibinfo{volume}{3}, \bibinfo{number}{CSCW} (\bibinfo{year}{2019}), \bibinfo{pages}{1--22}.
\newblock


\bibitem[\protect\citeauthoryear{Haraway}{Haraway}{1988}]%
        {haraway1988situated}
\bibfield{author}{\bibinfo{person}{Donna Haraway}.} \bibinfo{year}{1988}\natexlab{}.
\newblock \showarticletitle{Situated knowledges: The science question in feminism and the privilege of partial perspective}.
\newblock \bibinfo{journal}{\emph{Feminist studies}} \bibinfo{volume}{14}, \bibinfo{number}{3} (\bibinfo{year}{1988}), \bibinfo{pages}{575--599}.
\newblock


\bibitem[\protect\citeauthoryear{Hardy, Geier, Vargas, Doll, and Howard}{Hardy et~al\mbox{.}}{2022}]%
        {hardy2022lgbtq}
\bibfield{author}{\bibinfo{person}{Jean Hardy}, \bibinfo{person}{Caitlin Geier}, \bibinfo{person}{Stefani Vargas}, \bibinfo{person}{Riley Doll}, {and} \bibinfo{person}{Amy~Lyn Howard}.} \bibinfo{year}{2022}\natexlab{}.
\newblock \showarticletitle{LGBTQ Futures and Participatory Design: Investigating Visibility, Community, and the Future of Future Workshops}.
\newblock \bibinfo{journal}{\emph{Proc. ACM Hum.-Comput. Interact.}} \bibinfo{volume}{6}, \bibinfo{number}{CSCW2}, Article \bibinfo{articleno}{525} (\bibinfo{date}{nov} \bibinfo{year}{2022}), \bibinfo{numpages}{25}~pages.
\newblock
\urldef\tempurl%
\url{https://doi.org/10.1145/3555638}
\showDOI{\tempurl}


\bibitem[\protect\citeauthoryear{Hardy and Lindtner}{Hardy and Lindtner}{2017}]%
        {hardy2017constructing}
\bibfield{author}{\bibinfo{person}{Jean Hardy} {and} \bibinfo{person}{Silvia Lindtner}.} \bibinfo{year}{2017}\natexlab{}.
\newblock \showarticletitle{Constructing a desiring user: Discourse, rurality, and design in location-based social networks}. In \bibinfo{booktitle}{\emph{Proceedings of the 2017 ACM Conference on Computer Supported Cooperative Work and Social Computing}}. \bibinfo{pages}{13--25}.
\newblock


\bibitem[\protect\citeauthoryear{Hardy and Vargas}{Hardy and Vargas}{2019}]%
        {hardy2019participatory}
\bibfield{author}{\bibinfo{person}{Jean Hardy} {and} \bibinfo{person}{Stefani Vargas}.} \bibinfo{year}{2019}\natexlab{}.
\newblock \showarticletitle{Participatory design and the future of rural LGBTQ communities}. In \bibinfo{booktitle}{\emph{Companion Publication of the 2019 on Designing Interactive Systems Conference 2019 Companion}}. \bibinfo{pages}{195--199}.
\newblock


\bibitem[\protect\citeauthoryear{Harrer, Nielsen, and Jarnfelt}{Harrer et~al\mbox{.}}{2019}]%
        {harrer2019mice}
\bibfield{author}{\bibinfo{person}{Sabine Harrer}, \bibinfo{person}{Simon Nielsen}, {and} \bibinfo{person}{Patrick Jarnfelt}.} \bibinfo{year}{2019}\natexlab{}.
\newblock \showarticletitle{Of Mice and Pants: Queering the Conventional Gamer Mouse for Cooperative Play}. In \bibinfo{booktitle}{\emph{Extended Abstracts of the 2019 CHI Conference on Human Factors in Computing Systems}}. \bibinfo{pages}{1--11}.
\newblock


\bibitem[\protect\citeauthoryear{Harrington, Klassen, and Rankin}{Harrington et~al\mbox{.}}{2022}]%
        {harrington2022all}
\bibfield{author}{\bibinfo{person}{Christina~N Harrington}, \bibinfo{person}{Shamika Klassen}, {and} \bibinfo{person}{Yolanda~A Rankin}.} \bibinfo{year}{2022}\natexlab{}.
\newblock \showarticletitle{“All that You Touch, You Change”: Expanding the Canon of Speculative Design Towards Black Futuring}. In \bibinfo{booktitle}{\emph{CHI Conference on Human Factors in Computing Systems}}. \bibinfo{pages}{1--10}.
\newblock


\bibitem[\protect\citeauthoryear{Harrison, Tatar, and Sengers}{Harrison et~al\mbox{.}}{2007}]%
        {harrison2007three}
\bibfield{author}{\bibinfo{person}{Steve Harrison}, \bibinfo{person}{Deborah Tatar}, {and} \bibinfo{person}{Phoebe Sengers}.} \bibinfo{year}{2007}\natexlab{}.
\newblock \showarticletitle{The three paradigms of HCI}. In \bibinfo{booktitle}{\emph{Alt. Chi. Session at the SIGCHI Conference on human factors in computing systems San Jose, California, USA}}. \bibinfo{pages}{1--18}.
\newblock


\bibitem[\protect\citeauthoryear{Holzer, Tintarev, Bendahan, Kocher, Greenup, and Gillet}{Holzer et~al\mbox{.}}{2018}]%
        {holzer2018digitally}
\bibfield{author}{\bibinfo{person}{Adrian Holzer}, \bibinfo{person}{Nava Tintarev}, \bibinfo{person}{Samuel Bendahan}, \bibinfo{person}{Bruno Kocher}, \bibinfo{person}{Shane Greenup}, {and} \bibinfo{person}{Denis Gillet}.} \bibinfo{year}{2018}\natexlab{}.
\newblock \showarticletitle{Digitally scaffolding debate in the classroom}. In \bibinfo{booktitle}{\emph{Extended Abstracts of the 2018 CHI Conference on Human Factors in Computing Systems}}. \bibinfo{pages}{1--6}.
\newblock


\bibitem[\protect\citeauthoryear{Homan, Lu, Tu, Lytle, and Silenzio}{Homan et~al\mbox{.}}{2014}]%
        {homan2014social}
\bibfield{author}{\bibinfo{person}{Christopher~M Homan}, \bibinfo{person}{Naiji Lu}, \bibinfo{person}{Xin Tu}, \bibinfo{person}{Megan~C Lytle}, {and} \bibinfo{person}{Vincent~MB Silenzio}.} \bibinfo{year}{2014}\natexlab{}.
\newblock \showarticletitle{Social structure and depression in TrevorSpace}. In \bibinfo{booktitle}{\emph{Proceedings of the 17th ACM conference on Computer supported cooperative work \& social computing}}. \bibinfo{pages}{615--625}.
\newblock


\bibitem[\protect\citeauthoryear{Huang, Dobreski, and Xia}{Huang et~al\mbox{.}}{2017}]%
        {huang2017human}
\bibfield{author}{\bibinfo{person}{Yun Huang}, \bibinfo{person}{Brian Dobreski}, {and} \bibinfo{person}{Huichuan Xia}.} \bibinfo{year}{2017}\natexlab{}.
\newblock \showarticletitle{Human library: Understanding experience sharing for community knowledge building}. In \bibinfo{booktitle}{\emph{Proceedings of the 2017 ACM conference on computer supported cooperative work and social computing}}. \bibinfo{pages}{1152--1165}.
\newblock


\bibitem[\protect\citeauthoryear{Hube, Fetahu, and Gadiraju}{Hube et~al\mbox{.}}{2019}]%
        {hube2019understanding}
\bibfield{author}{\bibinfo{person}{Christoph Hube}, \bibinfo{person}{Besnik Fetahu}, {and} \bibinfo{person}{Ujwal Gadiraju}.} \bibinfo{year}{2019}\natexlab{}.
\newblock \showarticletitle{Understanding and mitigating worker biases in the crowdsourced collection of subjective judgments}. In \bibinfo{booktitle}{\emph{Proceedings of the 2019 CHI Conference on Human Factors in Computing Systems}}. \bibinfo{pages}{1--12}.
\newblock


\bibitem[\protect\citeauthoryear{Introne, Munoz, and Semaan}{Introne et~al\mbox{.}}{2021}]%
        {introne2021narrative}
\bibfield{author}{\bibinfo{person}{Joshua Introne}, \bibinfo{person}{Isabel Munoz}, {and} \bibinfo{person}{Bryan Semaan}.} \bibinfo{year}{2021}\natexlab{}.
\newblock \showarticletitle{The Narrative Tapestry Design Process: Weaving Online Social Support from Stories of Stigma}.
\newblock \bibinfo{journal}{\emph{Proceedings of the ACM on Human-Computer Interaction}} \bibinfo{volume}{5}, \bibinfo{number}{CSCW2} (\bibinfo{year}{2021}), \bibinfo{pages}{1--35}.
\newblock


\bibitem[\protect\citeauthoryear{Irani, Hayes, and Dourish}{Irani et~al\mbox{.}}{2008}]%
        {irani2008situated}
\bibfield{author}{\bibinfo{person}{Lilly~C Irani}, \bibinfo{person}{Gillian~R Hayes}, {and} \bibinfo{person}{Paul Dourish}.} \bibinfo{year}{2008}\natexlab{}.
\newblock \showarticletitle{Situated practices of looking: visual practice in an online world}. In \bibinfo{booktitle}{\emph{Proceedings of the 2008 ACM conference on Computer supported cooperative work}}. \bibinfo{pages}{187--196}.
\newblock


\bibitem[\protect\citeauthoryear{J.~Rogerson, A.~Sparrow, and R.~Gibbs}{J.~Rogerson et~al\mbox{.}}{2021}]%
        {j2021unpacking}
\bibfield{author}{\bibinfo{person}{Melissa J.~Rogerson}, \bibinfo{person}{Lucy A.~Sparrow}, {and} \bibinfo{person}{Martin R.~Gibbs}.} \bibinfo{year}{2021}\natexlab{}.
\newblock \showarticletitle{Unpacking “Boardgames with Apps”: The Hybrid Digital Boardgame Model}. In \bibinfo{booktitle}{\emph{Proceedings of the 2021 CHI Conference on Human Factors in Computing Systems}}. \bibinfo{pages}{1--17}.
\newblock


\bibitem[\protect\citeauthoryear{Jaroszewski, Lottridge, Haimson, and Quehl}{Jaroszewski et~al\mbox{.}}{2018}]%
        {jaroszewski2018genderfluid}
\bibfield{author}{\bibinfo{person}{Samantha Jaroszewski}, \bibinfo{person}{Danielle Lottridge}, \bibinfo{person}{Oliver~L Haimson}, {and} \bibinfo{person}{Katie Quehl}.} \bibinfo{year}{2018}\natexlab{}.
\newblock \showarticletitle{" Genderfluid" or" Attack Helicopter" Responsible HCI Research Practice with Non-binary Gender Variation in Online Communities}. In \bibinfo{booktitle}{\emph{Proceedings of the 2018 CHI conference on human factors in computing systems}}. \bibinfo{pages}{1--15}.
\newblock


\bibitem[\protect\citeauthoryear{Jhaver, Ghoshal, Bruckman, and Gilbert}{Jhaver et~al\mbox{.}}{2018}]%
        {jhaver2018online}
\bibfield{author}{\bibinfo{person}{Shagun Jhaver}, \bibinfo{person}{Sucheta Ghoshal}, \bibinfo{person}{Amy Bruckman}, {and} \bibinfo{person}{Eric Gilbert}.} \bibinfo{year}{2018}\natexlab{}.
\newblock \showarticletitle{Online harassment and content moderation: The case of blocklists}.
\newblock \bibinfo{journal}{\emph{ACM Transactions on Computer-Human Interaction (TOCHI)}} \bibinfo{volume}{25}, \bibinfo{number}{2} (\bibinfo{year}{2018}), \bibinfo{pages}{1--33}.
\newblock


\bibitem[\protect\citeauthoryear{Johannes~Dechant, Frommel, and Mandryk}{Johannes~Dechant et~al\mbox{.}}{2021}]%
        {johannes2021assessing}
\bibfield{author}{\bibinfo{person}{Martin Johannes~Dechant}, \bibinfo{person}{Julian Frommel}, {and} \bibinfo{person}{Regan Mandryk}.} \bibinfo{year}{2021}\natexlab{}.
\newblock \showarticletitle{Assessing social anxiety through digital biomarkers embedded in a gaming task}. In \bibinfo{booktitle}{\emph{Proceedings of the 2021 CHI Conference on Human Factors in Computing Systems}}. \bibinfo{pages}{1--15}.
\newblock


\bibitem[\protect\citeauthoryear{Jurafsky, Chahuneau, Routledge, and Smith}{Jurafsky et~al\mbox{.}}{2014}]%
        {jurafsky2014narrative}
\bibfield{author}{\bibinfo{person}{Dan Jurafsky}, \bibinfo{person}{Victor Chahuneau}, \bibinfo{person}{Bryan~R Routledge}, {and} \bibinfo{person}{Noah~A Smith}.} \bibinfo{year}{2014}\natexlab{}.
\newblock \showarticletitle{Narrative framing of consumer sentiment in online restaurant reviews}.
\newblock \bibinfo{journal}{\emph{First Monday}} (\bibinfo{year}{2014}).
\newblock


\bibitem[\protect\citeauthoryear{Kao, Ratan, Mousas, Joshi, and Melcer}{Kao et~al\mbox{.}}{2022}]%
        {kao2022audio}
\bibfield{author}{\bibinfo{person}{Dominic Kao}, \bibinfo{person}{Rabindra Ratan}, \bibinfo{person}{Christos Mousas}, \bibinfo{person}{Amogh Joshi}, {and} \bibinfo{person}{Edward~F Melcer}.} \bibinfo{year}{2022}\natexlab{}.
\newblock \showarticletitle{Audio Matters Too: How Audial Avatar Customization Enhances Visual Avatar Customization}. In \bibinfo{booktitle}{\emph{CHI Conference on Human Factors in Computing Systems}}. \bibinfo{pages}{1--27}.
\newblock


\bibitem[\protect\citeauthoryear{Karizat, Delmonaco, Eslami, and Andalibi}{Karizat et~al\mbox{.}}{2021}]%
        {karizat2021algorithmic}
\bibfield{author}{\bibinfo{person}{Nadia Karizat}, \bibinfo{person}{Daniel Delmonaco}, \bibinfo{person}{Motahhare Eslami}, {and} \bibinfo{person}{Nazanin Andalibi}.} \bibinfo{year}{2021}\natexlab{}.
\newblock \showarticletitle{Algorithmic Folk Theories and Identity: How TikTok Users Co-Produce Knowledge of Identity and Engage in Algorithmic Resistance}.
\newblock \bibinfo{journal}{\emph{Proceedings of the ACM on Human-Computer Interaction CSCW2 (2021), forthcoming}} (\bibinfo{year}{2021}).
\newblock


\bibitem[\protect\citeauthoryear{Keyes}{Keyes}{2018}]%
        {keyes2018misgendering}
\bibfield{author}{\bibinfo{person}{Os Keyes}.} \bibinfo{year}{2018}\natexlab{}.
\newblock \showarticletitle{The misgendering machines: Trans/HCI implications of automatic gender recognition}.
\newblock \bibinfo{journal}{\emph{Proceedings of the ACM on human-computer interaction}} \bibinfo{volume}{2}, \bibinfo{number}{CSCW} (\bibinfo{year}{2018}), \bibinfo{pages}{1--22}.
\newblock


\bibitem[\protect\citeauthoryear{Khayatt}{Khayatt}{2002}]%
        {khayatt2002toward}
\bibfield{author}{\bibinfo{person}{Didi Khayatt}.} \bibinfo{year}{2002}\natexlab{}.
\newblock \showarticletitle{Toward a queer identity}.
\newblock \bibinfo{journal}{\emph{Sexualities}} \bibinfo{volume}{5}, \bibinfo{number}{4} (\bibinfo{year}{2002}), \bibinfo{pages}{487--501}.
\newblock


\bibitem[\protect\citeauthoryear{Kingsley, Sinha, Wang, Eslami, and Hong}{Kingsley et~al\mbox{.}}{2022}]%
        {Kingsley2022Give}
\bibfield{author}{\bibinfo{person}{Sara Kingsley}, \bibinfo{person}{Proteeti Sinha}, \bibinfo{person}{Clara Wang}, \bibinfo{person}{Motahhare Eslami}, {and} \bibinfo{person}{Jason~I. Hong}.} \bibinfo{year}{2022}\natexlab{}.
\newblock \showarticletitle{"Give Everybody [..] a Little Bit More Equity": Content Creator Perspectives and Responses to the Algorithmic Demonetization of Content Associated with Disadvantaged Groups}.
\newblock \bibinfo{journal}{\emph{Proc. ACM Hum.-Comput. Interact.}} \bibinfo{volume}{6}, \bibinfo{number}{CSCW2}, Article \bibinfo{articleno}{424} (\bibinfo{date}{nov} \bibinfo{year}{2022}), \bibinfo{numpages}{37}~pages.
\newblock
\urldef\tempurl%
\url{https://doi.org/10.1145/3555149}
\showDOI{\tempurl}


\bibitem[\protect\citeauthoryear{Kinnaird, Romero, and Abowd}{Kinnaird et~al\mbox{.}}{2010}]%
        {kinnaird2010connect}
\bibfield{author}{\bibinfo{person}{Peter Kinnaird}, \bibinfo{person}{Mario Romero}, {and} \bibinfo{person}{Gregory Abowd}.} \bibinfo{year}{2010}\natexlab{}.
\newblock \showarticletitle{Connect 2 congress: visual analytics for civic oversight}.
\newblock In \bibinfo{booktitle}{\emph{CHI'10 Extended Abstracts on Human Factors in Computing Systems}}. \bibinfo{pages}{2853--2862}.
\newblock


\bibitem[\protect\citeauthoryear{Kinnee, Rosner, and Desjardins}{Kinnee et~al\mbox{.}}{2022}]%
        {kinnee2022sonic}
\bibfield{author}{\bibinfo{person}{Brian Kinnee}, \bibinfo{person}{Daniela~K Rosner}, {and} \bibinfo{person}{Audrey Desjardins}.} \bibinfo{year}{2022}\natexlab{}.
\newblock \showarticletitle{Sonic Technologies of a Queer Breakup}. In \bibinfo{booktitle}{\emph{Designing Interactive Systems Conference}}. \bibinfo{pages}{1377--1393}.
\newblock


\bibitem[\protect\citeauthoryear{Klassen, Kingsley, McCall, Weinberg, and Fiesler}{Klassen et~al\mbox{.}}{2021}]%
        {klassen2021more}
\bibfield{author}{\bibinfo{person}{Shamika Klassen}, \bibinfo{person}{Sara Kingsley}, \bibinfo{person}{Kalyn McCall}, \bibinfo{person}{Joy Weinberg}, {and} \bibinfo{person}{Casey Fiesler}.} \bibinfo{year}{2021}\natexlab{}.
\newblock \showarticletitle{More than a Modern Day Green Book: Exploring the Online Community of Black Twitter}.
\newblock \bibinfo{journal}{\emph{Proceedings of the ACM on Human-Computer Interaction}} \bibinfo{volume}{5}, \bibinfo{number}{CSCW2} (\bibinfo{year}{2021}), \bibinfo{pages}{1--29}.
\newblock


\bibitem[\protect\citeauthoryear{Kriplean, Bonnar, Borning, Kinney, and Gill}{Kriplean et~al\mbox{.}}{2014}]%
        {kriplean2014integrating}
\bibfield{author}{\bibinfo{person}{Travis Kriplean}, \bibinfo{person}{Caitlin Bonnar}, \bibinfo{person}{Alan Borning}, \bibinfo{person}{Bo Kinney}, {and} \bibinfo{person}{Brian Gill}.} \bibinfo{year}{2014}\natexlab{}.
\newblock \showarticletitle{Integrating on-demand fact-checking with public dialogue}. In \bibinfo{booktitle}{\emph{Proceedings of the 17th ACM conference on Computer supported cooperative work \& social computing}}. \bibinfo{pages}{1188--1199}.
\newblock


\bibitem[\protect\citeauthoryear{Kulick}{Kulick}{1998}]%
        {kulick1998travesti}
\bibfield{author}{\bibinfo{person}{Don Kulick}.} \bibinfo{year}{1998}\natexlab{}.
\newblock \bibinfo{booktitle}{\emph{Travesti: Sex, gender, and culture among Brazilian transgendered prostitutes}}.
\newblock \bibinfo{publisher}{University of Chicago Press}.
\newblock


\bibitem[\protect\citeauthoryear{Kulshrestha, Eslami, Messias, Zafar, Ghosh, Gummadi, and Karahalios}{Kulshrestha et~al\mbox{.}}{2017}]%
        {kulshrestha2017quantifying}
\bibfield{author}{\bibinfo{person}{Juhi Kulshrestha}, \bibinfo{person}{Motahhare Eslami}, \bibinfo{person}{Johnnatan Messias}, \bibinfo{person}{Muhammad~Bilal Zafar}, \bibinfo{person}{Saptarshi Ghosh}, \bibinfo{person}{Krishna~P Gummadi}, {and} \bibinfo{person}{Karrie Karahalios}.} \bibinfo{year}{2017}\natexlab{}.
\newblock \showarticletitle{Quantifying search bias: Investigating sources of bias for political searches in social media}. In \bibinfo{booktitle}{\emph{Proceedings of the 2017 ACM Conference on Computer Supported Cooperative Work and Social Computing}}. \bibinfo{pages}{417--432}.
\newblock


\bibitem[\protect\citeauthoryear{Kumar and Karusala}{Kumar and Karusala}{2021}]%
        {kumar2021braving}
\bibfield{author}{\bibinfo{person}{Neha Kumar} {and} \bibinfo{person}{Naveena Karusala}.} \bibinfo{year}{2021}\natexlab{}.
\newblock \showarticletitle{Braving citational justice in human-computer interaction}. In \bibinfo{booktitle}{\emph{Extended Abstracts of the 2021 CHI Conference on Human Factors in Computing Systems}}. \bibinfo{pages}{1--9}.
\newblock


\bibitem[\protect\citeauthoryear{Kuznetsov, Novotny, Klein, Saez-Trumper, and Kittur}{Kuznetsov et~al\mbox{.}}{2022}]%
        {kuznetsov2022templates}
\bibfield{author}{\bibinfo{person}{Andrew Kuznetsov}, \bibinfo{person}{Margeigh Novotny}, \bibinfo{person}{Jessica Klein}, \bibinfo{person}{Diego Saez-Trumper}, {and} \bibinfo{person}{Aniket Kittur}.} \bibinfo{year}{2022}\natexlab{}.
\newblock \showarticletitle{Templates and Trust-o-meters: Towards a widely deployable indicator of trust in Wikipedia}. In \bibinfo{booktitle}{\emph{CHI Conference on Human Factors in Computing Systems}}. \bibinfo{pages}{1--17}.
\newblock


\bibitem[\protect\citeauthoryear{Le, Boynton, Mejova, Shafiq, and Srinivasan}{Le et~al\mbox{.}}{2017}]%
        {le2017revisiting}
\bibfield{author}{\bibinfo{person}{Huyen~T Le}, \bibinfo{person}{GR Boynton}, \bibinfo{person}{Yelena Mejova}, \bibinfo{person}{Zubair Shafiq}, {and} \bibinfo{person}{Padmini Srinivasan}.} \bibinfo{year}{2017}\natexlab{}.
\newblock \showarticletitle{Revisiting the american voter on twitter}. In \bibinfo{booktitle}{\emph{Proceedings of the 2017 CHI Conference on Human Factors in Computing Systems}}. \bibinfo{pages}{4507--4519}.
\newblock


\bibitem[\protect\citeauthoryear{Lee, Kiesler, and Forlizzi}{Lee et~al\mbox{.}}{2010}]%
        {lee2010receptionist}
\bibfield{author}{\bibinfo{person}{Min~Kyung Lee}, \bibinfo{person}{Sara Kiesler}, {and} \bibinfo{person}{Jodi Forlizzi}.} \bibinfo{year}{2010}\natexlab{}.
\newblock \showarticletitle{Receptionist or information kiosk: how do people talk with a robot?}. In \bibinfo{booktitle}{\emph{Proceedings of the 2010 ACM conference on Computer supported cooperative work}}. \bibinfo{pages}{31--40}.
\newblock


\bibitem[\protect\citeauthoryear{Lerner, He, Kawakami, Zeamer, and Hoyle}{Lerner et~al\mbox{.}}{2020}]%
        {Lerner2020Activism}
\bibfield{author}{\bibinfo{person}{Ada Lerner}, \bibinfo{person}{Helen~Yuxun He}, \bibinfo{person}{Anna Kawakami}, \bibinfo{person}{Silvia~Catherine Zeamer}, {and} \bibinfo{person}{Roberto Hoyle}.} \bibinfo{year}{2020}\natexlab{}.
\newblock \showarticletitle{Privacy and Activism in the Transgender Community}. In \bibinfo{booktitle}{\emph{Proceedings of the 2020 CHI Conference on Human Factors in Computing Systems}} (Honolulu, HI, USA) \emph{(\bibinfo{series}{CHI '20})}. \bibinfo{publisher}{Association for Computing Machinery}, \bibinfo{address}{New York, NY, USA}, \bibinfo{pages}{1–13}.
\newblock
\showISBNx{9781450367080}
\urldef\tempurl%
\url{https://doi.org/10.1145/3313831.3376339}
\showDOI{\tempurl}


\bibitem[\protect\citeauthoryear{Li, Spektor, Xia, Huh, Cederberg, Gong, Shinohara, and Carrington}{Li et~al\mbox{.}}{2022}]%
        {li2022feels}
\bibfield{author}{\bibinfo{person}{Franklin~Mingzhe Li}, \bibinfo{person}{Franchesca Spektor}, \bibinfo{person}{Meng Xia}, \bibinfo{person}{Mina Huh}, \bibinfo{person}{Peter Cederberg}, \bibinfo{person}{Yuqi Gong}, \bibinfo{person}{Kristen Shinohara}, {and} \bibinfo{person}{Patrick Carrington}.} \bibinfo{year}{2022}\natexlab{}.
\newblock \showarticletitle{“It Feels Like Taking a Gamble”: Exploring Perceptions, Practices, and Challenges of Using Makeup and Cosmetics for People with Visual Impairments}. In \bibinfo{booktitle}{\emph{CHI Conference on Human Factors in Computing Systems}}. \bibinfo{pages}{1--15}.
\newblock


\bibitem[\protect\citeauthoryear{Li, Uttarapong, Freeman, and Wohn}{Li et~al\mbox{.}}{2020}]%
        {li2020spontaneous}
\bibfield{author}{\bibinfo{person}{Lingyuan Li}, \bibinfo{person}{Jirassaya Uttarapong}, \bibinfo{person}{Guo Freeman}, {and} \bibinfo{person}{Donghee~Yvette Wohn}.} \bibinfo{year}{2020}\natexlab{}.
\newblock \showarticletitle{Spontaneous, Yet Studious: Esports Commentators' Live Performance and Self-Presentation Practices}.
\newblock \bibinfo{journal}{\emph{Proceedings of the ACM on Human-Computer Interaction}} \bibinfo{volume}{4}, \bibinfo{number}{CSCW2} (\bibinfo{year}{2020}), \bibinfo{pages}{1--25}.
\newblock


\bibitem[\protect\citeauthoryear{Liang, Munson, and Kientz}{Liang et~al\mbox{.}}{2021}]%
        {liang2021embracing}
\bibfield{author}{\bibinfo{person}{Calvin~A Liang}, \bibinfo{person}{Sean~A Munson}, {and} \bibinfo{person}{Julie~A Kientz}.} \bibinfo{year}{2021}\natexlab{}.
\newblock \showarticletitle{Embracing Four Tensions in Human-Computer Interaction Research with Marginalized People}.
\newblock \bibinfo{journal}{\emph{ACM Transactions on Computer-Human Interaction (TOCHI)}} \bibinfo{volume}{28}, \bibinfo{number}{2} (\bibinfo{year}{2021}), \bibinfo{pages}{1--47}.
\newblock


\bibitem[\protect\citeauthoryear{Light}{Light}{2011}]%
        {light2011hci}
\bibfield{author}{\bibinfo{person}{Ann Light}.} \bibinfo{year}{2011}\natexlab{}.
\newblock \showarticletitle{HCI as heterodoxy: Technologies of identity and the queering of interaction with computers}.
\newblock \bibinfo{journal}{\emph{Interacting with computers}} \bibinfo{volume}{23}, \bibinfo{number}{5} (\bibinfo{year}{2011}), \bibinfo{pages}{430--438}.
\newblock


\bibitem[\protect\citeauthoryear{Light}{Light}{1971}]%
        {light1971measures}
\bibfield{author}{\bibinfo{person}{Richard~J Light}.} \bibinfo{year}{1971}\natexlab{}.
\newblock \showarticletitle{Measures of response agreement for qualitative data: some generalizations and alternatives.}
\newblock \bibinfo{journal}{\emph{Psychological bulletin}} \bibinfo{volume}{76}, \bibinfo{number}{5} (\bibinfo{year}{1971}), \bibinfo{pages}{365}.
\newblock


\bibitem[\protect\citeauthoryear{Lin and Margot~Lindtner}{Lin and Margot~Lindtner}{2021}]%
        {lin2021techniques}
\bibfield{author}{\bibinfo{person}{Cindy Lin} {and} \bibinfo{person}{Silvia Margot~Lindtner}.} \bibinfo{year}{2021}\natexlab{}.
\newblock \showarticletitle{Techniques of use: Confronting value systems of productivity, progress, and usefulness in computing and design}. In \bibinfo{booktitle}{\emph{Proceedings of the 2021 CHI Conference on Human Factors in Computing Systems}}. \bibinfo{pages}{1--16}.
\newblock


\bibitem[\protect\citeauthoryear{Logas, Schlesinger, Li, and Das}{Logas et~al\mbox{.}}{2022}]%
        {logas2022image}
\bibfield{author}{\bibinfo{person}{Jacob Logas}, \bibinfo{person}{Ari Schlesinger}, \bibinfo{person}{Zhouyu Li}, {and} \bibinfo{person}{Sauvik Das}.} \bibinfo{year}{2022}\natexlab{}.
\newblock \showarticletitle{Image DePO: Towards Gradual Decentralization of Online Social Networks using Decentralized Privacy Overlays}.
\newblock \bibinfo{journal}{\emph{Proceedings of the ACM on Human-Computer Interaction}} \bibinfo{volume}{6}, \bibinfo{number}{CSCW1} (\bibinfo{year}{2022}), \bibinfo{pages}{1--28}.
\newblock


\bibitem[\protect\citeauthoryear{Macbeth, Adeyema, Lieberman, and Fry}{Macbeth et~al\mbox{.}}{2013}]%
        {macbeth2013script}
\bibfield{author}{\bibinfo{person}{Jamie Macbeth}, \bibinfo{person}{Hanna Adeyema}, \bibinfo{person}{Henry Lieberman}, {and} \bibinfo{person}{Christopher Fry}.} \bibinfo{year}{2013}\natexlab{}.
\newblock \showarticletitle{Script-based story matching for cyberbullying prevention}.
\newblock In \bibinfo{booktitle}{\emph{CHI'13 Extended Abstracts on Human Factors in Computing Systems}}. \bibinfo{pages}{901--906}.
\newblock


\bibitem[\protect\citeauthoryear{Maestre, MacLeod, Connelly, Dunbar, Beck, Siek, and Shih}{Maestre et~al\mbox{.}}{2018}]%
        {maestre2018defining}
\bibfield{author}{\bibinfo{person}{Juan~F Maestre}, \bibinfo{person}{Haley MacLeod}, \bibinfo{person}{Ciabhan~L Connelly}, \bibinfo{person}{Julia~C Dunbar}, \bibinfo{person}{Jordan Beck}, \bibinfo{person}{Katie~A Siek}, {and} \bibinfo{person}{Patrick~C Shih}.} \bibinfo{year}{2018}\natexlab{}.
\newblock \showarticletitle{Defining through expansion: conducting asynchronous remote communities (arc) research with stigmatized groups}. In \bibinfo{booktitle}{\emph{Proceedings of the 2018 CHI Conference on Human Factors in Computing Systems}}. \bibinfo{pages}{1--13}.
\newblock


\bibitem[\protect\citeauthoryear{Matias, Hounsel, and Feamster}{Matias et~al\mbox{.}}{2022}]%
        {matias2022software}
\bibfield{author}{\bibinfo{person}{J~Nathan Matias}, \bibinfo{person}{Austin Hounsel}, {and} \bibinfo{person}{Nick Feamster}.} \bibinfo{year}{2022}\natexlab{}.
\newblock \showarticletitle{Software-Supported Audits of Decision-Making Systems: Testing Google and Facebook's Political Advertising Policies}.
\newblock \bibinfo{journal}{\emph{Proceedings of the ACM on Human-Computer Interaction}} \bibinfo{volume}{6}, \bibinfo{number}{CSCW1} (\bibinfo{year}{2022}), \bibinfo{pages}{1--19}.
\newblock


\bibitem[\protect\citeauthoryear{McDonald and Forte}{McDonald and Forte}{2020}]%
        {mcdonald2020politics}
\bibfield{author}{\bibinfo{person}{Nora McDonald} {and} \bibinfo{person}{Andrea Forte}.} \bibinfo{year}{2020}\natexlab{}.
\newblock \showarticletitle{The politics of privacy theories: Moving from norms to vulnerabilities}. In \bibinfo{booktitle}{\emph{Proceedings of the 2020 CHI Conference on Human Factors in Computing Systems}}. \bibinfo{pages}{1--14}.
\newblock


\bibitem[\protect\citeauthoryear{McNee, Riedl, and Konstan}{McNee et~al\mbox{.}}{2006}]%
        {mcnee2006making}
\bibfield{author}{\bibinfo{person}{Sean~M McNee}, \bibinfo{person}{John Riedl}, {and} \bibinfo{person}{Joseph~A Konstan}.} \bibinfo{year}{2006}\natexlab{}.
\newblock \showarticletitle{Making recommendations better: an analytic model for human-recommender interaction}. In \bibinfo{booktitle}{\emph{CHI'06 extended abstracts on Human factors in computing systems}}. \bibinfo{pages}{1103--1108}.
\newblock


\bibitem[\protect\citeauthoryear{Mohanty}{Mohanty}{1988}]%
        {mohanty1988under}
\bibfield{author}{\bibinfo{person}{Chandra~Talpade Mohanty}.} \bibinfo{year}{1988}\natexlab{}.
\newblock \showarticletitle{Under Western Eyes: Feminist Scholarship and Colonial Discourses}.
\newblock \bibinfo{journal}{\emph{Feminist Review}} \bibinfo{volume}{30}, \bibinfo{number}{1} (\bibinfo{year}{1988}), \bibinfo{pages}{61--88}.
\newblock


\bibitem[\protect\citeauthoryear{Moitra, Ahmed, and Chandra}{Moitra et~al\mbox{.}}{2021a}]%
        {moitra2021parsing}
\bibfield{author}{\bibinfo{person}{Aparna Moitra}, \bibinfo{person}{Syed~Ishtiaque Ahmed}, {and} \bibinfo{person}{Priyank Chandra}.} \bibinfo{year}{2021}\natexlab{a}.
\newblock \showarticletitle{Parsing the'Me'in\# MeToo: Sexual Harassment, Social Media, and Justice Infrastructures}.
\newblock \bibinfo{journal}{\emph{Proceedings of the ACM on Human-Computer Interaction}} \bibinfo{volume}{5}, \bibinfo{number}{CSCW1} (\bibinfo{year}{2021}), \bibinfo{pages}{1--34}.
\newblock


\bibitem[\protect\citeauthoryear{Moitra, Marathe, Ahmed, and Chandra}{Moitra et~al\mbox{.}}{2021b}]%
        {moitra2021negotiating}
\bibfield{author}{\bibinfo{person}{Aparna Moitra}, \bibinfo{person}{Megh Marathe}, \bibinfo{person}{Syed~Ishtiaque Ahmed}, {and} \bibinfo{person}{Priyank Chandra}.} \bibinfo{year}{2021}\natexlab{b}.
\newblock \showarticletitle{Negotiating intersectional non-Normative queer identities in India}. In \bibinfo{booktitle}{\emph{Extended Abstracts of the 2021 CHI Conference on Human Factors in Computing Systems}}. \bibinfo{pages}{1--6}.
\newblock


\bibitem[\protect\citeauthoryear{Morag~Yaar, Grossman, Kimchi, Nash, Hatan, and Erel}{Morag~Yaar et~al\mbox{.}}{2022}]%
        {morag2022tobe}
\bibfield{author}{\bibinfo{person}{Noa Morag~Yaar}, \bibinfo{person}{Eden Grossman}, \bibinfo{person}{Noam Kimchi}, \bibinfo{person}{Ofir Nash}, \bibinfo{person}{Sagi Hatan}, {and} \bibinfo{person}{Hadas Erel}.} \bibinfo{year}{2022}\natexlab{}.
\newblock \showarticletitle{Tobe: A Virtual Keyboard and an Animated Character for Individual and Educational Cyberbullying Intervention}. In \bibinfo{booktitle}{\emph{CHI Conference on Human Factors in Computing Systems Extended Abstracts}}. \bibinfo{pages}{1--6}.
\newblock


\bibitem[\protect\citeauthoryear{Muller, Wharton, McIver~Jr, and Laux}{Muller et~al\mbox{.}}{1997}]%
        {muller1997toward}
\bibfield{author}{\bibinfo{person}{Michael~J Muller}, \bibinfo{person}{Cathleen Wharton}, \bibinfo{person}{William~J McIver~Jr}, {and} \bibinfo{person}{Lila Laux}.} \bibinfo{year}{1997}\natexlab{}.
\newblock \showarticletitle{Toward an HCI research and practice agenda based on human needs and social responsibility}. In \bibinfo{booktitle}{\emph{Proceedings of the ACM SIGCHI Conference on Human factors in computing systems}}. \bibinfo{pages}{155--161}.
\newblock


\bibitem[\protect\citeauthoryear{Musgrave, Cummings, and Schoenebeck}{Musgrave et~al\mbox{.}}{2022}]%
        {musgrave2022experiences}
\bibfield{author}{\bibinfo{person}{Tyler Musgrave}, \bibinfo{person}{Alia Cummings}, {and} \bibinfo{person}{Sarita Schoenebeck}.} \bibinfo{year}{2022}\natexlab{}.
\newblock \showarticletitle{Experiences of Harm, Healing, and Joy among Black Women and Femmes on Social Media}. In \bibinfo{booktitle}{\emph{CHI Conference on Human Factors in Computing Systems}}. \bibinfo{pages}{1--17}.
\newblock


\bibitem[\protect\citeauthoryear{Nanda}{Nanda}{1990}]%
        {nanda1990neither}
\bibfield{author}{\bibinfo{person}{Serena Nanda}.} \bibinfo{year}{1990}\natexlab{}.
\newblock \bibinfo{booktitle}{\emph{Neither man nor woman: The hijras of India}}.
\newblock \bibinfo{publisher}{Wadsworth Pub. Co}.
\newblock


\bibitem[\protect\citeauthoryear{Nissenbaum and Shifman}{Nissenbaum and Shifman}{2017}]%
        {nissenbaum_internet_2017}
\bibfield{author}{\bibinfo{person}{Asaf Nissenbaum} {and} \bibinfo{person}{Limor Shifman}.} \bibinfo{year}{2017}\natexlab{}.
\newblock \showarticletitle{Internet memes as contested cultural capital: {The} case of 4chan’s /b/ board}.
\newblock \bibinfo{journal}{\emph{New Media \& Society}} \bibinfo{volume}{19}, \bibinfo{number}{4} (\bibinfo{date}{April} \bibinfo{year}{2017}), \bibinfo{pages}{483--501}.
\newblock
\showISSN{1461-4448}
\urldef\tempurl%
\url{https://doi.org/10.1177/1461444815609313}
\showDOI{\tempurl}
\newblock
\shownote{Publisher: SAGE Publications.}


\bibitem[\protect\citeauthoryear{Nova, DeVito, Saha, Rashid, Roy~Turzo, Afrin, and Guha}{Nova et~al\mbox{.}}{2020}]%
        {Nova2020Understanding}
\bibfield{author}{\bibinfo{person}{Fayika~Farhat Nova}, \bibinfo{person}{Michael~Ann DeVito}, \bibinfo{person}{Pratyasha Saha}, \bibinfo{person}{Kazi~Shohanur Rashid}, \bibinfo{person}{Shashwata Roy~Turzo}, \bibinfo{person}{Sadia Afrin}, {and} \bibinfo{person}{Shion Guha}.} \bibinfo{year}{2020}\natexlab{}.
\newblock \showarticletitle{Understanding How Marginalized Hijra in Bangladesh Navigate Complex Social Media Ecosystem}. In \bibinfo{booktitle}{\emph{Conference Companion Publication of the 2020 on Computer Supported Cooperative Work and Social Computing}} (Virtual Event, USA) \emph{(\bibinfo{series}{CSCW '20 Companion})}. \bibinfo{publisher}{Association for Computing Machinery}, \bibinfo{address}{New York, NY, USA}, \bibinfo{pages}{353–358}.
\newblock
\showISBNx{9781450380591}
\urldef\tempurl%
\url{https://doi.org/10.1145/3406865.3418317}
\showDOI{\tempurl}


\bibitem[\protect\citeauthoryear{Nova, DeVito, Saha, Rashid, Roy~Turzo, Afrin, and Guha}{Nova et~al\mbox{.}}{2021}]%
        {nova2021facebook}
\bibfield{author}{\bibinfo{person}{Fayika~Farhat Nova}, \bibinfo{person}{Michael~Ann DeVito}, \bibinfo{person}{Pratyasha Saha}, \bibinfo{person}{Kazi~Shohanur Rashid}, \bibinfo{person}{Shashwata Roy~Turzo}, \bibinfo{person}{Sadia Afrin}, {and} \bibinfo{person}{Shion Guha}.} \bibinfo{year}{2021}\natexlab{}.
\newblock \showarticletitle{" Facebook Promotes More Harassment" Social Media Ecosystem, Skill and Marginalized Hijra Identity in Bangladesh}.
\newblock \bibinfo{journal}{\emph{Proceedings of the ACM on Human-Computer Interaction}} \bibinfo{volume}{5}, \bibinfo{number}{CSCW1} (\bibinfo{year}{2021}), \bibinfo{pages}{1--35}.
\newblock


\bibitem[\protect\citeauthoryear{Odom}{Odom}{2015}]%
        {odom2015understanding}
\bibfield{author}{\bibinfo{person}{William Odom}.} \bibinfo{year}{2015}\natexlab{}.
\newblock \showarticletitle{Understanding long-term interactions with a slow technology: an investigation of experiences with FutureMe}. In \bibinfo{booktitle}{\emph{Proceedings of the 33rd Annual ACM Conference on Human Factors in Computing Systems}}. \bibinfo{pages}{575--584}.
\newblock


\bibitem[\protect\citeauthoryear{Offenwanger, Milligan, Chang, Bullard, and Yoon}{Offenwanger et~al\mbox{.}}{2021}]%
        {offenwanger2021diagnosing}
\bibfield{author}{\bibinfo{person}{Anna Offenwanger}, \bibinfo{person}{Alan~John Milligan}, \bibinfo{person}{Minsuk Chang}, \bibinfo{person}{Julia Bullard}, {and} \bibinfo{person}{Dongwook Yoon}.} \bibinfo{year}{2021}\natexlab{}.
\newblock \showarticletitle{Diagnosing bias in the gender representation of HCI research participants: how it happens and where we are}. In \bibinfo{booktitle}{\emph{Proceedings of the 2021 CHI Conference on Human Factors in Computing Systems}}. \bibinfo{pages}{1--18}.
\newblock


\bibitem[\protect\citeauthoryear{Ogbonnaya-Ogburu, Smith, To, and Toyama}{Ogbonnaya-Ogburu et~al\mbox{.}}{2020}]%
        {ogbonnaya2020critical}
\bibfield{author}{\bibinfo{person}{Ihudiya~Finda Ogbonnaya-Ogburu}, \bibinfo{person}{Angela~DR Smith}, \bibinfo{person}{Alexandra To}, {and} \bibinfo{person}{Kentaro Toyama}.} \bibinfo{year}{2020}\natexlab{}.
\newblock \showarticletitle{Critical race theory for HCI}. In \bibinfo{booktitle}{\emph{Proceedings of the 2020 CHI Conference on Human Factors in Computing Systems}}. \bibinfo{pages}{1--16}.
\newblock


\bibitem[\protect\citeauthoryear{Otterbacher}{Otterbacher}{2015}]%
        {Otterbacher2015Crowdsourcing}
\bibfield{author}{\bibinfo{person}{Jahna Otterbacher}.} \bibinfo{year}{2015}\natexlab{}.
\newblock \showarticletitle{Crowdsourcing Stereotypes: Linguistic Bias in Metadata Generated via GWAP}. In \bibinfo{booktitle}{\emph{Proceedings of the 33rd Annual ACM Conference on Human Factors in Computing Systems}} (Seoul, Republic of Korea) \emph{(\bibinfo{series}{CHI '15})}. \bibinfo{publisher}{Association for Computing Machinery}, \bibinfo{address}{New York, NY, USA}, \bibinfo{pages}{1955–1964}.
\newblock
\showISBNx{9781450331456}
\urldef\tempurl%
\url{https://doi.org/10.1145/2702123.2702151}
\showDOI{\tempurl}


\bibitem[\protect\citeauthoryear{Pei and Nardi}{Pei and Nardi}{2019}]%
        {pei2019we}
\bibfield{author}{\bibinfo{person}{Lucy Pei} {and} \bibinfo{person}{Bonnie Nardi}.} \bibinfo{year}{2019}\natexlab{}.
\newblock \showarticletitle{We did it right, but it was still wrong: Toward assets-based design}. In \bibinfo{booktitle}{\emph{Extended Abstracts of the 2019 CHI Conference on Human Factors in Computing Systems}}. \bibinfo{pages}{1--11}.
\newblock


\bibitem[\protect\citeauthoryear{Pinch, Birnholtz, Kraus, Macapagal, and A.~Moskowitz}{Pinch et~al\mbox{.}}{2021}]%
        {pinch2021s}
\bibfield{author}{\bibinfo{person}{Annika Pinch}, \bibinfo{person}{Jeremy Birnholtz}, \bibinfo{person}{Ashley Kraus}, \bibinfo{person}{Kathryn Macapagal}, {and} \bibinfo{person}{David A.~Moskowitz}.} \bibinfo{year}{2021}\natexlab{}.
\newblock \showarticletitle{“It's not exactly prominent or direct, but it's there”: Understanding Strategies for Sensitive Disclosure Online}. In \bibinfo{booktitle}{\emph{Companion Publication of the 2021 Conference on Computer Supported Cooperative Work and Social Computing}}. \bibinfo{pages}{149--152}.
\newblock


\bibitem[\protect\citeauthoryear{Pinter, Scheuerman, and Brubaker}{Pinter et~al\mbox{.}}{2021}]%
        {pinter2021entering}
\bibfield{author}{\bibinfo{person}{Anthony~T Pinter}, \bibinfo{person}{Morgan~Klaus Scheuerman}, {and} \bibinfo{person}{Jed~R Brubaker}.} \bibinfo{year}{2021}\natexlab{}.
\newblock \showarticletitle{Entering doors, evading traps: Benefits and risks of visibility during transgender coming outs}.
\newblock \bibinfo{journal}{\emph{Proceedings of the ACM on Human-Computer Interaction}} \bibinfo{volume}{4}, \bibinfo{number}{CSCW3} (\bibinfo{year}{2021}), \bibinfo{pages}{1--27}.
\newblock


\bibitem[\protect\citeauthoryear{Pyle, Roosevelt, Lacombe-Duncan, and Andalibi}{Pyle et~al\mbox{.}}{2021}]%
        {pyle2021lgbtq}
\bibfield{author}{\bibinfo{person}{Cassidy Pyle}, \bibinfo{person}{Lee Roosevelt}, \bibinfo{person}{Ashley Lacombe-Duncan}, {and} \bibinfo{person}{Nazanin Andalibi}.} \bibinfo{year}{2021}\natexlab{}.
\newblock \showarticletitle{LGBTQ Persons' Pregnancy Loss Disclosures to Known Ties on Social Media: Disclosure Decisions and Ideal Disclosure Environments}. In \bibinfo{booktitle}{\emph{Proceedings of the 2021 CHI Conference on Human Factors in Computing Systems}}. \bibinfo{pages}{1--17}.
\newblock


\bibitem[\protect\citeauthoryear{Rankin and Thomas}{Rankin and Thomas}{2019}]%
        {rankin2019straighten}
\bibfield{author}{\bibinfo{person}{Yolanda~A Rankin} {and} \bibinfo{person}{Jakita~O Thomas}.} \bibinfo{year}{2019}\natexlab{}.
\newblock \showarticletitle{Straighten up and fly right: Rethinking intersectionality in HCI research}.
\newblock \bibinfo{journal}{\emph{Interactions}} \bibinfo{volume}{26}, \bibinfo{number}{6} (\bibinfo{year}{2019}), \bibinfo{pages}{64--68}.
\newblock


\bibitem[\protect\citeauthoryear{Rao, Hurlbutt, Nass, and JanakiRam}{Rao et~al\mbox{.}}{2009}]%
        {rao2009my}
\bibfield{author}{\bibinfo{person}{Shailendra Rao}, \bibinfo{person}{Tom Hurlbutt}, \bibinfo{person}{Clifford Nass}, {and} \bibinfo{person}{Nundu JanakiRam}.} \bibinfo{year}{2009}\natexlab{}.
\newblock \showarticletitle{My Dating Site Thinks I'm a Loser: effects of personal photos and presentation intervals on perceptions of recommender systems}. In \bibinfo{booktitle}{\emph{Proceedings of the SIGCHI Conference on Human Factors in Computing Systems}}. \bibinfo{pages}{221--224}.
\newblock


\bibitem[\protect\citeauthoryear{Richter-Lunn and Krishna~Kumar}{Richter-Lunn and Krishna~Kumar}{2022}]%
        {richter2022acu}
\bibfield{author}{\bibinfo{person}{Katarina Richter-Lunn} {and} \bibinfo{person}{Ila Krishna~Kumar}.} \bibinfo{year}{2022}\natexlab{}.
\newblock \showarticletitle{Acu. ation: Real Time Acupoint Stimulation To Mediate The Urge To Smoke.}. In \bibinfo{booktitle}{\emph{CHI Conference on Human Factors in Computing Systems Extended Abstracts}}. \bibinfo{pages}{1--7}.
\newblock


\bibitem[\protect\citeauthoryear{Rinc\'{o}n, Keyes, and Cath}{Rinc\'{o}n et~al\mbox{.}}{2021}]%
        {Rincon2021Speaking}
\bibfield{author}{\bibinfo{person}{Cami Rinc\'{o}n}, \bibinfo{person}{Os Keyes}, {and} \bibinfo{person}{Corinne Cath}.} \bibinfo{year}{2021}\natexlab{}.
\newblock \showarticletitle{Speaking from Experience: Trans/Non-Binary Requirements for Voice-Activated AI}.
\newblock \bibinfo{journal}{\emph{Proc. ACM Hum.-Comput. Interact.}} \bibinfo{volume}{5}, \bibinfo{number}{CSCW1}, Article \bibinfo{articleno}{132} (\bibinfo{date}{apr} \bibinfo{year}{2021}), \bibinfo{numpages}{27}~pages.
\newblock
\urldef\tempurl%
\url{https://doi.org/10.1145/3449206}
\showDOI{\tempurl}


\bibitem[\protect\citeauthoryear{Rivest, Shamir, and Adleman}{Rivest et~al\mbox{.}}{1978}]%
        {rivest1978method}
\bibfield{author}{\bibinfo{person}{Ronald~L Rivest}, \bibinfo{person}{Adi Shamir}, {and} \bibinfo{person}{Leonard Adleman}.} \bibinfo{year}{1978}\natexlab{}.
\newblock \showarticletitle{A method for obtaining digital signatures and public-key cryptosystems}.
\newblock \bibinfo{journal}{\emph{Commun. ACM}} \bibinfo{volume}{21}, \bibinfo{number}{2} (\bibinfo{year}{1978}), \bibinfo{pages}{120--126}.
\newblock


\bibitem[\protect\citeauthoryear{Rizvi, Casanova-Perez, Ramaswamy, Dirks, Bascom, and Weibel}{Rizvi et~al\mbox{.}}{2022}]%
        {rizvi2022qtbipoc}
\bibfield{author}{\bibinfo{person}{Naba Rizvi}, \bibinfo{person}{Reggie Casanova-Perez}, \bibinfo{person}{Harshini Ramaswamy}, \bibinfo{person}{Lisa Dirks}, \bibinfo{person}{Emily Bascom}, {and} \bibinfo{person}{Nadir Weibel}.} \bibinfo{year}{2022}\natexlab{}.
\newblock \showarticletitle{QTBIPOC PD: Exploring the Intersections of Race, Gender, and Sexual Orientation in Participatory Design}. In \bibinfo{booktitle}{\emph{CHI Conference on Human Factors in Computing Systems Extended Abstracts}}. \bibinfo{pages}{1--4}.
\newblock


\bibitem[\protect\citeauthoryear{Rzeszotarski and Kittur}{Rzeszotarski and Kittur}{2012}]%
        {rzeszotarski2012learning}
\bibfield{author}{\bibinfo{person}{Jeffrey Rzeszotarski} {and} \bibinfo{person}{Aniket Kittur}.} \bibinfo{year}{2012}\natexlab{}.
\newblock \showarticletitle{Learning from history: predicting reverted work at the word level in wikipedia}. In \bibinfo{booktitle}{\emph{Proceedings of the ACM 2012 Conference on Computer Supported Cooperative Work}}. \bibinfo{pages}{437--440}.
\newblock


\bibitem[\protect\citeauthoryear{Sambasivan, Batool, Ahmed, Matthews, Thomas, Gayt{\'a}n-Lugo, Nemer, Bursztein, Churchill, and Consolvo}{Sambasivan et~al\mbox{.}}{2019}]%
        {sambasivan2019they}
\bibfield{author}{\bibinfo{person}{Nithya Sambasivan}, \bibinfo{person}{Amna Batool}, \bibinfo{person}{Nova Ahmed}, \bibinfo{person}{Tara Matthews}, \bibinfo{person}{Kurt Thomas}, \bibinfo{person}{Laura~Sanely Gayt{\'a}n-Lugo}, \bibinfo{person}{David Nemer}, \bibinfo{person}{Elie Bursztein}, \bibinfo{person}{Elizabeth Churchill}, {and} \bibinfo{person}{Sunny Consolvo}.} \bibinfo{year}{2019}\natexlab{}.
\newblock \showarticletitle{" They Don't Leave Us Alone Anywhere We Go" Gender and Digital Abuse in South Asia}. In \bibinfo{booktitle}{\emph{proceedings of the 2019 CHI Conference on Human Factors in Computing Systems}}. \bibinfo{pages}{1--14}.
\newblock


\bibitem[\protect\citeauthoryear{Sannon and Forte}{Sannon and Forte}{2022}]%
        {sannon_litt_review_privacy}
\bibfield{author}{\bibinfo{person}{Shruti Sannon} {and} \bibinfo{person}{Andrea Forte}.} \bibinfo{year}{2022}\natexlab{}.
\newblock \showarticletitle{Privacy Research with Marginalized Groups: What We Know, What's Needed, and What's Next}.
\newblock \bibinfo{journal}{\emph{Proc. ACM Hum.-Comput. Interact.}} \bibinfo{volume}{6}, \bibinfo{number}{CSCW2}, Article \bibinfo{articleno}{455} (\bibinfo{date}{nov} \bibinfo{year}{2022}), \bibinfo{numpages}{33}~pages.
\newblock
\urldef\tempurl%
\url{https://doi.org/10.1145/3555556}
\showDOI{\tempurl}


\bibitem[\protect\citeauthoryear{Scheuerman, Branham, and Hamidi}{Scheuerman et~al\mbox{.}}{2018}]%
        {scheuerman2018safe}
\bibfield{author}{\bibinfo{person}{Morgan~Klaus Scheuerman}, \bibinfo{person}{Stacy~M Branham}, {and} \bibinfo{person}{Foad Hamidi}.} \bibinfo{year}{2018}\natexlab{}.
\newblock \showarticletitle{Safe spaces and safe places: Unpacking technology-mediated experiences of safety and harm with transgender people}.
\newblock \bibinfo{journal}{\emph{Proceedings of the ACM on Human-computer Interaction}} \bibinfo{volume}{2}, \bibinfo{number}{CSCW} (\bibinfo{year}{2018}), \bibinfo{pages}{1--27}.
\newblock


\bibitem[\protect\citeauthoryear{Scheuerman, Jiang, Spiel, and Brubaker}{Scheuerman et~al\mbox{.}}{2021}]%
        {scheuerman2021revisiting}
\bibfield{author}{\bibinfo{person}{Morgan~Klaus Scheuerman}, \bibinfo{person}{Aaron Jiang}, \bibinfo{person}{Katta Spiel}, {and} \bibinfo{person}{Jed~R Brubaker}.} \bibinfo{year}{2021}\natexlab{}.
\newblock \showarticletitle{Revisiting gendered web forms: An evaluation of gender inputs with (non-) binary people}. In \bibinfo{booktitle}{\emph{Proceedings of the 2021 CHI conference on human factors in computing systems}}. \bibinfo{pages}{1--18}.
\newblock


\bibitem[\protect\citeauthoryear{Scheuerman, Paul, and Brubaker}{Scheuerman et~al\mbox{.}}{2019}]%
        {scheuerman2019computers}
\bibfield{author}{\bibinfo{person}{Morgan~Klaus Scheuerman}, \bibinfo{person}{Jacob~M Paul}, {and} \bibinfo{person}{Jed~R Brubaker}.} \bibinfo{year}{2019}\natexlab{}.
\newblock \showarticletitle{How computers see gender: An evaluation of gender classification in commercial facial analysis services}.
\newblock \bibinfo{journal}{\emph{Proceedings of the ACM on Human-Computer Interaction}} \bibinfo{volume}{3}, \bibinfo{number}{CSCW} (\bibinfo{year}{2019}), \bibinfo{pages}{1--33}.
\newblock


\bibitem[\protect\citeauthoryear{Scheuerman, Spiel, Haimson, Hamidi, and Branham}{Scheuerman et~al\mbox{.}}{2020a}]%
        {scheuerman2020GenderGuidelines}
\bibfield{author}{\bibinfo{person}{Morgan~Klaus Scheuerman}, \bibinfo{person}{Katta Spiel}, \bibinfo{person}{Oliver~L Haimson}, \bibinfo{person}{Foad Hamidi}, {and} \bibinfo{person}{Stacy~M Branham}.} \bibinfo{year}{2020}\natexlab{a}.
\newblock \showarticletitle{HCI guidelines for gender equity and inclusivity}.
\newblock \bibinfo{journal}{\emph{UMBC Faculty Collection}} (\bibinfo{year}{2020}).
\newblock
\urldef\tempurl%
\url{https://www.morgan-klaus.com/gender-guidelines.html}
\showURL{%
\tempurl}


\bibitem[\protect\citeauthoryear{Scheuerman, Wade, Lustig, and Brubaker}{Scheuerman et~al\mbox{.}}{2020b}]%
        {scheuerman2020we}
\bibfield{author}{\bibinfo{person}{Morgan~Klaus Scheuerman}, \bibinfo{person}{Kandrea Wade}, \bibinfo{person}{Caitlin Lustig}, {and} \bibinfo{person}{Jed~R Brubaker}.} \bibinfo{year}{2020}\natexlab{b}.
\newblock \showarticletitle{How we've taught algorithms to see identity: Constructing race and gender in image databases for facial analysis}.
\newblock \bibinfo{journal}{\emph{Proceedings of the ACM on Human-computer Interaction}} \bibinfo{volume}{4}, \bibinfo{number}{CSCW1} (\bibinfo{year}{2020}), \bibinfo{pages}{1--35}.
\newblock


\bibitem[\protect\citeauthoryear{Seaborn and Frank}{Seaborn and Frank}{2022}]%
        {Seaborn2022PronounsPepper}
\bibfield{author}{\bibinfo{person}{Katie Seaborn} {and} \bibinfo{person}{Alexa Frank}.} \bibinfo{year}{2022}\natexlab{}.
\newblock \showarticletitle{What Pronouns for Pepper? A Critical Review of Gender/Ing in Research}. In \bibinfo{booktitle}{\emph{Proceedings of the 2022 CHI Conference on Human Factors in Computing Systems}} (New Orleans, LA, USA) \emph{(\bibinfo{series}{CHI '22})}. \bibinfo{publisher}{Association for Computing Machinery}, \bibinfo{address}{New York, NY, USA}, Article \bibinfo{articleno}{239}, \bibinfo{numpages}{15}~pages.
\newblock
\showISBNx{9781450391573}
\urldef\tempurl%
\url{https://doi.org/10.1145/3491102.3501996}
\showDOI{\tempurl}


\bibitem[\protect\citeauthoryear{Seaborn, Pennefather, and Kotani}{Seaborn et~al\mbox{.}}{2022}]%
        {seaborn2022exploring}
\bibfield{author}{\bibinfo{person}{Katie Seaborn}, \bibinfo{person}{Peter Pennefather}, {and} \bibinfo{person}{Haruki Kotani}.} \bibinfo{year}{2022}\natexlab{}.
\newblock \showarticletitle{Exploring Gender-Expansive Categorization Options for Robots}. In \bibinfo{booktitle}{\emph{CHI Conference on Human Factors in Computing Systems Extended Abstracts}}. \bibinfo{pages}{1--6}.
\newblock


\bibitem[\protect\citeauthoryear{Seberger, Shklovski, Swiatek, and Patil}{Seberger et~al\mbox{.}}{2022}]%
        {seberger2022still}
\bibfield{author}{\bibinfo{person}{John~S Seberger}, \bibinfo{person}{Irina Shklovski}, \bibinfo{person}{Emily Swiatek}, {and} \bibinfo{person}{Sameer Patil}.} \bibinfo{year}{2022}\natexlab{}.
\newblock \showarticletitle{Still Creepy After All These Years: The Normalization of Affective Discomfort in App Use}. In \bibinfo{booktitle}{\emph{CHI Conference on Human Factors in Computing Systems}}. \bibinfo{pages}{1--19}.
\newblock


\bibitem[\protect\citeauthoryear{Sedgwick}{Sedgwick}{1993}]%
        {sedgwick1993tendencies}
\bibfield{author}{\bibinfo{person}{Eve~Kosofsky Sedgwick}.} \bibinfo{year}{1993}\natexlab{}.
\newblock \bibinfo{booktitle}{\emph{Tendencies}}.
\newblock \bibinfo{publisher}{Duke University Press}.
\newblock


\bibitem[\protect\citeauthoryear{Sengers, McCarthy, and Dourish}{Sengers et~al\mbox{.}}{2006}]%
        {sengers2006reflective}
\bibfield{author}{\bibinfo{person}{Phoebe Sengers}, \bibinfo{person}{John McCarthy}, {and} \bibinfo{person}{Paul Dourish}.} \bibinfo{year}{2006}\natexlab{}.
\newblock \showarticletitle{Reflective HCI: articulating an agenda for critical practice}. In \bibinfo{booktitle}{\emph{CHI'06 extended abstracts on Human factors in computing systems}}. \bibinfo{pages}{1683--1686}.
\newblock


\bibitem[\protect\citeauthoryear{Shami, Yang, Panc, Dugan, Ratchford, Rasmussen, Assogba, Steier, Soule, Lupushor, et~al\mbox{.}}{Shami et~al\mbox{.}}{2014}]%
        {shami2014understanding}
\bibfield{author}{\bibinfo{person}{N~Sadat Shami}, \bibinfo{person}{Jiang Yang}, \bibinfo{person}{Laura Panc}, \bibinfo{person}{Casey Dugan}, \bibinfo{person}{Tristan Ratchford}, \bibinfo{person}{Jamie~C Rasmussen}, \bibinfo{person}{Yannick~M Assogba}, \bibinfo{person}{Tal Steier}, \bibinfo{person}{Todd Soule}, \bibinfo{person}{Stela Lupushor}, {et~al\mbox{.}}} \bibinfo{year}{2014}\natexlab{}.
\newblock \showarticletitle{Understanding employee social media chatter with enterprise social pulse}. In \bibinfo{booktitle}{\emph{Proceedings of the 17th ACM conference on Computer supported cooperative work \& social computing}}. \bibinfo{pages}{379--392}.
\newblock


\bibitem[\protect\citeauthoryear{Shamseer, Moher, Clarke, Ghersi, Liberati, Petticrew, Shekelle, and Stewart}{Shamseer et~al\mbox{.}}{2015}]%
        {shamseer2015preferred}
\bibfield{author}{\bibinfo{person}{Larissa Shamseer}, \bibinfo{person}{David Moher}, \bibinfo{person}{Mike Clarke}, \bibinfo{person}{Davina Ghersi}, \bibinfo{person}{Alessandro Liberati}, \bibinfo{person}{Mark Petticrew}, \bibinfo{person}{Paul Shekelle}, {and} \bibinfo{person}{Lesley~A Stewart}.} \bibinfo{year}{2015}\natexlab{}.
\newblock \showarticletitle{Preferred reporting items for systematic review and meta-analysis protocols (PRISMA-P) 2015: elaboration and explanation}.
\newblock \bibinfo{journal}{\emph{Bmj}}  \bibinfo{volume}{349} (\bibinfo{year}{2015}).
\newblock


\bibitem[\protect\citeauthoryear{Shen, DeVos, Eslami, and Holstein}{Shen et~al\mbox{.}}{2021}]%
        {shen2021everyday}
\bibfield{author}{\bibinfo{person}{Hong Shen}, \bibinfo{person}{Alicia DeVos}, \bibinfo{person}{Motahhare Eslami}, {and} \bibinfo{person}{Kenneth Holstein}.} \bibinfo{year}{2021}\natexlab{}.
\newblock \showarticletitle{Everyday algorithm auditing: Understanding the power of everyday users in surfacing harmful algorithmic behaviors}.
\newblock \bibinfo{journal}{\emph{Proceedings of the ACM on Human-Computer Interaction}} \bibinfo{volume}{5}, \bibinfo{number}{CSCW2} (\bibinfo{year}{2021}), \bibinfo{pages}{1--29}.
\newblock


\bibitem[\protect\citeauthoryear{Shklovski and Mainwaring}{Shklovski and Mainwaring}{2005}]%
        {shklovski2005exploring}
\bibfield{author}{\bibinfo{person}{Irina~A Shklovski} {and} \bibinfo{person}{Scott~D Mainwaring}.} \bibinfo{year}{2005}\natexlab{}.
\newblock \showarticletitle{Exploring technology adoption and use through the lens of residential mobility}. In \bibinfo{booktitle}{\emph{Proceedings of the SIGCHI Conference on Human Factors in Computing Systems}}. \bibinfo{pages}{621--630}.
\newblock


\bibitem[\protect\citeauthoryear{Simpson and Semaan}{Simpson and Semaan}{2021}]%
        {simpson2021you}
\bibfield{author}{\bibinfo{person}{Ellen Simpson} {and} \bibinfo{person}{Bryan Semaan}.} \bibinfo{year}{2021}\natexlab{}.
\newblock \showarticletitle{For You, or For" You"? Everyday LGBTQ+ Encounters with TikTok}.
\newblock \bibinfo{journal}{\emph{Proceedings of the ACM on Human-Computer Interaction}} \bibinfo{volume}{4}, \bibinfo{number}{CSCW3} (\bibinfo{year}{2021}), \bibinfo{pages}{1--34}.
\newblock


\bibitem[\protect\citeauthoryear{Skeba and Baumer}{Skeba and Baumer}{2020}]%
        {skeba2020informational}
\bibfield{author}{\bibinfo{person}{Patrick Skeba} {and} \bibinfo{person}{Eric~PS Baumer}.} \bibinfo{year}{2020}\natexlab{}.
\newblock \showarticletitle{Informational Friction as a Lens for Studying Algorithmic Aspects of Privacy}.
\newblock \bibinfo{journal}{\emph{Proceedings of the ACM on Human-Computer Interaction}} \bibinfo{volume}{4}, \bibinfo{number}{CSCW2} (\bibinfo{year}{2020}), \bibinfo{pages}{1--22}.
\newblock


\bibitem[\protect\citeauthoryear{Soden, Ribes, Avle, and Sutherland}{Soden et~al\mbox{.}}{2021}]%
        {soden_historicism}
\bibfield{author}{\bibinfo{person}{Robert Soden}, \bibinfo{person}{David Ribes}, \bibinfo{person}{Seyram Avle}, {and} \bibinfo{person}{Will Sutherland}.} \bibinfo{year}{2021}\natexlab{}.
\newblock \showarticletitle{Time for Historicism in CSCW: An Invitation}.
\newblock \bibinfo{journal}{\emph{Proc. ACM Hum.-Comput. Interact.}} \bibinfo{volume}{5}, \bibinfo{number}{CSCW2}, Article \bibinfo{articleno}{459} (\bibinfo{date}{oct} \bibinfo{year}{2021}), \bibinfo{numpages}{18}~pages.
\newblock
\urldef\tempurl%
\url{https://doi.org/10.1145/3479603}
\showDOI{\tempurl}


\bibitem[\protect\citeauthoryear{Song, Baba, Nakanishi, Yoshikawa, and Ishiguro}{Song et~al\mbox{.}}{2020}]%
        {song2020mind}
\bibfield{author}{\bibinfo{person}{Sichao Song}, \bibinfo{person}{Jun Baba}, \bibinfo{person}{Junya Nakanishi}, \bibinfo{person}{Yuichiro Yoshikawa}, {and} \bibinfo{person}{Hiroshi Ishiguro}.} \bibinfo{year}{2020}\natexlab{}.
\newblock \showarticletitle{Mind The Voice!: Effect of Robot Voice Pitch, Robot Voice Gender, and User Gender on User Perception of Teleoperated Robots}. In \bibinfo{booktitle}{\emph{Extended Abstracts of the 2020 CHI Conference on Human Factors in Computing Systems}}. \bibinfo{pages}{1--8}.
\newblock


\bibitem[\protect\citeauthoryear{Sontag}{Sontag}{2018}]%
        {sontag2018notes}
\bibfield{author}{\bibinfo{person}{Susan Sontag}.} \bibinfo{year}{2018}\natexlab{}.
\newblock \bibinfo{booktitle}{\emph{Notes on camp}}.
\newblock \bibinfo{publisher}{Penguin UK}.
\newblock


\bibitem[\protect\citeauthoryear{Spiel}{Spiel}{2021}]%
        {spiel2021they}
\bibfield{author}{\bibinfo{person}{Katta Spiel}.} \bibinfo{year}{2021}\natexlab{}.
\newblock \showarticletitle{” Why are they all obsessed with Gender?”—(Non) binary Navigations through Technological Infrastructures}. In \bibinfo{booktitle}{\emph{Designing Interactive Systems Conference 2021}}. \bibinfo{pages}{478--494}.
\newblock


\bibitem[\protect\citeauthoryear{Spiel, Haimson, and Lottridge}{Spiel et~al\mbox{.}}{2019a}]%
        {spiel2019better}
\bibfield{author}{\bibinfo{person}{Katta Spiel}, \bibinfo{person}{Oliver~L Haimson}, {and} \bibinfo{person}{Danielle Lottridge}.} \bibinfo{year}{2019}\natexlab{a}.
\newblock \showarticletitle{How to do better with gender on surveys: a guide for HCI researchers}.
\newblock \bibinfo{journal}{\emph{Interactions}} \bibinfo{volume}{26}, \bibinfo{number}{4} (\bibinfo{year}{2019}), \bibinfo{pages}{62--65}.
\newblock


\bibitem[\protect\citeauthoryear{Spiel, Keyes, and Barlas}{Spiel et~al\mbox{.}}{2019b}]%
        {spiel2019patching}
\bibfield{author}{\bibinfo{person}{Katta Spiel}, \bibinfo{person}{Os Keyes}, {and} \bibinfo{person}{P{\i}nar Barlas}.} \bibinfo{year}{2019}\natexlab{b}.
\newblock \showarticletitle{Patching gender: Non-binary utopias in HCI}. In \bibinfo{booktitle}{\emph{Extended Abstracts of the 2019 CHI Conference on Human Factors in Computing Systems}}. \bibinfo{pages}{1--11}.
\newblock


\bibitem[\protect\citeauthoryear{Spiel, Keyes, Walker, DeVito, Birnholtz, Brul{\'e}, Light, Barlas, Hardy, Ahmed, et~al\mbox{.}}{Spiel et~al\mbox{.}}{2019c}]%
        {spiel2019queer}
\bibfield{author}{\bibinfo{person}{Katta Spiel}, \bibinfo{person}{Os Keyes}, \bibinfo{person}{Ashley~Marie Walker}, \bibinfo{person}{Michael~Ann DeVito}, \bibinfo{person}{Jeremy Birnholtz}, \bibinfo{person}{Emeline Brul{\'e}}, \bibinfo{person}{Ann Light}, \bibinfo{person}{P{\i}nar Barlas}, \bibinfo{person}{Jean Hardy}, \bibinfo{person}{Alex Ahmed}, {et~al\mbox{.}}} \bibinfo{year}{2019}\natexlab{c}.
\newblock \showarticletitle{Queer (ing) HCI: Moving forward in theory and practice}. In \bibinfo{booktitle}{\emph{Extended Abstracts of the 2019 CHI Conference on Human Factors in Computing Systems}}. \bibinfo{pages}{1--4}.
\newblock


\bibitem[\protect\citeauthoryear{Starks, Dillahunt, and Haimson}{Starks et~al\mbox{.}}{2019}]%
        {starks2019designing}
\bibfield{author}{\bibinfo{person}{Denny~L Starks}, \bibinfo{person}{Tawanna Dillahunt}, {and} \bibinfo{person}{Oliver~L Haimson}.} \bibinfo{year}{2019}\natexlab{}.
\newblock \showarticletitle{Designing technology to support safety for transgender women \& non-binary people of color}. In \bibinfo{booktitle}{\emph{Companion Publication of the 2019 on Designing Interactive Systems Conference 2019 Companion}}. \bibinfo{pages}{289--294}.
\newblock


\bibitem[\protect\citeauthoryear{State and Adamic}{State and Adamic}{2015}]%
        {state2015diffusion}
\bibfield{author}{\bibinfo{person}{Bogdan State} {and} \bibinfo{person}{Lada Adamic}.} \bibinfo{year}{2015}\natexlab{}.
\newblock \showarticletitle{The diffusion of support in an online social movement: Evidence from the adoption of equal-sign profile pictures}. In \bibinfo{booktitle}{\emph{Proceedings of the 18th ACM Conference on Computer Supported Cooperative Work \& Social Computing}}. \bibinfo{pages}{1741--1750}.
\newblock


\bibitem[\protect\citeauthoryear{Suchman}{Suchman}{1987}]%
        {suchman1987plans}
\bibfield{author}{\bibinfo{person}{Lucy~A Suchman}.} \bibinfo{year}{1987}\natexlab{}.
\newblock \bibinfo{booktitle}{\emph{Plans and situated actions: The problem of human-machine communication}}.
\newblock \bibinfo{publisher}{Cambridge university press}.
\newblock


\bibitem[\protect\citeauthoryear{Taylor, Hutson, and Alicea}{Taylor et~al\mbox{.}}{2017}]%
        {taylor2017social}
\bibfield{author}{\bibinfo{person}{Samuel~Hardman Taylor}, \bibinfo{person}{Jevan~Alexander Hutson}, {and} \bibinfo{person}{Tyler~Richard Alicea}.} \bibinfo{year}{2017}\natexlab{}.
\newblock \showarticletitle{Social consequences of Grindr use: Extending the internet-enhanced self-disclosure hypothesis}. In \bibinfo{booktitle}{\emph{Proceedings of the 2017 CHI Conference on Human Factors in Computing Systems}}. \bibinfo{pages}{6645--6657}.
\newblock


\bibitem[\protect\citeauthoryear{Terveen, Hill, and Amento}{Terveen et~al\mbox{.}}{1999}]%
        {terveen1999constructing}
\bibfield{author}{\bibinfo{person}{Loren Terveen}, \bibinfo{person}{Will Hill}, {and} \bibinfo{person}{Brian Amento}.} \bibinfo{year}{1999}\natexlab{}.
\newblock \showarticletitle{Constructing, organizing, and visualizing collections of topically related web resources}.
\newblock \bibinfo{journal}{\emph{ACM Transactions on Computer-Human Interaction (TOCHI)}} \bibinfo{volume}{6}, \bibinfo{number}{1} (\bibinfo{year}{1999}), \bibinfo{pages}{67--94}.
\newblock


\bibitem[\protect\citeauthoryear{To, Smith, Showkat, Adjagbodjou, and Harrington}{To et~al\mbox{.}}{2023}]%
        {to_everyday_dis_2023}
\bibfield{author}{\bibinfo{person}{Alexandra To}, \bibinfo{person}{Angela D.~R. Smith}, \bibinfo{person}{Dilruba Showkat}, \bibinfo{person}{Adinawa Adjagbodjou}, {and} \bibinfo{person}{Christina Harrington}.} \bibinfo{year}{2023}\natexlab{}.
\newblock \showarticletitle{Flourishing in the Everyday: Moving Beyond Damage-Centered Design in HCI for BIPOC Communities}. In \bibinfo{booktitle}{\emph{Proceedings of the 2023 ACM Designing Interactive Systems Conference}} (Pittsburgh, PA, USA) \emph{(\bibinfo{series}{DIS '23})}. \bibinfo{publisher}{Association for Computing Machinery}, \bibinfo{address}{New York, NY, USA}, \bibinfo{pages}{917–933}.
\newblock
\showISBNx{9781450398930}
\urldef\tempurl%
\url{https://doi.org/10.1145/3563657.3596057}
\showDOI{\tempurl}


\bibitem[\protect\citeauthoryear{Tseng, Freed, Engel, Ristenpart, and Dell}{Tseng et~al\mbox{.}}{2021}]%
        {tseng2021digital}
\bibfield{author}{\bibinfo{person}{Emily Tseng}, \bibinfo{person}{Diana Freed}, \bibinfo{person}{Kristen Engel}, \bibinfo{person}{Thomas Ristenpart}, {and} \bibinfo{person}{Nicola Dell}.} \bibinfo{year}{2021}\natexlab{}.
\newblock \showarticletitle{A digital safety dilemma: Analysis of computer-mediated computer security interventions for intimate partner violence during COVID-19}. In \bibinfo{booktitle}{\emph{Proceedings of the 2021 CHI Conference on Human Factors in Computing Systems}}. \bibinfo{pages}{1--17}.
\newblock


\bibitem[\protect\citeauthoryear{Tseng, Sabet, Bellini, Sodhi, Ristenpart, and Dell}{Tseng et~al\mbox{.}}{2022}]%
        {tseng2022care}
\bibfield{author}{\bibinfo{person}{Emily Tseng}, \bibinfo{person}{Mehrnaz Sabet}, \bibinfo{person}{Rosanna Bellini}, \bibinfo{person}{Harkiran~Kaur Sodhi}, \bibinfo{person}{Thomas Ristenpart}, {and} \bibinfo{person}{Nicola Dell}.} \bibinfo{year}{2022}\natexlab{}.
\newblock \showarticletitle{Care Infrastructures for Digital Security in Intimate Partner Violence}. In \bibinfo{booktitle}{\emph{CHI Conference on Human Factors in Computing Systems}}. \bibinfo{pages}{1--20}.
\newblock


\bibitem[\protect\citeauthoryear{Tuck and Yang}{Tuck and Yang}{2014}]%
        {tuck2014refusingresearch}
\bibfield{author}{\bibinfo{person}{Eve Tuck} {and} \bibinfo{person}{K~Wayne Yang}.} \bibinfo{year}{2014}\natexlab{}.
\newblock \showarticletitle{R-words: Refusing research}.
\newblock \bibinfo{journal}{\emph{Humanizing research: Decolonizing qualitative inquiry with youth and communities}}  \bibinfo{volume}{223} (\bibinfo{year}{2014}), \bibinfo{pages}{248}.
\newblock


\bibitem[\protect\citeauthoryear{Uttarapong, Bonifacio, Jereza, and Wohn}{Uttarapong et~al\mbox{.}}{2022}]%
        {uttarapong2022social}
\bibfield{author}{\bibinfo{person}{Jirassaya Uttarapong}, \bibinfo{person}{Ross Bonifacio}, \bibinfo{person}{Rae Jereza}, {and} \bibinfo{person}{Donghee~Yvette Wohn}.} \bibinfo{year}{2022}\natexlab{}.
\newblock \showarticletitle{Social Support in Digital Patronage: OnlyFans Adult Content Creators as an Online Community}. In \bibinfo{booktitle}{\emph{CHI Conference on Human Factors in Computing Systems Extended Abstracts}}. \bibinfo{pages}{1--7}.
\newblock


\bibitem[\protect\citeauthoryear{Vaccaro, Xiao, Hamilton, and Karahalios}{Vaccaro et~al\mbox{.}}{2021}]%
        {vaccaro2021contestability}
\bibfield{author}{\bibinfo{person}{Kristen Vaccaro}, \bibinfo{person}{Ziang Xiao}, \bibinfo{person}{Kevin Hamilton}, {and} \bibinfo{person}{Karrie Karahalios}.} \bibinfo{year}{2021}\natexlab{}.
\newblock \showarticletitle{Contestability For Content Moderation}.
\newblock \bibinfo{journal}{\emph{Proceedings of the ACM on Human-Computer Interaction}} \bibinfo{volume}{5}, \bibinfo{number}{CSCW2} (\bibinfo{year}{2021}), \bibinfo{pages}{1--28}.
\newblock


\bibitem[\protect\citeauthoryear{Van~Gelder}{Van~Gelder}{[n.d.]}]%
        {van1996strange}
\bibfield{author}{\bibinfo{person}{Lindsy Van~Gelder}.} \bibinfo{year}{[n.d.]}\natexlab{}.
\newblock \bibinfo{title}{The Strange Case of the Electronic Lover.}
\newblock
\newblock


\bibitem[\protect\citeauthoryear{Vashistha, Garg, Anderson, and Raza}{Vashistha et~al\mbox{.}}{2019}]%
        {vashistha2019threats}
\bibfield{author}{\bibinfo{person}{Aditya Vashistha}, \bibinfo{person}{Abhinav Garg}, \bibinfo{person}{Richard Anderson}, {and} \bibinfo{person}{Agha~Ali Raza}.} \bibinfo{year}{2019}\natexlab{}.
\newblock \showarticletitle{Threats, abuses, flirting, and blackmail: Gender inequity in social media voice forums}. In \bibinfo{booktitle}{\emph{Proceedings of the 2019 CHI Conference on Human Factors in Computing Systems}}. \bibinfo{pages}{1--13}.
\newblock


\bibitem[\protect\citeauthoryear{Walker and DeVito}{Walker and DeVito}{2020}]%
        {walker_more_2020}
\bibfield{author}{\bibinfo{person}{Ashley~Marie Walker} {and} \bibinfo{person}{Michael~Ann DeVito}.} \bibinfo{year}{2020}\natexlab{}.
\newblock \showarticletitle{"'{More} gay' fits in better": {Intracommunity} {Power} {Dynamics} and {Harms} in {Online} {LGBTQ}+ {Spaces}}. In \bibinfo{booktitle}{\emph{Proceedings of the 2020 {CHI} {Conference} on {Human} {Factors} in {Computing} {Systems}}}. \bibinfo{publisher}{ACM}, \bibinfo{address}{Honolulu HI USA}, \bibinfo{pages}{1--15}.
\newblock
\showISBNx{978-1-4503-6708-0}
\urldef\tempurl%
\url{https://doi.org/10.1145/3313831.3376497}
\showDOI{\tempurl}


\bibitem[\protect\citeauthoryear{Wang, Chen, Thirunarayan, and Sheth}{Wang et~al\mbox{.}}{2014}]%
        {wang2014cursing}
\bibfield{author}{\bibinfo{person}{Wenbo Wang}, \bibinfo{person}{Lu Chen}, \bibinfo{person}{Krishnaprasad Thirunarayan}, {and} \bibinfo{person}{Amit~P Sheth}.} \bibinfo{year}{2014}\natexlab{}.
\newblock \showarticletitle{Cursing in english on twitter}. In \bibinfo{booktitle}{\emph{Proceedings of the 17th ACM conference on Computer supported cooperative work \& social computing}}. \bibinfo{pages}{415--425}.
\newblock


\bibitem[\protect\citeauthoryear{Wang, Lu, and Wattenhofer}{Wang et~al\mbox{.}}{2022}]%
        {Wang2022GayDating}
\bibfield{author}{\bibinfo{person}{Ye Wang}, \bibinfo{person}{Zhicong Lu}, {and} \bibinfo{person}{Roger Wattenhofer}.} \bibinfo{year}{2022}\natexlab{}.
\newblock \showarticletitle{Gay Dating on Non-Dating Platforms: The Case of Online Dating Activities of Gay Men on a Q\&A Platform}.
\newblock \bibinfo{journal}{\emph{Proc. ACM Hum.-Comput. Interact.}} \bibinfo{volume}{6}, \bibinfo{number}{CSCW2}, Article \bibinfo{articleno}{298} (\bibinfo{date}{nov} \bibinfo{year}{2022}), \bibinfo{numpages}{23}~pages.
\newblock
\urldef\tempurl%
\url{https://doi.org/10.1145/3555189}
\showDOI{\tempurl}


\bibitem[\protect\citeauthoryear{Wang and Mark}{Wang and Mark}{2017}]%
        {wang2017engaging}
\bibfield{author}{\bibinfo{person}{Yiran Wang} {and} \bibinfo{person}{Gloria Mark}.} \bibinfo{year}{2017}\natexlab{}.
\newblock \showarticletitle{Engaging with political and social issues on Facebook in college life}. In \bibinfo{booktitle}{\emph{Proceedings of the 2017 ACM Conference on Computer Supported Cooperative Work and Social Computing}}. \bibinfo{pages}{433--445}.
\newblock


\bibitem[\protect\citeauthoryear{Warner, Gutmann, Sasse, and Blandford}{Warner et~al\mbox{.}}{2018}]%
        {warner2018privacy}
\bibfield{author}{\bibinfo{person}{Mark Warner}, \bibinfo{person}{Andreas Gutmann}, \bibinfo{person}{M~Angela Sasse}, {and} \bibinfo{person}{Ann Blandford}.} \bibinfo{year}{2018}\natexlab{}.
\newblock \showarticletitle{Privacy unraveling around explicit HIV status disclosure fields in the online geosocial hookup app Grindr}.
\newblock \bibinfo{journal}{\emph{Proceedings of the ACM on human-computer interaction}} \bibinfo{volume}{2}, \bibinfo{number}{CSCW} (\bibinfo{year}{2018}), \bibinfo{pages}{1--22}.
\newblock


\bibitem[\protect\citeauthoryear{Warner, Kitkowska, Gibbs, Maestre, and Blandford}{Warner et~al\mbox{.}}{2020}]%
        {warner2020evaluating}
\bibfield{author}{\bibinfo{person}{Mark Warner}, \bibinfo{person}{Agnieszka Kitkowska}, \bibinfo{person}{Jo Gibbs}, \bibinfo{person}{Juan~F Maestre}, {and} \bibinfo{person}{Ann Blandford}.} \bibinfo{year}{2020}\natexlab{}.
\newblock \showarticletitle{Evaluating'Prefer not to say'Around Sensitive Disclosures}. In \bibinfo{booktitle}{\emph{Proceedings of the 2020 CHI Conference on Human Factors in Computing Systems}}. \bibinfo{pages}{1--13}.
\newblock


\bibitem[\protect\citeauthoryear{Warner, Maestre, Gibbs, Chung, and Blandford}{Warner et~al\mbox{.}}{2019}]%
        {warner2019signal}
\bibfield{author}{\bibinfo{person}{Mark Warner}, \bibinfo{person}{Juan~F Maestre}, \bibinfo{person}{Jo Gibbs}, \bibinfo{person}{Chia-Fang Chung}, {and} \bibinfo{person}{Ann Blandford}.} \bibinfo{year}{2019}\natexlab{}.
\newblock \showarticletitle{Signal appropriation of explicit HIV status disclosure fields in sex-social apps used by gay and bisexual men}. In \bibinfo{booktitle}{\emph{Proceedings of the 2019 CHI Conference on Human Factors in Computing Systems}}. \bibinfo{pages}{1--15}.
\newblock


\bibitem[\protect\citeauthoryear{Williamson, Glassner, McLaughlin, Chase, and Smith}{Williamson et~al\mbox{.}}{1998}]%
        {williamson1998constructing}
\bibfield{author}{\bibinfo{person}{Mary~B. Williamson}, \bibinfo{person}{Andrew Glassner}, \bibinfo{person}{Margaret McLaughlin}, \bibinfo{person}{Cheryl Chase}, {and} \bibinfo{person}{Marc Smith}.} \bibinfo{year}{1998}\natexlab{}.
\newblock \showarticletitle{Constructing Community in Cyberspace}. In \bibinfo{booktitle}{\emph{CHI 98 Conference Summary on Human Factors in Computing Systems}} (Los Angeles, California, USA) \emph{(\bibinfo{series}{CHI '98})}. \bibinfo{publisher}{Association for Computing Machinery}, \bibinfo{address}{New York, NY, USA}, \bibinfo{pages}{84–85}.
\newblock
\showISBNx{1581130287}
\urldef\tempurl%
\url{https://doi.org/10.1145/286498.286541}
\showDOI{\tempurl}


\bibitem[\protect\citeauthoryear{Wisniewski, Xu, Rosson, and Carroll}{Wisniewski et~al\mbox{.}}{2014}]%
        {wisniewski2014adolescent}
\bibfield{author}{\bibinfo{person}{Pamela~J Wisniewski}, \bibinfo{person}{Heng Xu}, \bibinfo{person}{Mary~Beth Rosson}, {and} \bibinfo{person}{John~M Carroll}.} \bibinfo{year}{2014}\natexlab{}.
\newblock \showarticletitle{Adolescent online safety: the" moral" of the story}. In \bibinfo{booktitle}{\emph{Proceedings of the 17th ACM conference on Computer supported cooperative work \& social computing}}. \bibinfo{pages}{1258--1271}.
\newblock


\bibitem[\protect\citeauthoryear{Wolfswinkel, Furtmueller, and Wilderom}{Wolfswinkel et~al\mbox{.}}{2013}]%
        {wolfswinkel2013using}
\bibfield{author}{\bibinfo{person}{Joost~F Wolfswinkel}, \bibinfo{person}{Elfi Furtmueller}, {and} \bibinfo{person}{Celeste~PM Wilderom}.} \bibinfo{year}{2013}\natexlab{}.
\newblock \showarticletitle{Using grounded theory as a method for rigorously reviewing literature}.
\newblock \bibinfo{journal}{\emph{European journal of information systems}} \bibinfo{volume}{22}, \bibinfo{number}{1} (\bibinfo{year}{2013}), \bibinfo{pages}{45--55}.
\newblock


\bibitem[\protect\citeauthoryear{Wong-Villacres, DiSalvo, Kumar, and DiSalvo}{Wong-Villacres et~al\mbox{.}}{2020}]%
        {wong2020culture}
\bibfield{author}{\bibinfo{person}{Marisol Wong-Villacres}, \bibinfo{person}{Carl DiSalvo}, \bibinfo{person}{Neha Kumar}, {and} \bibinfo{person}{Betsy DiSalvo}.} \bibinfo{year}{2020}\natexlab{}.
\newblock \showarticletitle{Culture in Action: Unpacking Capacities to Inform Assets-Based Design}. In \bibinfo{booktitle}{\emph{Proceedings of the 2020 CHI Conference on Human Factors in Computing Systems}}. \bibinfo{pages}{1--14}.
\newblock


\bibitem[\protect\citeauthoryear{Wood, Long, Feltwell, Rowland, Brooker, Mahoney, Vines, Barnett, and Lawson}{Wood et~al\mbox{.}}{2018}]%
        {wood2018rethinking}
\bibfield{author}{\bibinfo{person}{Gavin Wood}, \bibinfo{person}{Kiel Long}, \bibinfo{person}{Tom Feltwell}, \bibinfo{person}{Scarlett Rowland}, \bibinfo{person}{Phillip Brooker}, \bibinfo{person}{Jamie Mahoney}, \bibinfo{person}{John Vines}, \bibinfo{person}{Julie Barnett}, {and} \bibinfo{person}{Shaun Lawson}.} \bibinfo{year}{2018}\natexlab{}.
\newblock \showarticletitle{Rethinking engagement with online news through social and visual co-annotation}. In \bibinfo{booktitle}{\emph{Proceedings of the 2018 CHI conference on human factors in computing systems}}. \bibinfo{pages}{1--12}.
\newblock


\bibitem[\protect\citeauthoryear{Wu, Edwards, and Das}{Wu et~al\mbox{.}}{2022}]%
        {wu2022reasonable}
\bibfield{author}{\bibinfo{person}{Yuxi Wu}, \bibinfo{person}{W~Keith Edwards}, {and} \bibinfo{person}{Sauvik Das}.} \bibinfo{year}{2022}\natexlab{}.
\newblock \showarticletitle{“A Reasonable Thing to Ask For”: Towards a Unified Voice in Privacy Collective Action}. In \bibinfo{booktitle}{\emph{CHI Conference on Human Factors in Computing Systems}}. \bibinfo{pages}{1--17}.
\newblock


\bibitem[\protect\citeauthoryear{Wyche, Schoenebeck, and Forte}{Wyche et~al\mbox{.}}{2013}]%
        {wyche2013facebook}
\bibfield{author}{\bibinfo{person}{Susan~P Wyche}, \bibinfo{person}{Sarita~Yardi Schoenebeck}, {and} \bibinfo{person}{Andrea Forte}.} \bibinfo{year}{2013}\natexlab{}.
\newblock \showarticletitle{" Facebook is a luxury" an exploratory study of social media use in rural Kenya}. In \bibinfo{booktitle}{\emph{Proceedings of the 2013 conference on Computer supported cooperative work}}. \bibinfo{pages}{33--44}.
\newblock


\bibitem[\protect\citeauthoryear{Yee, Tantipongpipat, and Mishra}{Yee et~al\mbox{.}}{2021}]%
        {yee2021image}
\bibfield{author}{\bibinfo{person}{Kyra Yee}, \bibinfo{person}{Uthaipon Tantipongpipat}, {and} \bibinfo{person}{Shubhanshu Mishra}.} \bibinfo{year}{2021}\natexlab{}.
\newblock \showarticletitle{Image cropping on twitter: Fairness metrics, their limitations, and the importance of representation, design, and agency}.
\newblock \bibinfo{journal}{\emph{Proceedings of the ACM on Human-Computer Interaction}} \bibinfo{volume}{5}, \bibinfo{number}{CSCW2} (\bibinfo{year}{2021}), \bibinfo{pages}{1--24}.
\newblock


\bibitem[\protect\citeauthoryear{Yoshino}{Yoshino}{2015}]%
        {yoshino2015new}
\bibfield{author}{\bibinfo{person}{Kenji Yoshino}.} \bibinfo{year}{2015}\natexlab{}.
\newblock \showarticletitle{A New Birth of Freedom?: Obergefell v. Hodges}.
\newblock \bibinfo{journal}{\emph{Harv. L. Rev.}}  \bibinfo{volume}{129} (\bibinfo{year}{2015}), \bibinfo{pages}{147}.
\newblock


\bibitem[\protect\citeauthoryear{Zhang and Counts}{Zhang and Counts}{2015}]%
        {zhang2015modeling}
\bibfield{author}{\bibinfo{person}{Amy~X Zhang} {and} \bibinfo{person}{Scott Counts}.} \bibinfo{year}{2015}\natexlab{}.
\newblock \showarticletitle{Modeling ideology and predicting policy change with social media: Case of same-sex marriage}. In \bibinfo{booktitle}{\emph{Proceedings of the 33rd Annual ACM Conference on Human Factors in Computing Systems}}. \bibinfo{pages}{2603--2612}.
\newblock


\bibitem[\protect\citeauthoryear{Zytko, Furlo, Carlin, and Archer}{Zytko et~al\mbox{.}}{2021}]%
        {zytko2021computer}
\bibfield{author}{\bibinfo{person}{Douglas Zytko}, \bibinfo{person}{Nicholas Furlo}, \bibinfo{person}{Bailey Carlin}, {and} \bibinfo{person}{Matthew Archer}.} \bibinfo{year}{2021}\natexlab{}.
\newblock \showarticletitle{Computer-mediated consent to sex: the context of Tinder}.
\newblock \bibinfo{journal}{\emph{Proceedings of the ACM on Human-Computer Interaction}} \bibinfo{volume}{5}, \bibinfo{number}{CSCW1} (\bibinfo{year}{2021}), \bibinfo{pages}{1--26}.
\newblock


\end{thebibliography}

 \clearpage

\appendix

\section{Codebook}

\begin{table}[h]
    
    \renewcommand{\arraystretch}{1.5}
    \begin{tabular}{| m{1cm} | m{3.5cm} | m{10cm} | }
        \hline
        \centering Code & 
        \centering Description & 
        \centering Examples \& Non-Examples \tabularnewline \hline
        
        \centering 4 & 
        \textbf{Exclusively involves} queer people & 
        
        Examples
        
        \begin{itemize}
            \setlength\itemsep{0.10em}
            \item All or nearly all participants are queer
            \item HIV paper with only queer people
            \item Entire workshop about queer issues
            \item Queer autoethnography
        \end{itemize} 
        \\ \hline
        
        \centering 3 & 
        \textbf{Significantly involves} queer people or issues but not the center of the research & 
        
        Examples
        
        \begin{itemize}
            \setlength\itemsep{0.10em}
            \item Queer people as one of multiple distinct case studies
            \item Empirical study with a substantial number of queer participants
            \item Entire paper built on queer rights as controversial social issue
            \item Panelist substantially mentions queer issues but isn't the entire panel
        \end{itemize} 
        \\ \hline
        
        \centering 2 & 
        \textbf{Discusses} queer people in some way even if not a primary part of the work & 
        
        Examples

            \begin{itemize}
                \setlength\itemsep{0.10em}
                \item Frames queer people in some way (e.g., marginalized or controversial)
                \item Positionality statement
                \item Queer word in data contributes to a frame (\textit{e.g.,} lesbian as sexual word)
            \end{itemize} 
        
        \vspace{0.10em}
        
        Non-Examples

            \begin{itemize}
                \setlength\itemsep{0.10em}
                \item 3: Whole paper relies on something like queer as controversy
                \item 1: Queer word shows up in dataset that doesn't contribute a frame
            \end{itemize} 
        
        \\ \hline
        
        \centering 1 & 
        \textbf{Briefly mentions} queer people & 
        
        Examples
        
        \begin{itemize}
            \setlength\itemsep{0.10em}
            \item Participant demographic listed but no discussion
            \item The word “gay” in a table with no discussion
        \end{itemize} 
        \\ \hline
        
        \centering 0 & 
        \vfill No Mention & 
    
        Examples 
        \vspace{0.10em}
        \begin{itemize}
            \setlength\itemsep{0.10em}
            \item Papers written by an author with "gay" in their name
            
            \item Non-binary classifier
        \end{itemize} 
        \\ \hline
        
    \end{tabular}
    \caption{Scale used to annotate papers in our dataset for the degree to which they involved LGBTQ+ people}
    \Description[A codebook with codes, descriptions of each code, and examples and non-examples.]{A codebook with codes, descriptions of each code, and examples and non-examples.}{}
    \label{tab:codebook}
\end{table}

\end{document}